\documentclass[11pt,a4paper]{article}
\pdfoutput=1
\usepackage{LMstyle}

\usepackage[]{amsmath}
	\numberwithin{equation}{section}
\usepackage[]{amssymb}
\usepackage{amsfonts}
\usepackage{mathrsfs}

\DeclareMathOperator{\R}{\mathbb{R}}
\DeclareMathOperator{\C}{\mathbb{C}}

\DeclareMathOperator{\cs}{\mathbb{S}}

\newcommand{\mz}{\mathcal{Z}}

\newcommand{\mN}{\mathcal{N}}
\newcommand{\mI}{\mathcal{I}}

\usepackage[utf8]{inputenc}
\usepackage[english]{babel}
\usepackage{enumerate}

\usepackage[toc,page]{appendix}

\usepackage{ytableau}
\usepackage{tikz}
		\usetikzlibrary{er,positioning,arrows.meta}
		\usetikzlibrary{calc}
		\usetikzlibrary{shapes.geometric}

	\title{Exact results and Schur expansions in quiver Chern--Simons-matter theories}
	
	\author[a]{Leonardo Santilli}
	\emailAdd{lsantilli@fc.ul.pt}
	
	\author[b,a]{and Miguel Tierz}
	\emailAdd{tierz@fc.ul.pt}
	
	\affiliation[a]{Grupo de F\'{i}sica Matem\'{a}tica, Departamento de Matem\'{a}tica, Faculdade de Ci\^{e}ncias, Universidade de Lisboa, Campo Grande, Edif\'{i}cio C6, 1749-016 Lisboa, Portugal.}
	
\affiliation[b]{Departamento de Matem\'{a}tica, Faculdade de Ci\^{e}ncias, ISCTE - Instituto Universit\'{a}rio de Lisboa, Avenida das For\c{c}as Armadas, 1649-026 Lisboa, Portugal.}

	\abstract{We study several quiver Chern--Simons-matter theories on the three-sphere, combining the matrix model formulation with a systematic use of Mordell's integral, computing partition functions and checking dualities. We also consider Wilson loops in ABJ(M) theories, distinguishing between typical (long) and atypical (short) representations and focusing on the former. Using the Berele--Regev factorization of supersymmetric Schur polynomials, we express the expectation value of the Wilson loops in terms of sums of observables of two factorized copies of $U(N)$ pure Chern--Simons theory on the sphere. Then, we use the Cauchy identity to study the partition functions of a number of quiver Chern--Simons-matter models and the result is interpreted as a perturbative expansion in the parameters  $t_j = - e^{2 \pi m_j}$, where $m_j$ are the masses. Through the paper, we incorporate different generalizations, such as deformations by real masses and/or Fayet--Iliopoulos parameters, the consideration of a Romans mass in the gravity dual, and adjoint matter.
}

\begin{document}

\maketitle

\clearpage		

	\section{Introduction}

	In the last decade, the combined use of random matrix techniques together with the application of the supersymmetric localization method \cite{Pestun:2007rz} has produced a wealth of analytical results in the study of supersymmetric gauge theories on compact manifolds, in a number of dimensions \cite{Pestun:2016zxk}. Both finite $N$ properties and large $N$ phenomena such as phase transitions have been elucidated by applying standard matrix model tools.
	
	A very tractable set of theories in this area corresponds to Chern--Simons theories with supersymmetric matter in three dimensions \cite{KapWilYaak} (see \cite{WillettReview} for an overview on localization in three dimensions and \cite{MarinoReview} for an early review of Chern--Simons-matter matrix models). While a large number of results have already been uncovered for these models, we develop here further exact analytical characterizations of such theories, using the matrix model formulation. For this, we will be supplementing the matrix model approach with other analytical tools, such as the consideration of the so-called Mordell integral \cite{Mordell} which, in spite of its deceptively simple appearance, unpacks a wealth of analytical and physical information.\par
	
	In Section \ref{S2} we will be presenting the necessary details, not only on Mordell's integral but on the other mathematical tools used. This Section will provide physics background as well and hence can also be used as Introduction, while we sketch now below the results and methods followed in a more qualitative and panoramic manner.\par
	
	In contrast to previous works following a similar approach to the one in our first part of the paper \cite{TRS,TGias,RussoSchaposnik}, our study will include quiver Chern--Simons-matter theories. In this way, in this first part, contained in Section \ref{sec:exactZ}, we compute exactly the partition functions of various examples of Chern--Simons-matter theories on the three-sphere, systematically exploiting and interpreting the above mentioned result by Mordell \cite{Mordell}. The theories to be studied will be mostly Abelian quiver models whose computation is nevertheless laborious, but made possible by Mordell's result.\par
	
	Some of these evaluations are actually duality checks. For example, we compute explicitly the partition function of the $U(1)^3$ theory, which, once particularized to Chern--Simons levels $(k_1,k_2,k_3) =(1,-1,1)$, becomes the so-called Model III of Jafferis and Yin \cite{Jafferis:2008em}, and we obtain, as expected by duality, the equality with the simpler to compute, and known, partition function of SQED with two fundamental hypermultiplets and no Chern--Simons coupling.\par
		Non-Abelian quiver theories, with the $U(1)_{k_1} \times U(2)_{k_2}$ theory as main example, are also briefly discussed and moreover we present the setup and sufficient conditions to evaluate the partition function of Abelian linear Chern--Simons quivers of arbitrary rank by iterative application of Mordell integrals.\par
		
		We shall also be studying ABJ theories \cite{ABJ}, that are $\mN=6$ $U(N_1)_k \times U(N_2)_{-k}$ Chern--Simons theories with two bi-fundamental hypermultiplets. As it is well-known, they generalize ABJM theory \cite{ABJM}, which is recovered when $N_1=N_2 =:N$.
	There are exact computations of observables in ABJM theory when $N=2$ in \cite{Okuyama}, and in \cite{RussoSilva} when mass and Fayet--Iliopoulos deformations are turned on. Besides, the partition function of ABJ theory with arbitrary rank has been evaluated in \cite{ABJexact} using the continuation from Chern--Simons theory on the lens space $L(2,1)$ as introduced in \cite{MaPutExact}. In \cite{HondaABJ}, the result was confirmed through a direct integral transformation. \par
	
	Thus, in Subsection \ref{sec:lnearquivers}, we complement these works by extending this type of analytical evaluations. We will give mass to the bi-fundamental hypermultiplets and add a Fayet--Iliopoulos parameter, and consider the deformed Abelian ABJM theory with Chern--Simons levels $k_1$ and $k_2$, reflecting the presence of a Romans mass in the dual gravitational theory \cite{GaiTom}.\par

	Then, in Section \ref{sec:WLABJ}, we will focus our attention on the vacuum expectation values of correlators of Wilson loops in ABJ(M) theory on $\mathbb{S}^3$. Among the various order loop operators in ABJ(M) theories \cite{RoadmapWL}, we will consider $\frac{1}{2}$-BPS Wilson loops \cite{DrukTranca}, whose expectation value is captured by a matrix model that corresponds to the insertion of  supersymmetric Schur polynomials in the ABJ(M) matrix model \cite{DrukTranca,MaPutExact}.
\par
	
	As a novel consideration in the context of the study of such averages, we distinguish between typical (long) and atypical (short) representations and focus on the former, using the so-called Berele--Regev factorization of supersymmetric Schur polynomials \cite{Berele} to give expressions in terms of sums of observables of $U(N)$ Chern--Simons theory on $\mathbb{S}^3$. The necessary background is given in the introductory Subsections \ref{sec:WLsuperalgebras} and \ref{sec:WLdeftypical}.\par

	As a matter of fact, the case of correlators is often simpler, with this approach, than the study of a single Wilson loop average. For example, by considering the case of two Wilson loops, we shall show that a consequence of the Berele--Regev factorization \cite{Berele}, is that the interacting term of the ABJ two-matrix model cancels out directly, and we immediately obtain the direct, disentangled product of two correlators of pairs of Wilson loops in $U(N)$ Chern--Simons on $\mathbb{S}^3$, each one computed independently, giving quantum dimensions. Furthermore, we will show in Subsection \ref{sec:WLnecklace} how this formalism extends to quivers.\par
	
	Then, in Section \ref{sec:Schurexpansion}, we discuss a broad class of quiver Chern--Simons theories and our main tool will be the Cauchy identity. Its use entails expanding the matter contribution in a basis of symmetric functions, the Schur basis.
	As we shall show, these Schur expansions in the matrix model are perturbative evaluations of the observables described by the matrix model representation.\par
	
	The perturbative meaning of the results has its roots in the nature of the Cauchy identity, reviewed in Subsection \ref{sec:Cauchyetc}. Importantly, the results are not perturbative in the gauge couplings, but in certain other variables playing the role of fugacities for the flavour symmetries. In Subsection \ref{sec:1WLSchurExp} we will combine the Cauchy identity with the Berele--Regev factorization and the results of Section \ref{sec:WLABJ} to study the expectation value of a single Wilson loop in ABJ theory.\par
	
	In the Outlook Section, we conclude by discussing possible avenues for further research. Technical details, including non-trivial aspects of the solution by Mordell are presented in Appendix \ref{app:mordellproof}, in pedagogical manner. Additionally, we also present further discussion of aspects of Section \ref{sec:Schurexpansion}. In particular, in Appendix \ref{app:noknot}, we provide commentary regarding eventual comparisons between the expansions obtained and generating function of knot invariants, while also giving explicit Schur expansions for selected quivers in Appendix \ref{app:AbelianquiverSchurexp}.

	\section{Physics background and mathematical setup}
	\label{S2}
	
	\subsection{Chern--Simons theories on $\mathbb{S}^3$}
	\label{sec:CSS3loc}
	We consider Chern--Simons-matter theories with $\mathcal{N} \ge 3$ supersymmetry in three dimensions. These theories are obtained deforming the action of $\mathcal{N}=4$ theories of vector and hypermultiplets by Chern--Simons (CS) couplings that preserve at least six of the eight supercharges. The resulting theories have a $SU(2)_{\text{R}}$ R-symmetry, but when the microscopic, non-conformal theory is put on $\mathbb{S}^3$, only a maximal torus $U(1)_{\mathrm{R}} \subset SU(2)_{\mathrm{R}}$ is manifest. On a practical level, this amount of supersymmetry guarantees that:
	\begin{itemize}
		\item the CS levels are not renormalized, and that
		\item we can identify the R-charges in the UV, where our computations are performed, with the R-charges in the IR, where the theory is strongly coupled.
	\end{itemize}\par
	There exists a vast literature describing the moduli spaces of vacua of these gauge theories, the most directly relevant for the present work being \cite{MartelliSparks:2008,Kimura:moduli,JafferisTom,BCC2}. All the theories we discuss can be engineered in type IIB string theory using, beyond D3, D5 and NS5 branes, also $(p,q)$-branes \cite{Kitao:1998mf,Bergman:1999na}.\par
	\medskip
	We first recall how to write the partition function of a $3d$ $\mN \ge 3$ theory on $\cs^3$ \cite{KapWilYaak}, which also serves as a presentation of our notation and conventions. The Chern--Simons theories we study have unitary gauge groups of the form 
	\begin{equation*}
		U(N_1) \times U(N_2) \times \cdots \times U(N_r) .
	\end{equation*}
	Besides, we mainly consider hypermultiplets in the fundamental representation of a gauge group factor $U(N_p)$, as well as in the bi-fundamental representation of $U(N_p) \times U(N_{p+1})$.\par
	In quiver notation, the number of nodes is $r$, with the node $p$ corresponding to a gauge factor $U(N_p)$, for $p=1, \dots, r$. Unoriented edges between two nodes represent the bi-fundamental hypermultiplets.\par
	The partition function receives the contributions \cite{KapWilYaak}:
	\begin{align*}
		\text{vector multiplet at node $p$:} & \quad \quad \quad \prod_{1 \le a < b \le N_p} \left(  2 \sinh \pi ( x_{p,a} - x_{p,b}  ) \right)^2 , \\
		\text{CS term at node $p$:} & \quad \quad \quad \prod_{a=1} ^{N_p} e^{ i \pi k_p (x_{p,a} )^2   } , \\
		\text{bi-fund. hypers between $p$ and $p^{\prime}$:} & \quad \quad \quad \prod_{a=1}^{N_p} \prod_{a^{\prime} =1}^{N_{p^{\prime}}}  \left(  2 \cosh \pi ( x_{p,a} - x_{p^{\prime}, a^{\prime}} ) \right)^{-1} .
	\end{align*}
	The CS levels are $k_p$, which are required to be integers when $N_p>1$ but can be rational for an Abelian gauge factor, $N_p=1$. The isomorphism $U(N) \simeq  [U(1) \times SU(N)]/\mathbb{Z}_N$ shows that each node yields an Abelian factor, to which there corresponds a topological global $U(1)_{\mathrm{top}}$ symmetry. Real Fayet--Iliopoulos (FI) parameters $\zeta_p$ are introduced as background values of a twisted Abelian vector multiplet for the $U(1)_{\mathrm{top},1} \times \cdots \times U(1)_{\mathrm{top},r}$ symmetry. Furthermore, if we attach $N_{f,p}$ fundamental hypermultiplets to the node $p$, we can turn on real masses $\vec{m} = (m_{p,j})_{j=1, \dots, N_{f,p}} ^{p=1, \dots, r}$ as background values of a vector multiplet for the global symmetry $PS[ U ( N_{f,1} ) \times \cdots \times U ( N_{f,r} ) ]$ rotating the fundamentals. The tracelessness condition constrains the masses $\vert \vec{m} \vert = 0$. The contributions of FI terms and massive hypermultiplets to the partition function are:
	\begin{align*}
		\text{FI term at node $p$:} & \quad \quad \quad \prod_{a=1} ^{N_p} e^{ i 2 \pi \zeta_p x_{p,a} } , \\
		\text{fund. hypers at node $p$:} & \quad \quad \quad \prod_{j=1} ^{N_{f,p}} \prod_{a=1}^{N_p}   \left( 2 \cosh \pi ( x_{p,a} + m_{p,j} )  \right)^{-1} .
	\end{align*}
	Eventually, we have to integrate over all the $x_{p,a}$. These variables parametrize the Cartan subalgebra of the gauge group, 
	\begin{equation*}
		\vec{x} = (x_{p,a}  )^{ p=1, \dots, r} _{a=1, \dots, N_p} \in \mathfrak{u}(1)^{N_1} \times \cdots \times \mathfrak{u}(1)^{N_r} \simeq \R^{N_1 + \cdots + N_r } .
	\end{equation*}
	If we let $R$ be the radius of $\cs^3$, these adimensional variables are $\vec{x}= R \sigma \vert_{\mathrm{loc}}$, where $\sigma \vert_{\mathrm{loc}} $ is the value of the real scalar $\sigma$ in the vector multiplet at the localization locus. The parameters $\vec{m}$ are adimensional as well, $\vec{m}=R \sigma_{\mathrm{b.g.}}$ for $\sigma_{\mathrm{b.g.}}$ the scalar in the background vector multiplet.\par

	\subsubsection{ABJ(M) theories and CS levels}
	\label{sec:GTRomans}
	    
ABJ(M) theories \cite{ABJM,ABJ} are $U(N_1) \times U(N_2)$ CS theories with $\mN=6$ supersymmetry, which forces the CS level to be $(k_1,k_2)=(k,-k)$. In quiver notation, these are extended $\widehat{\mathsf{A}}_1$ quiver theories. They have their origin in string/M-theory, and have been conceived as the gauge theory dual to the $\mN=6$ gravity solution on $\text{AdS}_4 \times \mathbb{CP}^3$.\par
A natural question on the gauge theory side is whether there exists a theory with  generic levels $(k_1,k_2)$. This point has been addressed in the early days of ABJM theory by Gaiotto and Tomasiello in \cite{GaiTom}. It is possible to deform ABJ(M) theories to arbitrary levels deforming the gravity dual solution by a Romans mass, commonly denoted $F_0$. There are different ways to do so \cite{GaiTom}, and we will only consider the $\mN=3$ supersymmetric solution. The resulting gauge theory has the same field content of the ABJ(M) theory, but with CS levels $(k_1,k_2)$ that obey $k_1+k_2=F_0$. For the deformation of other Chern--Simons-matter theories by a Romans mass in the gravity dual, see for example \cite{GuarinoJafferis1,GuarinoTarrio1}.\par
We remark that, by mass deformations, we will \emph{not} refer to a Romans mass, and instead always mean the procedure described above to give mass to the hypermultiplets promoting the masses to background scalar fields.

	\subsubsection{$\frac{1}{2}$-BPS Wilson loops}
		In $\mathcal{N} \ge 3$ supersymmetric Chern--Simons theories, supersymmetry-preserving Wilson loops in a representation $\mathcal{R}$ of the gauge group wrap a great circle in $\cs^3$. Their vacuum expectation value (vev) is computed by localization \cite{KapWilYaak}:
		\begin{equation*}
			\left\langle W_{\mathcal{R}} \right\rangle = \left\langle \mathrm{Tr}_{\mathcal{R}} e^{2 \pi R \sigma\vert_{\mathrm{loc}} } \right\rangle .
		\end{equation*}
		In the formula, $2 \pi R$ is the length of the great circle, $\sigma \vert_{\mathrm{loc}}$ is the value of the real scalar $\sigma$ at the localization locus as explained above, $\mathrm{Tr}_{\mathcal{R}}$ is the normalized trace in the representation $\mathcal{R}$ and $\langle \cdots \rangle$ means the average of the quantity in the ensemble obtained from localization, which of course depends on the theory under study.\par
		In quiver CS theories it is possible to construct Wilson loops charged under a $U(N)$ factor of the gauge group that preserve (at least) two supercharges. For the special case of ABJ theories, however, it is possible to consider Wilson loops in a representation $\mathcal{R}$ of the supergroup $U(N_1 \vert N_2)$ that preserve half of the $\mN=6$ supersymmetry, that is $\frac{1}{2}$-BPS Wilson loops \cite{DrukTranca,Lee:2010hk}.

		\subsubsection{Unknot invariant in pure Chern-Simons theory}
		
			The vev of a Wilson loop in bosonic pure Chern--Simons theory computes the unknot invariant \cite{WittenJones}. It was first evaluated with the CS matrix model in \cite{TDoli}, giving: 
		\begin{equation}
		\label{eq:WLCSk}
			 \langle W_{\mu} \rangle_{\mathrm{CS} (N)}  = (\dim_q \mu) ~q^{- \frac{1}{2} \mathsf{C}_{2;N} (\mu) } .
		\end{equation}
		In the computation in \cite{TDoli} it was shown that the integration of a Schur polynomial in a Stieltjes--Wigert ensemble (and equivalently, in the CS matrix model \cite{T0}) gives the principal specialization of the Schur polynomial, leading to the expression \eqref{eq:WLCSk}. This property has been discussed, later on, in a broader sense in \cite{Mironov:2017och,Mironov:2018ekq} (see also \cite{tierz2020matrix} for a general discussion of two Schur polynomial insertions).\par
	
		In \eqref{eq:WLCSk} the $q$-parameter was taken to be real $0<q=e^{-g} <1$ and is related to the $q$-parameter of CS theory at level $k$ through the analytic continuation 
		\begin{equation}
    	\label{gto2pik}
		g \mapsto \frac{i 2 \pi }{k} ,
    	\end{equation}
		$\dim_q \mu$ is the quantum dimension of the representation $\mu$  and $\mathsf{C}_{2;N} (\mu)$ is the quadratic Casimir of $U(N)$ in the representation $\mu$. Knot and link invariants computed in Chern--Simons theory come equipped with a framing \cite{WittenJones}, and we stress that \eqref{eq:WLCSk} is computed not in canonical framing but in the matrix model framing, which is a specific case of Seifert framing.
	
	\subsection{Mordell integrals}
	
	The two integrals we will exploit are \cite{Mordell}:
	\begin{align}
		\Psi_{+} (\xi, \lambda ; \kappa, \varrho) & := \int_{- \infty } ^{+ \infty} dx \frac{ e^{i \pi \frac{\kappa}{\varrho} x^2 - 2 \pi x \xi }  }{ e^{2 \pi x} - e^{i 2 \pi \lambda } } \notag	 \\
			& = \frac{ e^{ - i \pi \lambda \left(  2 + 2 \xi + \frac{\kappa}{\varrho} \lambda \right)  } }{ e^{ i \pi \varrho \left(  2 \xi + 2 \frac{\kappa}{\varrho} \lambda - \kappa \right) }  - 1  } \left[ - \sqrt{ \frac{ i \varrho}{\kappa} } \sum_{\alpha=1} ^{\kappa} e^{i \pi \frac{\varrho}{\kappa} \left( \xi + \frac{\kappa}{\varrho} \lambda + \alpha \right)^2   }  + i \sum_{\beta=1}^{\varrho} e^{ i \pi \beta \left( 2 \xi + 2 \frac{\kappa}{\varrho} \lambda - \frac{\kappa}{\varrho} \beta  \right)  }  \right]  \label{MordellG+}
	\end{align}
	and
	\begin{align}
		\Psi_{-} (\xi, \lambda ;  \kappa, \varrho  ) & := \int_{- \infty } ^{+ \infty} dx \frac{ e^{- i \pi \frac{\kappa}{\varrho} x^2 - 2 \pi x \xi } }{ e^{ 2 \pi x}  - e^{i 2 \pi \lambda } }	 \notag \\
			& = \frac{ e^{- i \pi \lambda \left( 2 + 2 \xi - \frac{\kappa}{\varrho} \lambda \right)  } }{ e^{i \pi \varrho \left( 2 \xi - 2 \frac{\kappa}{\varrho} \lambda - \kappa \right) } -1  } \left[ \sqrt{- \frac{ i \varrho}{ \kappa} } \sum_{\alpha= 0} ^{\kappa-1} e^{- i \pi \frac{\varrho}{\kappa} \left( \xi - \frac{\kappa}{\varrho} \lambda - \alpha \right)^2   }  + i \sum_{\beta=1}^{ \varrho} e^{ i \pi \beta \left( 2 \xi - 2 \frac{\kappa}{\varrho} \lambda + \frac{\kappa}{\varrho} \beta  \right)  }   \right] ,  \label{MordellG-} 
	\end{align}
	valid for $\kappa, \varrho \in \mathbb{Z}_{>0}$ and $0 < \Re \lambda <1 $. Note that the left-hand side only depends on the ratio $\kappa/\varrho$. Strictly speaking, these identities only appear in \cite[Eq. (8.1)-(8.2)]{Mordell} for $\lambda =0$, but can be easily extended mimicking the manipulations that lead to \cite[Eq. (3.8)]{Mordell}. Since, in doing so, there is a subtlety in the choice of integration contour, we spell the details in Appendix \ref{app:mordellproof} for completeness.\par
	The building blocks of our solutions will be the integrals
	\begin{equation}
	\label{defIk}
		\mI_{k} (y, \check{\xi} ) := \int_{- \infty} ^{+ \infty} dx \frac{ e^{i \pi k x^2 + 2 \pi x \check{\xi} } }{ e^{2 \pi x} + e^{2 \pi y} } ,
	\end{equation}
	for rational $k$. Comparing with \eqref{MordellG+} and \eqref{MordellG-}, it is clear that
	\begin{equation}
	\label{IkMord}
		\mI_{k} (y, \check{\xi} ) = \begin{cases} \Psi_{+} ( \xi = - \check{\xi}, \lambda= \frac{1}{2} - i y ; \kappa, \varrho)  & \text{ if } k= + \frac{\kappa}{\varrho} ; \\  \Psi_{-} ( \xi = - \check{\xi}, \lambda= \frac{1}{2} - i y ; \kappa, \varrho)  & \text{ if } k= - \frac{\kappa}{\varrho}  , \end{cases}
	\end{equation}
	for $\kappa , \varrho \in \mathbb{Z}_{>0}$.\par

	\subsection{Moments of the log-normal}
	\label{sec:momentsSW}
	
	We introduce the moments of the log-normal distribution, which will appear in our computations. Using a change of variables of the form $X_a=e^{2\pi x_a}$, it was shown in \cite{T0} that the partition function of $U(N)$ Chern--Simons theory on $\cs^3$, analytically continued to $q= e^{-g}$, $g>0$, is proportional the partition function of the Stieltjes--Wigert (SW) ensemble. Hence, pure Chern--Simons is solved by polynomials orthogonal with respect to the measure
	\begin{equation*}
		e^{- \frac{1}{2g} (\log X)^2 } dX ,
	\end{equation*}
	on $0<X< \infty$, after the continuation $g \mapsto \frac{i 2 \pi }{k}$ introduced in \eqref{gto2pik}. 
	The moments of the log-normal measure are 
	\begin{equation*}
		\mu_{\alpha} (g) = \int_0 ^{+ \infty}  \frac{ d X}{ 2 \pi X} X^{\alpha} e^{- \frac{1}{2g} \left(  \log X \right)^2} = \int_{- \infty}  ^{\infty} dx e^{ - \frac{4 \pi}{ g} x^2 + 2 \pi \alpha  x } ,
	\end{equation*}
	defined for $\Re g >0$ and $\alpha \in \mathbb{Z}$. We immediately find
	\begin{equation*}
		\mu_{\alpha} (g) = \sqrt{ \frac{g}{2 \pi}}   e^{g\frac{ \alpha^2}{2}} = \sqrt{ \frac{g}{2 \pi}}   q^{- \frac{ \alpha^2}{2}} .
	\end{equation*}
	We can collect these moments into a formal generating series:
	\begin{equation}
	\label{genseriesmug}
		P (z; g) = \sum_{\alpha \in \mathbb{Z}}  z^{\alpha }  \mu_{\alpha}  (g) .
	\end{equation}
	In the present work, as usual for CS theories, we are interested in the analytic continuation \eqref{gto2pik}, and we write
	\begin{equation}
		\widetilde{\mu}_{\alpha} (k) = \sqrt{  \frac{i}{k} }  q^{- \frac{ \alpha^2}{2}} 
	\end{equation}
	to denote the moment continued as prescribed in \eqref{gto2pik}. When $q$ is a $\kappa^{\mathrm{th}}$ root of unity, namely $\vert k \vert = \frac{\kappa}{\varrho} $, the generating series \eqref{genseriesmug} only contains $\vert \kappa \vert$ different terms, hence we can resum it:
	\begin{align}
		\widetilde{P} (z;k) & =  \sum_{n \in \mathbb{Z}} \sum_{\alpha=1}^{\kappa } z^{n \kappa + \alpha }  \widetilde{\mu}_{\alpha} (k) = \sum_{\alpha=1}^{\kappa } \left[  \sum_{n=0} ^{\infty}z^{n \kappa + \alpha }  \widetilde{\mu}_{\alpha} (k) +  \sum_{n=0} ^{\infty}z^{- n \kappa + \alpha }  \widetilde{\mu}_{\alpha} (k)  \right]  \notag \\
					& = \sum_{\alpha=1}^{\kappa } z^{\alpha }  \widetilde{\mu}_{\alpha} (k)	. \label{genpolymuk}
	\end{align}\par
	This type of resummation is the reason \cite{Naculich:2007nc} why only integrable $U(N)$ representations contribute to the partition function of Chern--Simons theory on $\mathbb{S}^3$ with $q$ root of unity \cite{WittenJones}, while all the unitary irreducible representations contribute when $q$ is analytically continued to $q= e^{-g}$.\par
	\medskip
	Looking back at the Mordell integrals \eqref{MordellG+}-\eqref{MordellG-} we notice that the first of the two sums in $\Psi_{\pm}$ when $\xi \in \mathbb{Z}$ gives precisely 
	\begin{equation*}
		e^{ i \pi k \lambda^2} \widetilde{P} \left( e^{i 2 \pi \lambda \text{sign} (k) } ;k \right) ,
	\end{equation*}
	up to an irrelevant shift in the range of the variable $\alpha$, now running on $1+ \xi, \dots, \kappa + \xi $ in $\Psi_+$ and on $- \xi, \dots, \kappa - \xi -1 $ in $\Psi_-$. The overall Gaussian coefficient is cancelled by a contribution from the numerator of the overall term in \eqref{MordellG+}-\eqref{MordellG-}. We will see that the fugacity $e^{i 2 \pi \lambda}$ will play a central role, as further discussed in Sections \ref{sec:Lessons} and \ref{sec:Schurexpansion}.

	\subsection{Cauchy identities, Gauss sums and notation}
	\label{sec:Cauchyetc}
	For later convenience, we state here relevant mathematical identities which we will exploit in the text.
		
		\subsubsection{Cauchy identity}
		The Cauchy identity \cite{Macdonaldbook,Bumpbook}: 
		\begin{equation}
		\label{eq:Cauchyid}
			\prod_{a=1}^{N_1} \prod_{\dot{a}=1}^{N_2} \frac{1}{1- X_a Y_{\dot{a} } } = \sum_{\nu } \mathfrak{s}_{\nu} (X_1, \dots, X_{N_1} ) \mathfrak{s}_{\nu} (Y_1, \dots, Y_{N_2})  
		\end{equation}
		where $\mathfrak{s}_{\nu}$ is the Schur polynomial \cite{Macdonaldbook,Bumpbook} labelled by the Young diagram $\nu$ satisfying 
		\begin{equation*}
			\mathrm{length} (\nu) \le  \min \left\{ N_1, N_2 \right\} .
		\end{equation*}
		
		This is a well-known identity, which has become increasingly familiar, and useful, in many contexts, especially in random matrix theory \cite{BumpGam,Morozov:2018lsn,tierz2020matrix,jonnadula2020symmetric}. However, it is deceptively simple and this becomes more manifest when approaching it from the point of view of representation theory. We mention this as it gives some insight into some of the results in our paper, as we will find expressions in terms of simple observables of $U(N)$ Chern-Simons theory on $\mathbb{S}^3$.
		
		The result can be proven by taking a trace of a representation of $GL_N (\mathbb{C})$ in two different ways \cite{Bumpbook,CauchyNotes}:
		
		\begin{enumerate}[(i)]

\item as the sum of traces of irreducible sub-representations, and

\item as sum of weight spaces, which are irreducible sub-representations for a maximal torus.

\end{enumerate}

This will be relevant in our setting because viewing the trace as (ii) leads to a determinantal expression which is of the type the localization method gives for hypermultiplets. For the case of ABJ(M) theories, the matrix model representation of the partition function is a two-matrix model and the interaction term can also be cast in such determinant form. This observation holds more generally for quiver theories.\par

First of all, the Schur polynomial $\mathfrak{s}_{\mu} (X_1, \dots, X_N)$ is the character $ \mathrm{Tr} \left( \pi _{\mu }\right)$
of the holomorphic irreducible representation $\pi _{\mu }$ of $GL_{N}(\mathbb{C})$ with highest weight $\mu $, evaluated on diagonal matrices with
entries $X_{1},...,X_{N}.$ Let $E$ be the collection of complex $N$-by-$N$ matrices, and consider the space $\C [E]$ of holomorphic polynomials on it. $GL_N (\C) \times GL_N (\C)$ has a well-known action on $\C [E]$ and the representation $\pi _{\mu }\otimes \pi _{\mu }$
occurs only once, giving the decomposition of $\C [E]$ in terms of irreducible $GL_{N}(\mathbb{C})$ representations as:
\begin{equation*}
    \C [E] \simeq \sum_{\mu : \mu_N \ge 0 } \pi_{\mu} \otimes \pi_{\mu} .
\end{equation*}
Taking the trace of this expression leads to the right-hand side of \eqref{eq:Cauchyid}. That is:
\begin{equation*}
    \mathrm{Tr} \C [E] = \sum_{\mu : \mu _{N} \ge 0} \mathrm{Tr} \left( \pi _{\mu }\otimes
\pi _{\mu }\right) =\sum_{\mu : \mu _{N} \ge 0} \mathrm{Tr}  \pi _{\mu } \cdot  \mathrm{Tr} 
\pi _{\mu }=\sum_{\mu : \mu _{N} \ge 0}  \mathfrak{s}_{\mu} \cdot \mathfrak{s}_{\mu}  .
\end{equation*}
The other way to do this counting, is to consider that the monomials in $\C [E]$ are weight vectors for the subgroup $D_N\times D_N$ of diagonal
matrices in $GL_N (\C) \times GL_N (\C) $. Thus, evaluating the trace on diagonal elements $X \times Y \in D_N \times D_N$, 
\begin{equation*}
     \mathrm{Tr} \C [E] (X \times Y) =  \prod_{a, \dot{a}=1} ^{N} \sum_{n_{a \dot{a}} \ge 0} (X_a Y_{\dot{a}} )^{n_{a \dot{a}}} = \prod_{a, \dot{a}=1} ^{N} \frac{1}{1- X_a Y_{\dot{a}} } = \frac{1}{\det \left( 1 - X \otimes Y \right)} .
\end{equation*}
Note that the second equality holds analytically if $\vert X_a \vert <1$ and $\vert Y_{\dot{a}} \vert <1$, and algebraically otherwise \cite{CauchyNotes,BumpGam}. This establishes the Cauchy identity for diagonal elements, but, with both
sides being conjugation-invariant, it actually holds for all diagonalizable
elements of $GL_N (\C) \times GL_N (\C) $. In turns, these are dense, therefore the Cauchy
identity holds for any $\mathbf{X}, \mathbf{Y} \in GL_N (\C) $ \cite{CauchyNotes}: 
\begin{equation*}
    \sum_{\mu : \mu _{N} \ge 0}  \mathfrak{s}_{\mu} ( \mathbf{X} ) \cdot \mathfrak{s}_{\mu}  (\mathbf{Y}) =   \mathrm{Tr} \C [E] (  \mathbf{X}  \times   \mathbf{Y} ) = \frac{1}{\det \left( 1 - \mathbf{X}  \otimes \mathbf{Y}  \right)}  .
\end{equation*}
\par
\medskip
When $N_2=1$, the Cauchy identity reduces to the generating function of the complete homogeneous symmetric polynomials $\mathfrak{h}_{\nu}$ \cite{Macdonaldbook}: 
		\begin{equation}
		\label{eq:genhomo}
			\prod_{a=1}^{N} \frac{1}{1- X_a Y } = \sum_{\nu=0 }^{\infty} Y^{\nu} \mathfrak{h}_{\nu} ( X_1, \dots, X_N ) ,
		\end{equation}
		where 
		\begin{equation*}
			\mathfrak{h}_{\nu} ( X_1, \dots, X_N )  = \sum_{1 \le a_1 \le _2 \le \cdots \le a_{\nu} \le N } X_{a_1} \cdots X_{a_{\nu}} .
		\end{equation*}
		Equivalently,
		\begin{equation*}
		    	\mathfrak{h}_{\nu} ( X_1, \dots, X_N )  = \sum_{\mu, \vert \mu \vert = \nu } \mathfrak{m}_{\mu} (x_1, \cdots, x_N ) , \qquad \forall \nu \in \mathbb{Z}_{\ge 0} ,
		\end{equation*}
		with $ \mathfrak{m}_{\mu} (x_1, \dots, x_N) = x_1 ^{\mu_1} \cdots x_N ^{\mu_N}$ being the monomials \cite{Macdonaldbook} and the sum running over all partitions $\mu$ of size $\vert \mu \vert = \nu$.\par
		There exists a related identity, known as \emph{dual} Cauchy identity \cite{Bumpbook}: 
		\begin{equation}
		\label{eq:dualCauchy}
			\prod_{a=1}^{N_1} \prod_{\dot{a}=1}^{N_2} \left( 1+ X_a Y_{\dot{a} } \right) = \sum_{\nu } \mathfrak{s}_{\nu^{\prime}} (X_1, \dots, X_{N_1} ) \mathfrak{s}_{\nu} (Y_1, \dots, Y_{N_2})  
		\end{equation}
		with $\nu^{\prime}$ the conjugate partition, obtained transposing rows and columns of the Young diagram of the partition $\nu$. An important aspect of \eqref{eq:dualCauchy} is that, differently from \eqref{eq:Cauchyid}, the sum contains a finite number of terms, due to the restriction 
		\begin{equation*}
		    \text{length} (\nu^{\prime}) = \nu_1 \le N \equiv \min \left\{ N_1 , N_2 \right\} .
		\end{equation*}
		Therefore the sum in \eqref{eq:dualCauchy} only involves partitions whose Young diagrams fit in a $N \times N $ square.\par
		When $N_2=1$, this latter Cauchy identity reduces to the generating function of the elementary symmetric polynomials $\mathfrak{e}_{\nu}$ \cite{Macdonaldbook}: 
		\begin{equation}
		\label{eq:genele}
			\prod_{a=1}^{N} \left( 1+ X_a Y \right) = \sum_{\nu=0 }^{N} Y^{\nu} \mathfrak{e}_{\nu} ( X_1, \dots, X_N ) ,
		\end{equation}
		where 
		\begin{equation*}
			\mathfrak{e}_{\nu} ( X_1, \dots, X_N )  = \sum_{1 \le  a_1 < _2 < \cdots < a_{\nu} \le N } X_{a_1} \cdots X_{a_{\nu}} .
		\end{equation*}\par

			\subsubsection{Gauss sums}
	The Gauss sum identity \cite{TRS}:
	\begin{equation}
	\label{Gaussid}
		\frac{1}{\sqrt{i \kappa}} \sum_{\alpha =0} ^{\kappa-1} e^{ \frac{ i \pi}{\kappa} \left( \alpha - \ell - \frac{\kappa}{2} \right)^2} = 1 ,
	\end{equation}
	valid for $\kappa \in \mathbb{Z}_{>0}$ and for every $\ell \in \mathbb{Z}$. This formula will be instrumental to obtain the massless limit of all the computations in Section \ref{sec:exactZ}.

		\subsubsection{Remarks on notation}
	To avoid clutter, whenever possible we will change the notation $x_{p,a}$ for a more suitable one. For example, for $r=2$, we will write $(x_a,y_{\dot{a}})$ instead of $(x_{1,a}, x_{2,a^{\prime}})$. Similarly, we will mostly denote the masses simply $\left\{ m_j  \right\}$, when it is clear from the context to which node each one is attached. Besides, throughout the work, we will sometimes switch to exponentiated variables, which we will denote with upper case letters. So, for example, we will use $X_a = e^{2 \pi x_a}$, $M_j = e^{2 \pi m_j}$ and so on. Moreover, for a given CS level $k$, we define as usual
	\begin{equation}
	\label{defq}
		q= \exp \left( - \frac{ i 2 \pi }{k} \right) .
	\end{equation}\par
	\medskip
	The field content of the theories we study is conveniently encoded in $\mathsf{A}$-type Dynkin diagrams or in affine $\widehat{\mathsf{A}}$-type Dynkin diagrams. We will interchangeably call the first class $\mathsf{A}$ quivers or linear quivers, and the second class $\widehat{\mathsf{A}}$ quivers, extended quivers or necklace quivers.\par 
	We draw such quivers in the $3d$ $\mN=4$ quiver notation, so the edges represent hypermultiplets in the bi-fundamental representations and are not directed. The CS levels will not be explicit in the quiver diagrams. Moreover, the Dynkin diagram notation will refer only to the gauged nodes, or, stated more formally, we will always refer to the quiver and not to the \emph{framed} quiver.

	\section{Evaluation of partition functions}
	\label{sec:exactZ}

	\subsection{$U(1)_k$ Chern--Simons theory with a fundamental hypermultiplet}
	\label{sec:AbelianA1Nf1}

\begin{figure}[th]
	\centering
	\begin{tikzpicture}[auto,node distance=1cm,square/.style={regular polygon,regular polygon sides=4}]
		\node[circle,draw] (gauge) at (-1,0) {$1$};
		\node[square,draw] (flavour) at (1,0) {$ 1 $};
		\node (1) at (0.5,0) {};
		\draw[-](gauge)--(flavour);
	\end{tikzpicture}
	\caption{$U(1)_k$ theory with $N_f=1$ fundamental flavour. This is an Abelian $\mathsf{A}_1$ quiver.}
	\label{fig:A1AbNf1}
	\end{figure}
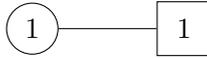

	We start our analysis revisiting the simplest Chern--Simons theory that includes matter: $U(1)_k$ Chern--Simons theory with a fundamental hypermultiplet, represented in Figure \ref{fig:A1AbNf1}.\par
	 \par
	The moduli space of vacua of the theory in flat space has been analyzed in \cite{Kapustin:1999ha}, with a focus on its S-duality properties. The CS term gives a topological mass to the vector multiplet, lifting the Coulomb branch. The moduli space has a non-compact one-dimensional Higgs branch, which is also lifted turning on a real mass deformation. In an Abelian CS theory, admitting rational $k$, S-duality acts as the S-matrix of the $SL(2, \mathbb{Z})$ group on the coupling while exchanging mass and FI terms
	\begin{equation}
	\label{eq:SdualU11}
		k \mapsto -\frac{1}{k} , \quad \zeta \mapsto m \quad m \mapsto - \zeta .
	\end{equation}
	The theory with gauge group $U(1)$ and $N_f=1$ is self-dual under S-duality \cite{Kapustin:1999ha}.\par
	The partition function at rational CS level $k$ and with mass and FI parameters turned on is
	\begin{equation}
	\label{eq:A1U1Nf1int}
		\mz_{U(1),1} \left( k,m,\zeta \right) = \int_{- \infty} ^{+ \infty} dx \frac{ e^{ i \pi x^2 k + i 2 \pi \zeta x } }{ 2 \cosh \pi (x+m) } .
	\end{equation}
	We do not need to consider both deformations: shifting variables $x^{\prime} = x+m$ we get 
	\begin{equation}
	\label{eq:A1Nf1ZmFIrel1}
		\mz_{U(1),1} \left( k,m,\zeta \right) = e^{i \pi k m^2 + i 2 \pi m \zeta } \mz_{U(1),1} \left( k,0, \zeta - k m\right) ,
	\end{equation}
	while shifting variables $x^{\prime} = x+ \zeta/k $ we get 
	\begin{equation}
	\label{eq:A1Nf1ZmFIrel2}
		\mz_{U(1),1} \left( k,m,\zeta \right) = e^{- i \frac{\pi}{k} \zeta^2 } \mz_{U(1),1} \left( k,m- \frac{ \zeta }{k} , 0\right) .
	\end{equation}
	Therefore, it is sufficient to take one of the two deformations, and the more general result follows immediately. Note how the prefactor suffers a change $k \mapsto - \frac{1}{k}$ when the roles of $m$ and $\zeta$ are exchanged, as well as the presence of the additional phase $e^{i 2 \pi \zeta m}$ in \eqref{eq:A1Nf1ZmFIrel1}, coupling the FI background twisted vector multiplet to the flavour background vector multiplet.\footnote{The minus sign $m \mapsto - \zeta$ in \eqref{eq:SdualU11} comes from our conventions, presented in Subsection \ref{sec:CSS3loc}. The necessity of that sign can be checked applying S-duality to \eqref{eq:A1Nf1ZmFIrel1} and \eqref{eq:A1Nf1ZmFIrel2}.}\par
	From the integral representation \eqref{eq:A1U1Nf1int} the self-duality is easily proven: 
	\begin{align*}
		\mz_{U(1),1} \left( k,m,\zeta \right) & =  \int_{- \infty} ^{+ \infty} dx  \int_{- \infty} ^{+ \infty} dy  \frac{ e^{ i \pi x^2 k -  i 2 \pi x ( y - \zeta ) } e^{- i 2 \pi y m} }{ 2 \cosh (\pi y) }  \\
			& = \sqrt{\frac{i}{k}} e^{- i 2 \pi m \zeta }   \int_{- \infty} ^{+ \infty} dy \frac{ e^{- i \frac{\pi}{k} y^2 - i 2 \pi y m } }{ 2 \cosh \pi (y+\zeta ) } = \sqrt{\frac{i}{k}} e^{- i 2 \pi m \zeta }   \mz_{U(1),1} \left( - k^{-1} ,\zeta , - m \right) ,
	\end{align*}
	where we have used the fact that $(\cosh \pi x)^{-1}$ is Fourier transformed into itself.\par
	We now use Mordell's formula to evaluate exactly the partition function. Starting with $m \ne 0$ and $\zeta=0$ in \eqref{eq:A1U1Nf1int} we have 
	\begin{equation}
		\mz_{U(1),1} \left( k,m,0 \right) = e^{-\pi m} \mI_{k} \left( -m, \frac{1}{2} \right) ,
	\label{eq:A1AbelNf1zeta0massive}
	\end{equation}
	given in terms of a Mordell integral. For rational $k$ with $\vert k \vert= \frac{\kappa}{\varrho}$, \eqref{IkMord} gives 
	\begin{align}
		\mz_{U(1),1} \left( k >0, m, 0 \right) &= \frac{ 1 }{1- (-1)^{\kappa \varrho - \kappa + \varrho } e^{- 2 \pi \kappa m}} \left\{ - e^{i \pi k \left(  m - \frac{i}{2} \right)^2  }\sum_{\beta=1} ^{\varrho} \left( - e^{- 2 \pi k m} \right)^{\beta} e^{- i \pi k \beta (\beta-1)} \right. \notag \\
			& \hspace{4.2cm} \left.  +  \sqrt{\frac{i}{k}} \sum_{\alpha=1}^{\kappa}  \left( - e^{- 2 \pi  m} \right)^{\alpha- \frac{1}{2}}  e^{i \frac{\pi}{k} \left( \alpha- \frac{1}{2} \right)^2}\right\} , \label{eq:ZA1U1Nf1Mord+} 
	\end{align}
	\begin{align}
		\mz_{U(1),1} \left( k <0,  m, 0 \right) &= \frac{ 1 }{1- (-1)^{\kappa \varrho - \kappa - \varrho } e^{2 \pi \kappa m}} \left\{ e^{i \pi k \left(  m - \frac{i}{2} \right)^2  }\sum_{\beta=1} ^{\varrho} \left( - e^{- 2 \pi k m} \right)^{\beta} e^{- i \pi k \beta (\beta-1)} \right. \notag \\
			& \hspace{4.2cm} \left.  +  \sqrt{\frac{i}{ \vert k \vert }} \sum_{\alpha=0}^{\kappa-1}  \left( - e^{ 2 \pi  m} \right)^{\alpha- \frac{1}{2}}  e^{i \frac{\pi}{k} \left( \alpha- \frac{1}{2} \right)^2}\right\} \label{eq:ZA1U1Nf1Mord-} .
	\end{align}
	The factor $e^{-\pi m}$ in \eqref{eq:A1AbelNf1zeta0massive} is cancelled against a contribution from the overall factor in the Mordell integrals \eqref{MordellG+}-\eqref{MordellG-}.\par
	When $k\in \mathbb{Z}$, hence $\varrho=1$, these latter two expressions reduce to 
	\begin{align*}
		\mz_{U(1),1} \left( k >0, m, 0 \right) &= \frac{e^{i \pi k \left( m^2 - \frac{1}{4} \right) + \pi m}}{2 \cosh (\pi k m)} +  \frac{1}{1+ e^{-2 \pi k m}} \sqrt{\frac{i}{k}} \sum_{\alpha=1}^{k}  \left( - e^{- 2 \pi  m} \right)^{\alpha- \frac{1}{2}}  q^{- \frac{1}{2} \left( \alpha- \frac{1}{2} \right)^2 } , \\
		\mz_{U(1),1} \left( k <0,  m, 0 \right) &= - \frac{e^{i \pi k \left( m^2 - \frac{1}{4} \right) + \pi m}}{2 \cosh (\pi k m)} +  \frac{1}{1+ e^{-2 \pi k m}} \sqrt{\frac{i}{k}} \sum_{\alpha=0}^{\vert k \vert -1}  \left( - e^{2 \pi  m} \right)^{\alpha- \frac{1}{2}}  q^{- \frac{1}{2} \left( \alpha- \frac{1}{2} \right)^2 } .
	\end{align*}
	The result is a real analytic function of $m$, and is holomorphic in the usual ``physical'' strip $-\frac{1}{2} < \Im m < \frac{1}{2}$. Note that, using the relation $\lambda = \frac{1}{2} + i m$ (see \eqref{IkMord}) between the physical variable $m$ and the variable $\lambda$ of \cite{Mordell}, the result is holomorphic in $0 < \Re \lambda < 1$, in agreement with the proof in Appendix \ref{app:mordellproof}, based on \cite{Mordell}.\par
	Setting instead $m=0, \zeta \ne 0$ in \eqref{eq:A1U1Nf1int} we have
	\begin{equation*}
		\mz_{U(1),1} \left( k,0,\zeta \right) = \mI_{k} \left( 0, \check{\xi} = \frac{1}{2} + i \zeta \right) .
	\end{equation*}
	The solution is read off from \eqref{IkMord} for any rational $k$,
	\begin{align}
		\mz_{U(1),1} \left( k >0, 0, \zeta \right) = & - \frac{e^{- i \frac{\pi}{4} k - \pi \zeta } }{1- (-1)^{ \kappa - \varrho - \kappa \varrho } e^{2 \pi \varrho \zeta } } \sum_{\beta=1} ^{\varrho} \left( - e^{2 \pi \zeta } \right)^{\beta} e^{- i \pi k \beta (\beta-1)}  \notag \\
			& - \frac{i e^{- i \frac{\pi}{k} \zeta^{2} } }{1- (-1)^{ \kappa - \varrho - \kappa \varrho } e^{2 \pi \varrho \zeta } } \sqrt{\frac{i}{k}} \sum_{\alpha=1}^{\kappa}  \left( - e^{2 \frac{\pi}{k} \zeta } \right)^{\alpha- \frac{1}{2}}  e^{i \frac{\pi}{k} \left( \alpha- \frac{1}{2} \right)^2}   , \label{eq:ZA1U1Nf1Mordmirror+} 
	\end{align}
	\begin{align}
		\mz_{U(1),1} \left( k <0,  0, \zeta \right) = &  - \frac{e^{- i \frac{\pi}{4} k - \pi \zeta } }{1- (-1)^{ \kappa + \varrho + \kappa \varrho } e^{2 \pi \varrho \zeta } } \sum_{\beta=1} ^{\varrho} \left( - e^{2 \pi \zeta } \right)^{\beta} e^{- i \pi k \beta (\beta-1)}  \notag \\  
		   & - \frac{i e^{- i \frac{\pi}{k} \zeta^{2} } }{1- (-1)^{ \kappa + \varrho + \kappa \varrho } e^{2 \pi \varrho \zeta } } \sqrt{\frac{i}{ \vert k \vert }}   \sum_{\alpha=1}^{\kappa}  \left( - e^{- 2 \frac{\pi}{k} \zeta } \right)^{\alpha- \frac{1}{2}}  e^{i \frac{\pi}{k} \left( \alpha- \frac{1}{2} \right)^2}   \label{eq:ZA1U1Nf1Mordmirror-}   .
	\end{align}
	When $k \in \mathbb{Z}$ it takes the simpler form:
	\begin{align*}
		\mz_{U(1),1} \left( k >0 \right) &= \frac{ e^{- i \pi \frac{k}{4} }}{ 2 \cosh (\pi \zeta) } - \frac{1 }{ e^{2 \pi \zeta} +1 } \sqrt{ \frac{i}{k}} \sum_{\alpha=1} ^{k} (-1)^{\alpha} q^{ - \frac{1}{2} \left( \alpha - \frac{1}{2} - i \zeta \right)^2 }   ,  \\
		\mz_{U(1),1} \left( k <0 \right) &= \frac{ e^{- i \pi \frac{ k  }{4} }}{ 2 \cosh (\pi \zeta) } -  \frac{1 }{ e^{2 \pi \zeta} +1 } \sqrt{ \frac{i}{k}} \sum_{\alpha=1} ^{\vert k \vert } (-1)^{\alpha} q^{ - \frac{1}{2} \left( \alpha - \frac{1}{2} + i \zeta \right)^2 } ,  
	\end{align*}
	where we recall that $q= e^{-i 2 \pi /k}$ from \eqref{defq}. The solution is real analytic in $\zeta \in \mathbb{R}$ and holomorphic in the strip $-\frac{1}{2} < \Im \zeta < \frac{1}{2}$.\par
	We recognize the generating polynomial of the moments of the SW distribution when $q$ is a $k^{\mathrm{th}}$ root of unity, $\widetilde{P} (z;k)$, evaluated at $z= - e^{- \text{sign} (k) 2 \pi m}$ for the theory with only mass term and at $z= - q^{\check{\xi}}$ for the theory with only FI term.\par
	Direct inspection shows that 
	\begin{equation*}
	\text{\eqref{eq:ZA1U1Nf1Mord+}} \ = \ \sqrt{\frac{i}{k}} \times \left[  \text{\eqref{eq:ZA1U1Nf1Mordmirror-}  with $-\zeta=m$ and $\kappa \leftrightarrow \varrho $} \right] , 
	\end{equation*}
	and likewise for \eqref{eq:ZA1U1Nf1Mord-} and \eqref{eq:ZA1U1Nf1Mordmirror+}. This together with the relations \eqref{eq:A1Nf1ZmFIrel1}-\eqref{eq:A1Nf1ZmFIrel2} gives a full check of the self S-duality of the solution.\par
\begin{figure}[ht]
	       \centering
	       \includegraphics[width=0.45\textwidth]{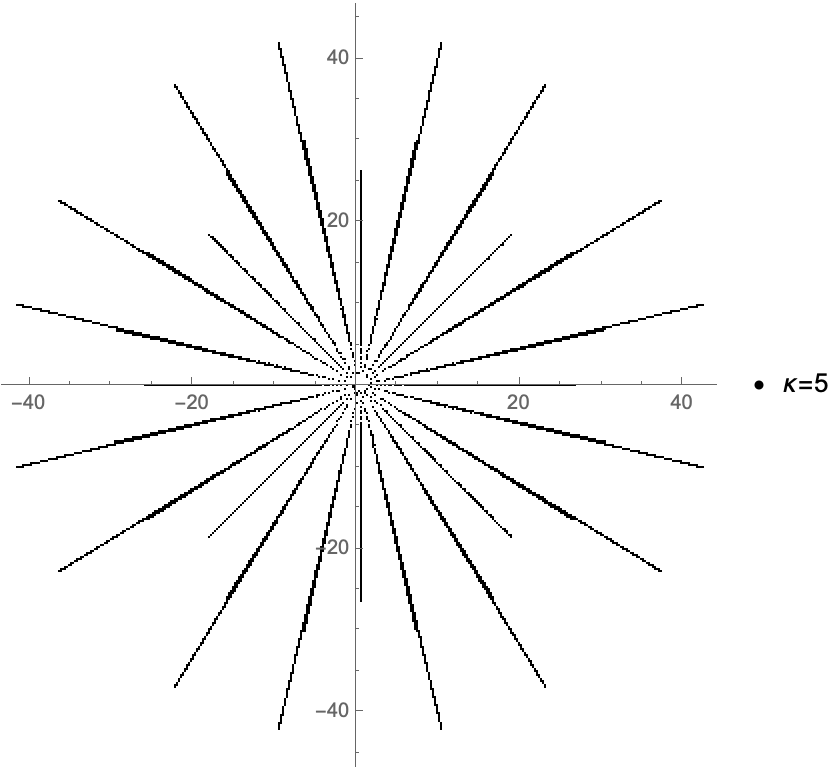}%
	       \hspace{0.05\textwidth}
	       \includegraphics[width=0.45\textwidth]{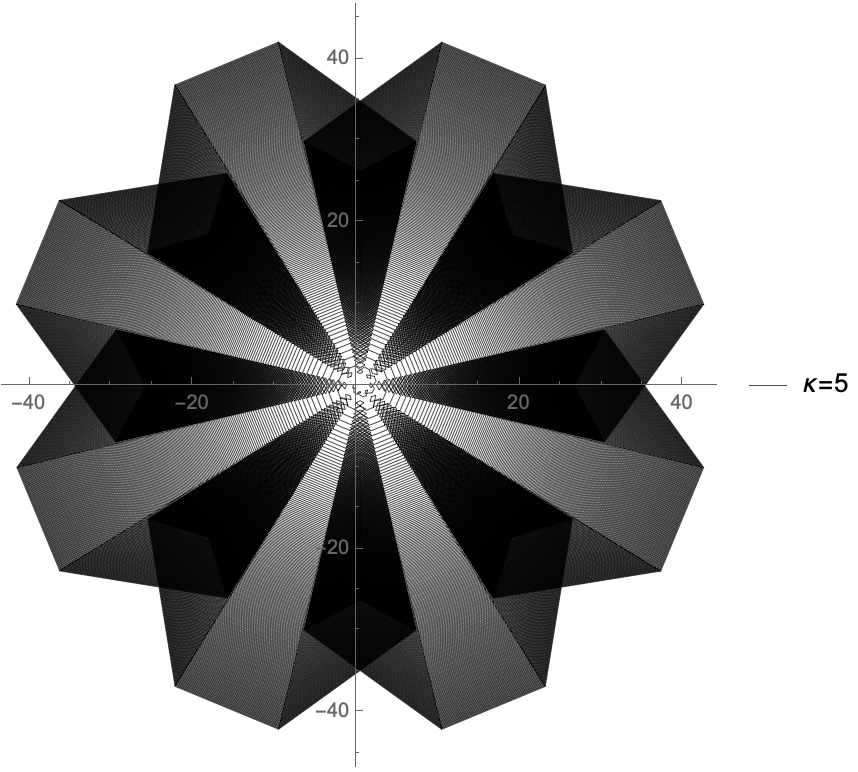}
	   \caption{Left: Plot of $\mathcal{Z}_{U(1),1}$ with $k=\frac{\kappa}{\varrho}$, $\zeta=0$, $m=0.2$ at fixed $\kappa=5$ and varying $\varrho=1, \dots, 10^4$. Right: Same plot, with points obtained from consecutive values of $\varrho$ joined by a segment.}
	   \label{fig:Ckapparho1}
	   \end{figure}
\begin{figure}[htb]
	       \centering
	       \includegraphics[width=0.45\textwidth]{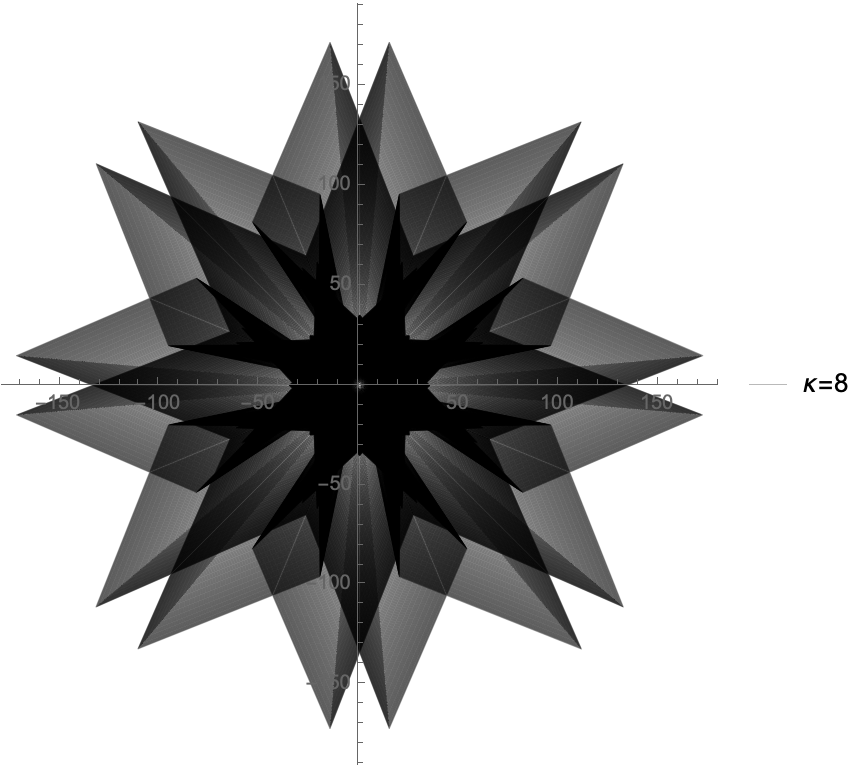}%
	       \hspace{0.05\textwidth}
	       \includegraphics[width=0.45\textwidth]{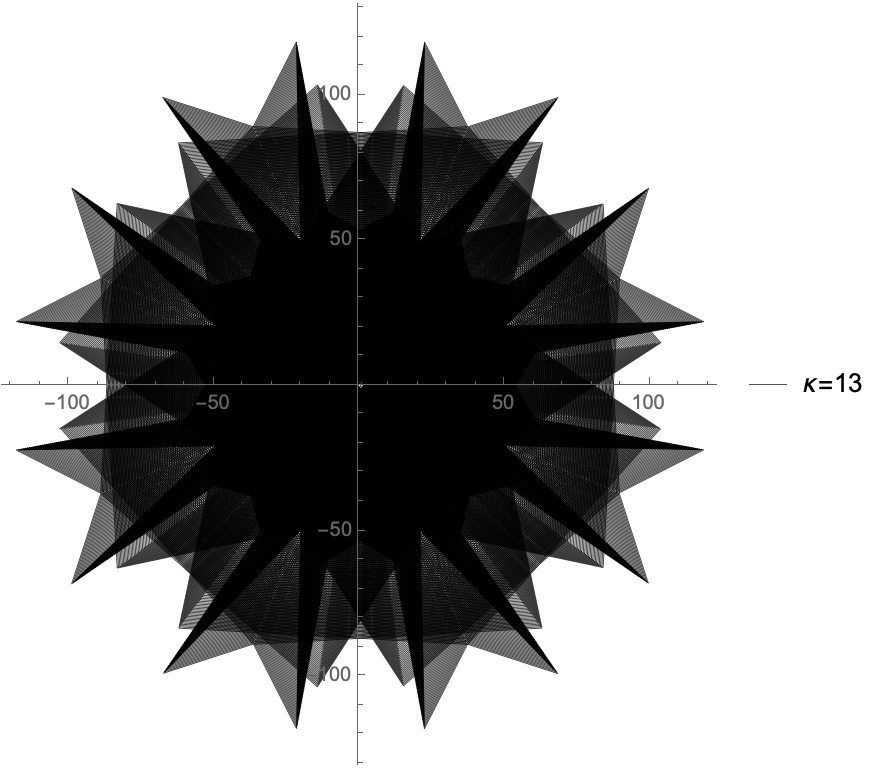}
	   \caption{Plot of $\mathcal{Z}_{U(1),1}$ with $k=\frac{\kappa}{\varrho}$, $\zeta=0$, $m=0.02$ at fixed $\kappa$ and varying $\varrho$. The points obtained from consecutive values of $\varrho$ are joined by a segment. Left: $\kappa=8$, $\varrho=1,\dots,2\times 10^4$. Right: $\kappa=13$, $\varrho=1,\dots,10^4$.}
	   \label{fig:Ckapparho2}
	   \end{figure}

	We plot the result \eqref{eq:ZA1U1Nf1Mord+} of $\mathcal{Z}_{U(1),1}$ with positive rational $k$ and $\zeta =0$ in Figure \ref{fig:Ckapparho1} and \ref{fig:Ckapparho2}. Being $q$ a $\kappa^{\mathrm{th}}$ root of unity, at fixed $\kappa$ and varying $\varrho$ the values of the partition function are placed along rays in $\mathbb{C}$. Increasing $\kappa$ increases the number of rays.\par
	
		\par
	   \par
\begin{figure}[htb]
	       \centering
	       \includegraphics[width=0.45\textwidth]{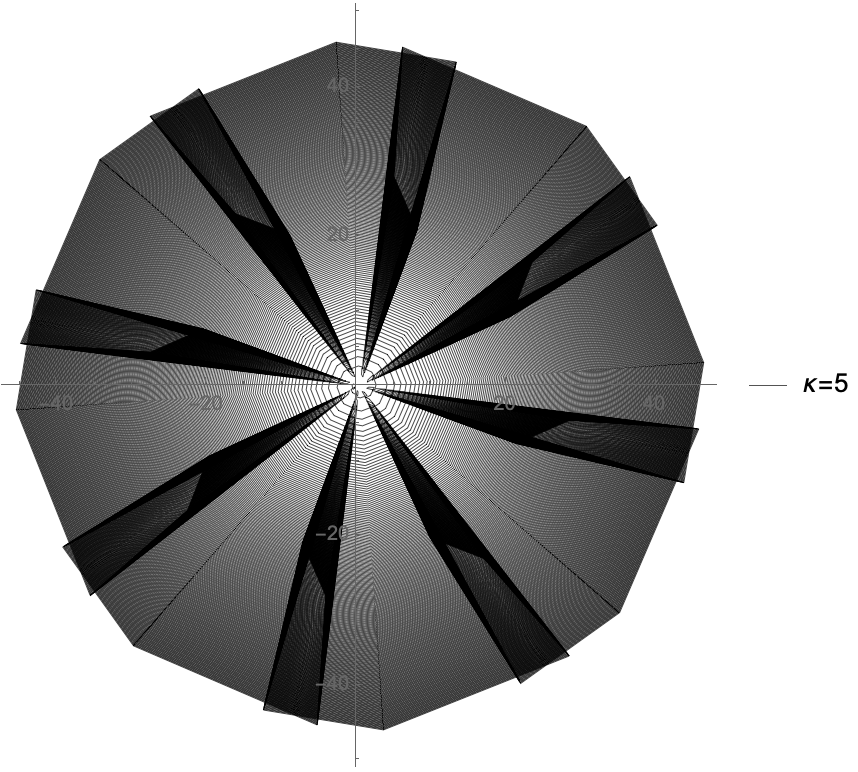}%
	       \hspace{0.05\textwidth}
	       \includegraphics[width=0.45\textwidth]{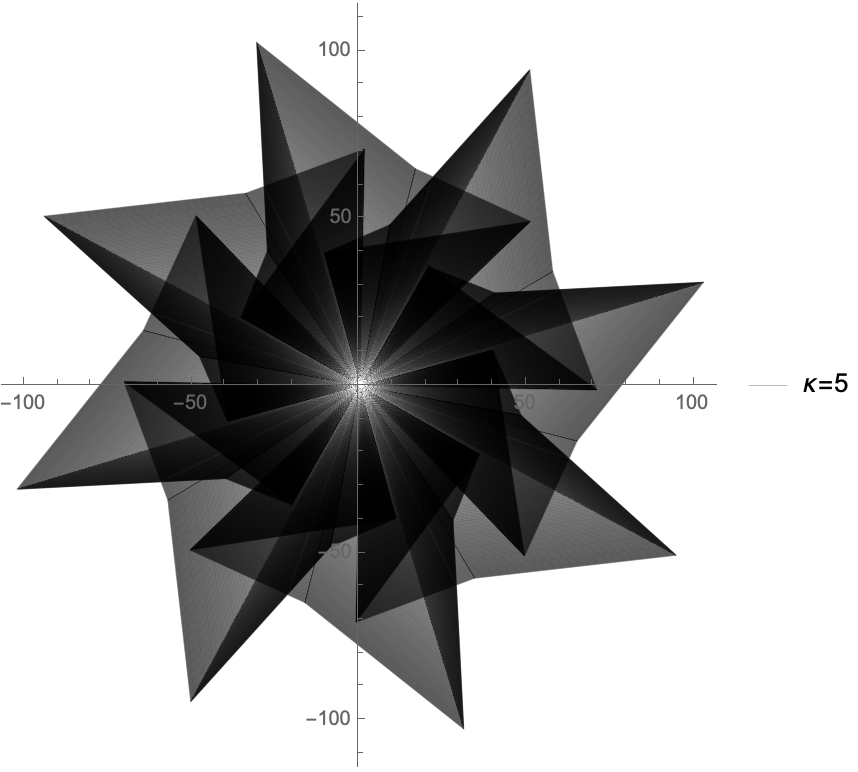}
	   \caption{Plot of $\mathcal{Z}_{U(1),1}$ with $k=\frac{\kappa}{\varrho}$, at fixed $\kappa=5$. Left: $m=\sqrt{0.03} +i 0.1$ and $\varrho=1, \dots, 10^4$. Right: $m=0.1 +i\sqrt{0.03}$ and $\varrho=1, \dots, 2 \times 10^4$.}
	   \label{fig:cplxmkappa5rho}
	   \end{figure}
\begin{figure}[htb]
	       \centering
	       \includegraphics[width=0.45\textwidth]{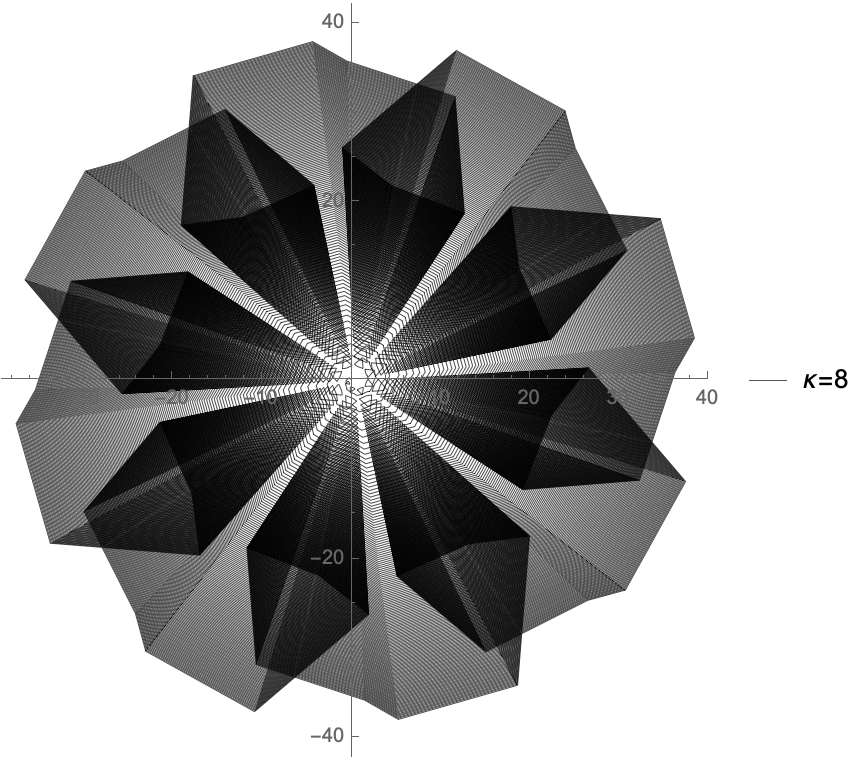}%
	       \hspace{0.05\textwidth}
	       \includegraphics[width=0.45\textwidth]{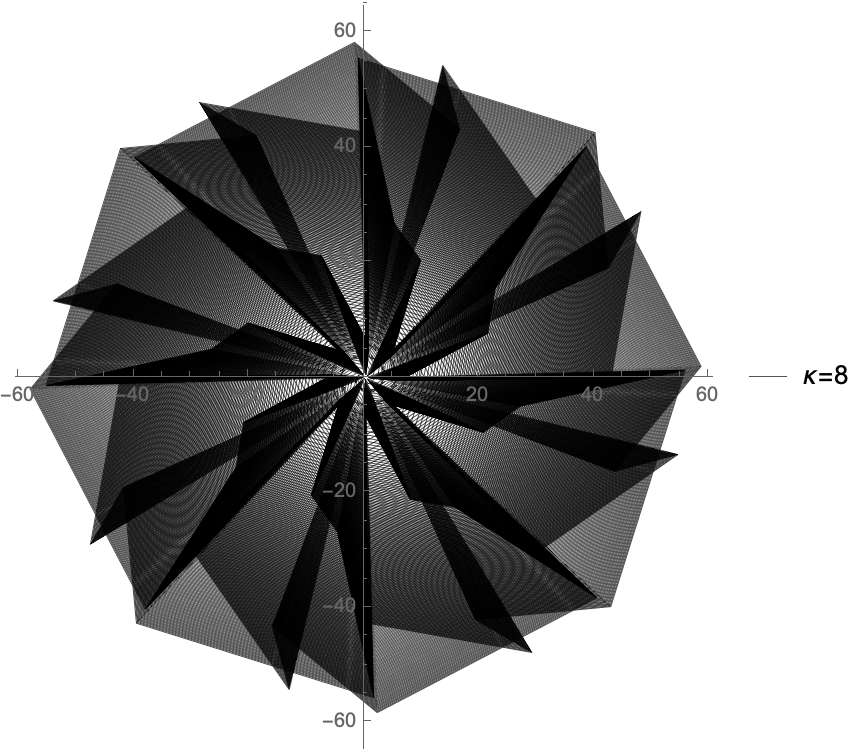}
	   \caption{Plot of $\mathcal{Z}_{U(1),1}$ with $k=\frac{\kappa}{\varrho}$, at fixed $\kappa=8$ and varying $\varrho=1, \dots, 10^4$. Left: $m=\sqrt{0.03} +i 0.1$. Right: $m=0.1 +i\sqrt{0.03}$.}
	   \label{fig:cplxmkappa8rho}
	   \end{figure}

	   As the result holds upon complexification of the mass with $\vert \Im m \vert < \frac{1}{2}$, it is instructive as well to plot the partition function at fixed $\kappa$ and increasing $\varrho$ for complex values of $m$, as we do in Figure \ref{fig:cplxmkappa5rho} (for $\kappa=5$) and Figure \ref{fig:cplxmkappa8rho} (for $\kappa=8$).\par

	   \par
	   \par

	\subsection{Single node quivers}

	\subsubsection{Abelian $\mathsf{A}_1$ theory with two flavours}
	\label{sec:A1AbelianNf2}

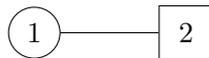
\begin{figure}[ht]
	\centering
	\begin{tikzpicture}[auto,node distance=1cm,square/.style={regular polygon,regular polygon sides=4}]
		\node[circle,draw] (gauge) at (-1,0){$1$};
		\node[square,draw] (flavour) at (1,0) {$ 2 $};
		\draw[-](gauge)--(flavour);
	\end{tikzpicture}
	\caption{Abelian $\mathsf{A}_1$ quiver with $N_f=2$ fundamental flavours.}
	\label{fig:A1Ab2}
	\end{figure}

	We consider a $U(1)$ CS theory with two massive hypermultiplets in the fundamental representation, see Figure \ref{fig:A1Ab2}. The result we present for this theory has been first derived in \cite{TRS}, and we revisit it here as a warm up. 
	\par
	
	The theory has $SU(2)$ flavour symmetry and the hypermultiplets have masses $(+m,-m)$. The partition function is
	\begin{align*}
		\mz_{U(1),2} (k,m) & = \int_{- \infty} ^{\infty} dx \frac{ e^{ i \pi k x^2} }{ 4 \cosh \pi \left( x-m \right) \cosh \pi \left( x+m \right) }   \\
				& = \frac{1}{ 2 \sinh ( 2 \pi m) } \int_{- \infty} ^{\infty} dx e^{ i \pi k x^2 + 2 \pi x} \left[ \frac{1}{ e^{ 2 \pi x}  + e^{- 2 \pi m}}  -  \frac{1}{ e^{ 2 \pi x}  + e^{2 \pi m}} \right] \\
				& = \frac{\mI_k (-m , 1)  - \mI_k (m, 1)}{ 2 \sinh ( 2 \pi m) } ,
	\end{align*}
	where in the last line we have recognized \eqref{defIk}. The solution in terms of the Mordell integrals \eqref{MordellG+} (when $k>0$) or \eqref{MordellG-} (when $k<0$) holds for any non-zero rational CS level. However, the expressions are clearer for $k \in \mathbb{Z}$. Under such assumption, from equation \eqref{IkMord} and simple manipulations, we obtain
	\begin{equation}
		\mz_{U(1),2} (k,m) = \frac{1}{ 2 \sinh ( 2 \pi m) }  \left[  - \frac{ i e^{ i \pi k \left( m^2 - \frac{1}{4} \right) } }{ \sinh ( \pi k m) }  - \frac{\widetilde{P} (- e^{- 2 \pi m}; k) }{ e^{- 2 \pi \vert k \vert m} -1 } + \frac{\widetilde{P} (- e^{2 \pi m}; k) }{ e^{2 \pi \vert k \vert  m} -1 }  \right] ,
	\label{eq:ZTSU2}
	\end{equation}
	with the polynomial $\widetilde{P} (z;k)$ defined in \eqref{genpolymuk}. We also have shifted the summation range hidden in $\widetilde{P} (z;k)$, so that the sum runs over $\alpha= 0, \dots, k-1$ if $k>0$ and $\alpha=1, \dots, \vert k \vert$ if $k<0$.\par
	The masses of the hypermultiplets have played a central role in the derivation, but we can take the massless limit of our final result \cite{TRS}. Despite each term being divergent, a careful analysis and the application of the Gauss sum identity \eqref{Gaussid} show that the result is finite and well defined, and reads 
	\begin{align*}
		\mz_{U(1),2} (k>0,m \to 0^{+}) & = \frac{e^{- i \pi \frac{k}{4}}}{(2 \pi m)^2 k} \left[  - i + i \frac{1}{\sqrt{ik}} \sum_{\alpha=0}^{k-1} e^{\frac{i \pi}{k} \left( \alpha + \frac{k}{2}\right)^2} \left( 1 + 2 \pi^2 m^2 \left( \alpha^2 + \frac{k^2}{6} - \alpha k \right)  \right) \right] \\
			& = \sqrt{\frac{i}{k}} \sum_{\alpha=0} ^{k-1} (-1)^{\alpha} q^{- \frac{ \alpha^2}{2} } \left[ \frac{1}{k}\left( \alpha - \frac{k}{2} \right)^2 - \frac{k}{12} \right]  ,
	\end{align*}
	where to go from the first to the second line we have used \eqref{Gaussid}. The analogous result when $k<0$ is derived by the same steps.\par
	The solution of the Mordell integrals $\Psi_{\pm}$ requires $0 < \Re \lambda < 1$, and we have used $\lambda = \frac{1}{2} \pm i m$. Therefore we can complexify the masses in the strip $- \frac{1}{2} < \Im m < \frac{1}{2}$, which is the usual ``physical'' region in which the integrals from localization do not develop singularities.

	\subsubsection{Abelian $\mathsf{A}_1$ theory with $N_{f}$ flavours}
	\label{sec:AbelianA1Nf}
	
	\begin{figure}[ht]
	\centering
	\begin{tikzpicture}[auto,node distance=1cm,square/.style={regular polygon,regular polygon sides=4}]
		\node[circle,draw] (gauge) at (-1,0) {$1$};
		\node[square,draw] (flavour) at (1,0) {\hspace{7pt} };
		\node[draw=none]  (aux) at (1,0) {$N_f$};
		\draw[-](gauge)--(flavour);
	\end{tikzpicture}
	\caption{Abelian $\mathsf{A}_1$ quiver with $N_f$ fundamental flavours.}
	\label{fig:A1AbNf}
	\end{figure}
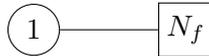\par

	The analysis of the Abelian $\mathsf{A}_1$ theory with two flavours is easily generalized to the case of $N_f$ flavours, represented in Figure \ref{fig:A1AbNf}. We assume the hypermultiplets have distinct masses, $m_s \ne m_j$ for $s \ne j$, $j = 1 , \dots, N_f$, and also turn on a FI parameter $\zeta \in \R$. Using the identity
	\begin{equation}
	\label{idmasses}
		\prod_{j=1} ^{N_f} \frac{1}{1+ M_j X } = \sum_{j=1} ^{N_f} \frac{ 1 }{ 1+ M_j X  } \prod_{s \ne j} \frac{1}{1 - \frac{M_s}{M_j}} ,
	\end{equation}
	we can rewrite the partition function of the theory as
	\begin{align*}
		\mz_{U(1), N_f} (k, \vec{m} , \zeta ) & = \int dx \frac{ e^{i \pi k x^2 + i 2 \pi \zeta x } }{ \prod_{j=1} ^{N_f} 2 \cosh \pi (x+m_j) } \\ 
			& = \sum_{j=1} ^{N_f}  \frac{ e^{2 \pi m_j (N_f-2) }   }{ \prod_{s \ne j} \left( e^{2 \pi m_j} - e^{2 \pi m_s} \right) }  \mI_k \left( - m_j, \frac{N_f}{2} + i \zeta  \right),
	\end{align*}
	where we have used $\sum_{j=1}^{N_f} m_j =0$. From \eqref{IkMord} we obtain an explicit solution in terms of a sum of Mordell integrals for every rational value of the CS level $k$: 
	\begin{align*}
		\mz_{U(1), N_f} (k>0, \vec{m} , \zeta=0 ) & = \sum_{j=1} ^{N_f}  \frac{ e^{ \pi m_j (N_f-2) + i \pi  \frac{ N_f}{2} }  }{ \prod_{s \ne j} \left( e^{2 \pi m_j} - e^{2 \pi m _s} \right) }  \frac{1}{1- (-1)^{\varrho \left( N_f +1 - \kappa  + \frac{\kappa}{\varrho} \right) } e^{- 2 \pi k m_j }  } \notag  \\
		& \times  \left[ i e^{- i \pi k \left( \frac{1}{2} + i m_j \right)^2 } \sum_{\beta=1}^{\varrho} (-1)^{\beta N_f} e^{- i \pi k \beta (\beta-1) - 2 \pi k m_j \beta } \right. \\
		& \qquad \left.  - \sqrt{ \frac{i}{k} } \sum_{\alpha=1} ^{\kappa}  \left( - e^{- 2 \pi m_j  } \right)^{\alpha + \frac{N_f}{2} }  q^{ - \frac{1}{2} \left(\alpha + \frac{N_f}{2}  \right)^2} \right] ,
	\end{align*}
	and 
	\begin{align*}
		\mz_{U(1), N_f} (k<0, \vec{m} , \zeta=0 ) & = \sum_{j=1} ^{N_f}  \frac{ e^{ \pi m_j (N_f-2) + i \pi  \frac{ N_f}{2} }  }{ \prod_{s \ne j} \left( e^{2 \pi m_j} - e^{2 \pi m _s} \right) }  \frac{1}{1- (-1)^{\varrho \left( N_f +1 - \kappa  - \frac{\kappa}{\varrho} \right) } e^{- 2 \pi k m_j }  } \notag  \\
		& \times  \left[ i e^{- i \pi k \left( \frac{1}{2} + i m_j \right)^2 } \sum_{\beta=1}^{\varrho} (-1)^{\beta N_f} e^{- i \pi k \beta (\beta-1) - 2 \pi k m_j \beta } \right. \\
		& \qquad \left.  + \sqrt{ \frac{i}{k} } \sum_{\alpha=0} ^{\kappa-1}  \left( - e^{2 \pi m_j  } \right)^{\alpha- \frac{N_f}{2} }  q^{ - \frac{1}{2} \left(\alpha - \frac{N_f}{2}  \right)^2} \right] ,
	\end{align*}
	When the number of flavours is even the sums in the last line of each expression become (cfr. \eqref{genpolymuk}) 
	\begin{equation*}
		\widetilde{P} \left( - e^{- \mathrm{sign} (k) 2 \pi m_j} ; k \right) ,
	\end{equation*}
	but with the summation range shifted by $ - \frac{ N_f}{2}$. As we have already pointed out in Subsection \ref{sec:momentsSW}, these are polynomials in the variable $e^{i 2 \pi \lambda \text{sign} (k)}$, hence are holomorphic in $\mathbb{C} \setminus \mathbb{R}_{\ge 0}$.\par
	The effect of reintroducing the FI parameter $\zeta$ can be reabsorbed in a change of variable, and the result is the same as above up to a shift of the masses, as in \eqref{eq:A1Nf1ZmFIrel2}. 
	Besides, the result holds upon complexification of the masses and FI parameters, as long as $ \left\vert \Im m_j - \frac{\Im \zeta }{k}  \right\vert  < \frac{1}{2}$.\par
	The solution relied on the assumption of generic masses, but the theory has a well defined confluent limit when two masses become equal. One approach to this case is based on a direct analysis of the cancellations in the formula above. An alternative and especially convenient approach is to interpret the partition function as the average of inverse characteristic polynomials in the Stieltjes--Wigert ensemble, expressing it then as a $N_f \times N_f$ determinant, whose limit is well known to give a Wronskian determinant \cite{Tchar}. We discuss this approach in Subsection \ref{sec:charpol}. A third approach, valid for all equal masses, was taken in \cite{TGias}.

	\subsubsection{Wilson loops from Mordell integrals: Abelian $\mathsf{A}_1$ theory}
				We consider again the Abelian CS theory with $N_f$ massive fundamental hypermultiplets and insert a circular Wilson loop in a complex irreducible $U(1)$ representation $\mu$, identified with an integer $ \mu \in \mathbb{Z}$. Its expectation value is:
				\begin{align*}
					\langle W_{\mu} \rangle_{U(1), N_f} & = \frac{1}{\mz_{U(1), N_f} } \int_{- \infty} ^{\infty} dx \frac{ e^{i \pi k x^2  + 2 \pi \mu x } }{ \prod_{j=1} ^{N_f} 2 \cosh \pi (x + m_j) } \\
						&= \frac{1}{\mz_{U(1), N_f} } \sum_{j=1} ^{N_f}  \frac{ e^{2 \pi m_j (N_f-2) }   }{ \prod_{s \ne j} \left( e^{2 \pi m_j} - e^{2 \pi m_s} \right) }  \mI_k \left( - m_j, \frac{N_f}{2} + \mu  \right) ,
				\end{align*}
				hence the result can be easily extracted from the above analysis or directly from \eqref{IkMord}. When $N_f =2$ we have the particularly simple relation
				\begin{equation*}
					\langle W_{\mu} \rangle_{U(1), N_f} = \frac{\mz_{U(1),2\mu+2}}{\mz_{U(1),2} } .
				\end{equation*}

	\subsubsection{Non-Abelian $\mathsf{A}_1$ theory with $N_f$ flavours}
	\label{sec:A1NonAb-Nf}
\begin{figure}[ht]
	\centering
	\begin{tikzpicture}[auto,node distance=1cm,square/.style={regular polygon,regular polygon sides=4}]
		\node[circle,draw] (gauge) at (-1,0) {$N$};
		\node[square,draw] (flavour) at (1,0) {\hspace{7pt} };
		\node[draw=none] (aux) at (1,0) {$N_f$};
		\draw[-](gauge)--(flavour);
	\end{tikzpicture}
	\caption{Non-Abelian $\mathsf{A}_1$ quiver with $N_f$ fundamental flavours.}
	\label{fig:A1NANf}
	\end{figure}
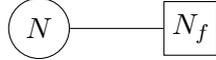

			The next example is the $U(N)$ theory with $N_f$ flavours, as in Figure \ref{fig:A1NANf}. The partition function at $N_f=2$ and no FI term has been solved in \cite{TRS}, using a change of variables of the form $X_a = e^{2 \pi ( x_a - c) }$ and writing the resulting expression as a Hankel determinant \cite{T0}. The crucial difference from \cite{TRS,TGias} is that we consider generic masses and also allow a FI term. In flat space, this choice lifts the Higgs branch and reduces the moduli space to isolated vacua.\par 
 
		We get \cite{TRS}: 
		\begin{align*}
			\mz_{U(N), N_f} (k, \vec{m} ) &= \prod_{a=1} ^{N} \int_{- \infty} ^{\infty}  d x_a  \frac{ e^{i \pi k x_a ^2  + i 2 \pi \zeta x_a } \ \prod_{ b \ne a} 2 \sinh \pi (x_{b}  - x_a ) }{ \prod_{j=1}^{N_f} 2  \cosh \pi \left( x_a+m_j \right)  }   \\
				&= e^{i \pi \frac{N}{k} \left[ N \left( N_f - 2N \right) - \zeta^2 \right] }  \det_{1 \le a,b \le N}  \left[ e^{  \frac{i \pi}{k}  \left( N_f -2N \right) (a+b-1) }  \int_{-\infty}  ^{+ \infty} dx \frac{ e^{ i \pi k x^2  + 2 \pi x \ell_{ab} } }{ \prod_{j=1} ^{N_f} \left( 1+e^{2 \pi (x- m^{\prime} _j)} \right) } \right] 
		\end{align*}
		where we defined for shortness $\ell_{ab} = a+b-1-N + \frac{N_f}{2}$ and $m^{\prime} _j = m_j - \frac{ \zeta}{k}$. Using \eqref{idmasses} each entry of the determinant is written as a sum of $N_f$ Mordell integrals:
		\begin{align}
			\mz_{U(N), N_f} (k, \vec{m} )  =  e^{i \pi \frac{N}{k} \left[ N \left( N_f -2N \right) - \zeta^2 \right] }  & \det_{1 \le a,b \le N}   \left[ e^{  \frac{i \pi}{k}  \left( N_f -2N \right) (a+b-1) } \right. \notag \\
			& \times \ \left. \sum_{j=1}^{N_f} \frac{ e^{2 \pi m^{\prime} _j (N_f-2) } }{ \prod_{s \ne j} \left( e^{2 \pi m^{\prime} _j} - e^{2 \pi m^{\prime} _s} \right) } \mI_k  \left( -m^{\prime} _j  , \ell_{ab} \right) \right].
		\end{align}
		This results extends \cite{TRS,TGias} to generic deformations, using a different approach than \cite{Tchar}. The massless limit can be taken, exploiting the identity \eqref{Gaussid} to see the cancellation of the singularities, cfr. \cite[Eq. 2.38]{TRS}, while the limit of coinciding masses is better understood in the formalism of \cite{Tchar}.\par

		\subsection{Lessons so far}
		\label{sec:Lessons}
			Before discussing quiver gauge theories, we pause to analyze the information that can be extracted by the exact solutions in terms of $\mI_k \left( -m, \frac{N_f}{2}  \right)$, as defined in \eqref{defIk}.\par
			A first observation is that the sums appearing in the right-hand side of the Mordell integrals are all of the form 
			\begin{align*}
				\sim & \sum_{\alpha=1} ^{k} \left( - e^{-2 \pi m} \right)^{\alpha} q^{- \frac{1}{2} \left( \alpha - \frac{N_f}{2}  \right)^2} \qquad (k>0) , \\
				\sim & \sum_{\alpha=0} ^{-k  -1} \left( - e^{2 \pi m} \right)^{\alpha} q^{- \frac{1}{2} \left( \alpha + \frac{N_f}{2}  \right)^2} \qquad (k<0) .
			\end{align*}
			Here we are considering $\zeta=0$ for clarity, but the argument goes through in exactly the same way turning on a real FI parameter. The shift in the Gaussian factor in each summand accounts for the shift $k \mapsto k- \frac{N_f}{2}$ from integrating out massive hypermultiplets.\par
			Another important aspect is that Mordell's solution is a holomorphic function of $\lambda= \frac{1}{2} + i m$ \cite{Mordell}. We notice that $\lambda$ is precisely the variable $\frac{\mathsf{t}}{2} + i m$ identified by Jafferis \cite{JafferisLoc} (see also the exhaustive discussion in \cite{Closset:2012vg}), with respect to which the partition function on $\mathbb{S}^3$ is holomorphic. Here $\mathsf{t}$ parametrizes the trial $U(1)_{\text{R}}$ R-charge of the hypermultiplet in the microscopic theory, and in our case is fixed to $\mathsf{t}= 1$ by the $\mN=3$ extended supersymmetry.\par
			Related to the just mentioned aspect, we stress the role of the numerator in the overall multiplicative term in \eqref{MordellG+} and \eqref{MordellG-}. This term always generates an overall factor $e^{- i \pi k \lambda^2}$, which is a CS coupling for the background vector multiplet of the global symmetry, precisely given in terms of the holomorphic variable $\lambda= \frac{1}{2} + i m$. On the other hand, a pure $U(1)_k$ CS theory coupled to a background vector multiplet generates an effective CS term $e^{i \frac{\pi}{k} \lambda^2 }$ \cite{Closset:2012vp,Closset:2012vg}, which emerges from the integrals. We also know that, for $N_f=1$, the theory must be self-dual under $k \mapsto -\frac{1}{k}$ \cite{Kapustin:1999ha}, as we have extensively discussed in Subsection \ref{sec:AbelianA1Nf1}. Therefore, the overall factor derived in \cite{Mordell} is essential to guarantee the invariance of the partition function under the S-duality when $N_f=1$, or more in general to reproduce the correct CS couplings for the background vector multiplets \cite{Closset:2012vp,Closset:2012vg}.\footnote{$\log \mathcal{Z}_{U(1),N_f}$ does not take a simple form, which prevents us from reading off the precise form of the mixed flavour-R CS couplings.}\par
			\medskip
			\subsubsection{$U(1)_k$ theory at rational $k$}
			\label{sec:Lessonkrational}
			
			As we have learned from the plots in Subsection \ref{sec:AbelianA1Nf} and the surrounding discussion, the study of the partition function $\mathcal{Z}_{U(1),N_f} \in \C $ at fixed $m$ as a function of $k= \frac{\kappa}{\varrho}$ uncovers a rich structure when $\varrho$ is increased keeping $\kappa$ fixed. This observation is compatible with the insight provided by the theory of Gauss sums \cite{coutsias1987disorder,berry1988renormalisation}. Pushing the analogy further, it may be interesting to understand the behaviour of $\mathcal{Z}_{U(1),N_f}$ when $k$ becomes irrational. This is not allowed in gauge theory for compact gauge group. However, the iterative application of the elementary Fourier transform identity 
			\begin{equation*}
				\int_{-\infty} ^{+\infty} dx \int_{-\infty} ^{+\infty} dy  \frac{ e^{i \pi a_1 x^2 + i \pi a_2 y^2 + i 2 \pi x y } }{ \prod_{j=1}^{N_f} 2 \cosh \pi (x+m_j) }  = \sqrt{\frac{i}{a_2}} \int_{-\infty} ^{+\infty} dx \frac{ e^{i \pi \left( a_1- \frac{1}{a_2} \right) x^2} }{ \prod_{j=1}^{N_f} 2 \cosh \pi (x+m_j) } 
			\end{equation*}
			allows to interpret the partition function at rational $k$, with continued fraction expansion 
			\begin{equation*}
			    k = \frac{\kappa}{\varrho} =  a_1 - \cfrac{1}{a_2 - \cfrac{1}{\dots  - \cfrac{1}{a_n} }} 
			\end{equation*}
			as a chain of $U(1)_{a_p}$ theories at integer CS levels $a_p$, $p=1, \dots, n$, with matter insertion only at the first node. Notice that this theory would correspond to a completely disconnected quiver (no bi-fundamentals), and the various nodes are coupled only through the mixed CS terms $k_{p,p+1}=1$. The CS level of the original theory attains an irrational value in the limit of infinitely many coupled CS theories.\par
			It would therefore be desirable to look further into the behaviour of $\mathcal{Z}_{U(1),N_f}$ when the number of integers $a_p$ in the continued fraction expansion of $k$ is increased, and eventually understand the $n \to \infty$ limit.

	\subsection{Abelian quivers}
	\label{sec:lnearquivers}
	We consider now Chern--Simons theories classified by Dynkin diagram of type $\mathsf{A}_r$, which correspond to linear quivers. We consider Abelian theories with gauge group $G= U(1)^{r}$.

		\subsubsection{Abelian $\mathsf{A}_2$ theory}
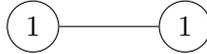
\begin{figure}[tbh]
	\centering
	\begin{tikzpicture}[auto,node distance=1cm]
		\node[circle,draw] (gauge1) at (1,0) {$1$};
		\node[circle,draw] (gauge2) at (-1,0) {$1$};
		\draw[-](gauge1)--(gauge2);
	\end{tikzpicture}
	\caption{Abelian $\mathsf{A}_2$ quiver.}
	\label{fig:A2Ab0}
	\end{figure}

		The first example is a two node quiver with Abelian gauge group, see Figure \ref{fig:A2Ab0}. The matter content consists only of the bi-fundamental hypermultiplet joining the nodes, to which we assign a mass $m$. We set the FI parameters to zero, as they can be reintroduced at the end by the usual shift of the masses and overall coefficient, as in \eqref{eq:A1Nf1ZmFIrel2}.\par
			 \par
		The partition function is:
		\begin{align*}
			\mz_{U(1)^2} (\vec{k}, m ) & = \int_{- \infty} ^{+ \infty} dx  \int_{- \infty} ^{+ \infty} dy \frac{ e^{ i \pi k_1 x^2  + i \pi k_2 y^2  } }{ 2 \cosh \pi (x-y+m) } \\
			& = \int_{- \infty} ^{+ \infty} dy e^{ i \pi k_2 y^2 +  \pi  (y-m) } \mI_{k_1} \left( y-m, \frac{1}{2} \right) .
		\end{align*}
		We take for concreteness $k_1  = \frac{\kappa_1}{\varrho_1}>0$ with either $\kappa_1$ even or $\varrho_1$ odd. This restriction is not necessary, but simplifies the expressions as we do not need to carry factors $(-1)^{\kappa_1 (\varrho_1-1)}$. From \eqref{MordellG+} and a rescaling of the integration variable, we get: 
		\begin{align*}
			\mz_{U(1)^2} (\vec{k}, m ) = \frac{1}{\kappa_1} & \left[ e^{i \pi k_1 (m^2 - \frac{1}{4}} \sum_{\beta=0} ^{\varrho_1} (- e^{-2 \pi k_1 m} )^{\beta} e^{i \pi k_1 \beta (\beta -1)} \mI_{k_{\text{eff}}} (0, \check{\xi}_1 (\beta) ) \right. \\
			& \left. + i \sqrt{\frac{i \varrho_1}{\kappa_1}} \sum_{\alpha=1} ^{\kappa_1} q_1 ^{-\frac{1}{2} \left( \alpha- \frac{1}{2} \right)^2} (- e^{-2 \pi m} )^{ \left( \alpha- \frac{1}{2} \right)} \mI_{k_{\text{eff}} ^{\prime}} \left( 0, \check{\xi}_2 (\alpha) \right) \right],
		\end{align*}
		where 
		\begin{align*}
		    k_{\text{eff}} & = \frac{k_2 \varrho_1 + \kappa_1}{\kappa_1 ^2} , \quad \check{\xi}_1 (\beta) = \frac{1}{\varrho_1} \left( -i m + \beta - \frac{1}{2} \right) , \\
		     k_{\text{eff}} ^{\prime} & = \frac{k_2}{\kappa_1 ^2} , \qquad \qquad \check{\xi}_2 (\alpha) = \frac{1}{\kappa_1} \left( \alpha - \frac{1}{2} \right) .
		\end{align*}
		The solution can be made explicit plugging \eqref{IkMord}. When $k_1+k_2=0$, it takes a much simpler form. We introduce both the mass and the FI parameter explicitly, assume $k>0$ without loss of generality, and write 
			\begin{align*}
			\mz_{U(1)^2} (\vec{k}, m, \zeta ) & = \int_{- \infty} ^{+ \infty} dx  \int_{- \infty} ^{+ \infty} dy \frac{ e^{ i \pi k (x^2 -y^) + i 2 \pi \zeta (x+y) } }{ 2 \cosh \pi (x-y+m) } \\
			& = \int_{- \infty} ^{+ \infty} dv \int_{- \infty} ^{+ \infty} dx  \frac{e^{-i \pi k v^2 + i 2 \pi \zeta v - i 2 \pi x (k v - \zeta ) } }{ 2 \cosh \pi (v-m) }  \\
			& = \frac{e^{i \pi \frac{\zeta^2}{k}} }{2 k \cosh \pi \left( m - \frac{\zeta}{k}\right) } ,
		\end{align*}
		where we have used the centre of mass variable $v=y-x$.\par
		In Subsection \ref{app:NonAbelianMordell} we present the computations of the lowest-rank non-Abelian $\mathsf{A}_2$ theory extending the ideas presented here.

		\subsubsection{Abelian $\mathsf{A}_3$ theory}
\begin{figure}[tbh]
	\centering
	\begin{tikzpicture}[auto,node distance=1cm]
		\node[circle,draw] (gauge1) at (0,0) {$1$};
		\node[circle,draw] (gauge2) at (-2,0) {$1$};
		\node[circle,draw] (gauge3) at (2,0) {$1$};
		\draw[-](gauge1)--(gauge2);
		\draw[-](gauge1)--(gauge3);
	\end{tikzpicture}
	\caption{Abelian $\mathsf{A}_3$ quiver.}
	\label{fig:A3Ab0}
	\end{figure}
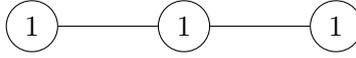

		The Abelian $\mathsf{A}_3$ quiver is depicted in Figure \ref{fig:A3Ab0}. We turn on a real FI parameter $\zeta$ in the middle node, and give masses $m_1$ and $m_2$ to the hypermultiplets. 
	 \par
		
		The partition function is:
		\begin{align*}
			\mz_{U(1)^3} (\vec{k}, \zeta, \vec{m} ) & = \int_{- \infty} ^{+ \infty} dv \int_{- \infty} ^{+ \infty} dx \int_{- \infty} ^{+ \infty} dy \frac{ e^{ i \pi \left( k_1 v^2 + k_2 x^2 + k_3 y^2 \right) + i 2 \pi \zeta x } }{ 2 \cosh \pi (v-x+m_1) 2 \cosh \pi (x-y+m_2)  }  .
		\end{align*}
		Instead of directly applying \eqref{IkMord}, we first use the change of variables 
		\begin{equation}
		    v^{\prime} = v-x, \quad y^{\prime} = y-x 
		\label{changevarcdm}
		\end{equation}
		(henceforth we drop the prime). We work under the assumption \cite{MartelliSparks:2008}
		\begin{equation*}
		    \sum_{p=1} ^{3} k_p =0 .
		\end{equation*}
		Integrating over $x$ we get 
		\begin{equation*}
		    	\mz_{U(1)^3} (\vec{k}, \zeta, \vec{m} ) = \frac{1}{\vert k_1 \vert} \int_{- \infty} ^{+ \infty} dv \frac{ e^{i \pi k_{\text{eff}} v^2 } }{ 2 \cosh \pi (v - m_2) 2 \cosh \pi \left( \frac{k_3}{k_1} v - m_2 + \frac{\zeta}{k_1} \right))  } ,
		\end{equation*}
		where we have defined the effective CS level
		\begin{equation*}
		     k_{\text{eff}}  = \frac{k_3}{k_1} (k_1 - k_3) .
		\end{equation*}
		At this point, from the denominator, we see that the tractable cases correspond to $k_3 = \pm k_1$. The first choice, $k_3=k_1$, means that we restrict to the one-parameter family of theories with CS levels 
		\begin{equation*}
		    (k_1, k_2, k_3) = (k, -2k, k) ,
		\end{equation*}
		in which case we get 
			\begin{equation*}
		    	\mz_{U(1)^3} ((k,-2k,k), \zeta, (m_1,m_2) ) = \frac{1}{\vert k \vert }\mz^{\text{SQED}}_{N_f=2} \left( - m_1 - \frac{\zeta}{k}, -m_2 \right) ,
		    \end{equation*}
		    where we have recognized the partition function of a single-node theory without CS term and two fundamental flavour of mass $- m_1 - \frac{\zeta}{k}$ and $-m_2$, respectively. Note that the hypermultiplet is off-shell, as it does not respect the $SU(2)$ flavour symmetry, unless we tune $ \frac{\zeta}{k}= -m_1 - m_2 $. We can safely turn off the FI parameter $\zeta$, as it only shifts $m_1$, and it is convenient to introduce an FI parameter $\tilde{\zeta}$ in the third node. We get \cite{SS}
		    \begin{equation*}
		        	\mz_{U(1)^3} ((k,-2k,k), \zeta, (m_1,m_2) ) = \frac{(e^{i 2 \pi m_2 \tilde{\zeta} } - e^{i 2 \pi m_1 \tilde{\zeta} })}{4 i \vert k \vert \sinh \pi (m_2 -m_1) \sinh (\pi \tilde{\zeta}) }  .
		    \end{equation*}\par
		    The other tractable case corresponds to the one-parameter family of CS theories with levels
		    \begin{equation*}
		        (k_1, k_2, k_3) = (k, 0 , -k) .
		    \end{equation*}
		    In this case $k_{\text{eff}}=-2k$, and the $U(1)^3$ partition function is given by 
		    \begin{equation*}
		    	\mz_{U(1)^3} ((k,0,-k), \zeta, (m_1,m_2) ) = \frac{1}{2 \vert k \vert \sinh \pi \left( m_2 + m_1 + \frac{\zeta}{k} \right)} \left[ \mI_{k_{\text{eff}}} (-m_1, 1 ) - \mI_{k_{\text{eff}}} (-m_2, 1 ) \right] ,
		    \end{equation*}
		    which, up to the factor $\vert k \vert^{-1}$, is the partition function of the $\mathsf{A}_1$ Abelian theory with $N_f=2$ studied in \cite{TRS} and in Subsection \ref{sec:A1AbelianNf2}, at level $k_{\text{eff}} = -2k$. \par
		    \medskip
		    A third instance in which the Abelian $U(1)^3$ theory is exactly solvable corresponds to the so-called Model III of Jafferis and Yin \cite{Jafferis:2008em}, with CS levels $\vec{k} =(1,-1,1)$. This theory is dual to SQED with two fundamental hypermultiplets and no CS couplings \cite{Jafferis:2008em}. The equality of the two partition functions, up to a phase, is easily proved from their integral representation, 
		    \begin{align}
			\mathcal{Z}^{\text{SQED}} _{N_f=2} (m^{\prime}, \zeta^{\prime}) & =  e^{i 2 \pi m^{\prime} \zeta^{\prime} } \int_{- \infty} ^{+\infty} dx \frac{ e^{i 2 \pi \zeta^{\prime} x }}{ [ 2 \cosh \pi (x+2 m^{\prime} )][ 2 \cosh (\pi x )]  } \notag  \\
			&= \frac{1}{\sqrt{i}} \left[ e^{- i \pi m_1 m_2} ~ \mz_{U(1)^3} ((1,-1,1), \vec{\zeta}=\vec{0}, m_1, m_2 ) \right]_{m_1= \zeta^{\prime}, m_2  = 2 m^{\prime} } \label{eq:ZA3fromSQEDNf2}
	    	\end{align}
	    	with the last equality following from the change of variables \eqref{changevarcdm}. The proof extends straightforwardly to the vev of a Wilson loop charged under one of the three $U(1)$'s.\par
		An exact evaluation of $\mathcal{Z}^{\text{SQED}} _{N_f=2}$ has been given in \cite{SS,TRusso}. In turn, we are able to evaluate the partition function on the $\mathsf{A}_3$ side using \eqref{IkMord}: 
		\begin{equation*}
			\mz_{U(1)^3} ((1,-1,1),  \vec{m} ) =  e^{\pi (-m_1 + m_2)} \int_{- \infty} ^{+ \infty} dx ~ e^{-i \pi x^2 + 2 \pi x}  \mI_{+1} \left(x-m_1, \frac{1}{2} \right)  \mI_{+1} \left(x+m_2, \frac{1}{2} \right) 
		\end{equation*}
		which, using \eqref{mIepshalf}, becomes 
		\begin{equation}
		    	\mz_{U(1)^3} ((1,-1,1),  \vec{m} ) = \frac{1}{2 \sinh \pi (m_1+ m_2)}  \left[  Z^{\mathrm{JY}}_1 (m_1,m_2) + Z^{\mathrm{JY}}_2 (m_1,m_2) + Z^{\mathrm{JY}}_3 (m_1,m_2) \right] ,
		\label{ZJaffYindescomp}
		\end{equation}
		where we have defined
		\begin{align*}
		Z^{\mathrm{JY}}_1 (m_1,m_2) & \equiv   \mI_{+1} (m_1, 1 -i m_1+ i m_2 ) -   \mI_{+1} (-m_2, 1 -i m_1+ i m_2 ) \\
		Z^{\mathrm{JY}}_2 (m_1,m_2) & \equiv i e^{i \pi \left( m_1 ^2 + m_2 ^2 \right) }  \left[  \mI_{+1} (m_1, 1 -i m_1+ i m_2 ) -   \mI_{+1} (-m_2, 1 -i m_1+ i m_2 ) \right] \\
		Z^{\mathrm{JY}}_3 (m_1,m_2) & \equiv- \sqrt{i}  \int_{- \infty} ^{+\infty} dx e^{2 \pi x } \left( e^{i \pi m_1 ^2 - i 2 \pi x m_1} + e^{i \pi m_2 ^2 + i 2 \pi x m_2} \right) \left[ \frac{ 1 }{ e^{2 \pi x } + e^{ 2 \pi m_1} }  - \frac{ 1 }{ e^{2 \pi x } + e^{- 2 \pi m_2} } \right]
        \end{align*}
	The first piece, which we have named $Z^{\mathrm{JY}}_1$, is given in \eqref{mIeps1} and contributes  
			\begin{equation*}
			  Z^{\mathrm{JY}}_1 (m_1, m_2 ) = \sqrt{-i} \left[ \frac{ 1 }{ 1- e^{2 \pi m_1} } - \frac{ 1 }{ 1- e^{- 2 \pi m_2} } + \frac{  e^{- i \pi m_1^2 }  }{ 2 \sinh ( \pi m_1) } - \frac{  e^{- i \pi m_2^2 }  }{ 2 \sinh ( \pi m_2) }  \right] .
		\end{equation*}
		The second piece is 
		\begin{equation*}
		Z^{\mathrm{JY}}_2 (m_1,m_2)= 	\sqrt{-i} e^{ i 2 \pi m_1 m_2 } \left[ \frac{  1 }{ e^{2 \pi m_2} -1 }   -   \frac{ e^{ - i \pi m_2 ^2} }{  2 \sinh ( \pi m_2) }   -   \frac{  1 }{ e^{-2 \pi m_2} -1 } -   \frac{ e^{ i \pi m_1 ^2} }{  2 \sinh ( \pi m_1) }   \right] .  
		\end{equation*}
		The last contribution is 
		\begin{equation*}
		    	Z^{\mathrm{JY}}_3 (m_1,m_2)= 	\sqrt{i} \left[  \frac{ e^{i \pi m_1 ^2 + i 2 \pi m_1 m_2 }  }{2 \sinh \left(  \pi m_1  \right)} - \frac{ e^{- i \pi m_1^2 }}{2 \sinh ( \pi  m_1 )} \right] + \sqrt{-i} \left[  \frac{ e^{i \pi m_2 ^2 + i 2 \pi m_1 m_2 }  }{2 \sinh \left(  \pi m_2  \right)} - \frac{ e^{- i \pi m_2 ^2 }}{2 \sinh ( \pi  m_2 )} \right] .
		\end{equation*}
		Plugging these three expressions back in \eqref{ZJaffYindescomp} and simplifying, we get 
		\begin{equation}
		\label{ZJafYinA3ex}
			\mz_{U(1)^3} ((1,-1,1),  \vec{m} ) = \sqrt{- i} \frac{ \left(  e^{- i 2 \pi m_1 m_2} - 1 \right) }{  [2 \sinh (\pi m_1)] [2 \sinh (\pi m_2)] }  .
		\end{equation}
		From \eqref{eq:ZA3fromSQEDNf2}, the result we find agrees with \cite{SS,TRusso}.

		\subsubsection{Abelian ABJM}
		\label{sec:AbelianABJM}
\begin{figure}[tbh]
	\centering
	\begin{tikzpicture}[auto,node distance=1cm]
		\node[circle,draw] (gauge1) {$1$};
		\node[circle,draw] (gauge2) [left = of gauge1]{$1$};
		\path (gauge1) edge [bend left] node {$\scriptstyle -m$} (gauge2);
		\path (gauge2) edge [bend left] node {$\scriptstyle +m$} (gauge1);
	\end{tikzpicture}
	\caption{Mass deformed Abelian ABJM theory.}
	\label{fig:ABJM1}
	\end{figure}
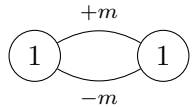

		We consider mass-deformed Abelian ABJM theory. This is $U(1)_k \times U(1)_{-k}$ Chern--Simons theory with two massive bi-fundamental hypermultiplets, represented in Figure \ref{fig:ABJM1}.\par

	 \par
	
		 The partition function of the theory is:
		 \begin{equation}
		 	\mz_{\mathrm{ABJ}(1 \vert 1)} (k, m) = \int_{- \infty} ^{+ \infty} dy  \int_{- \infty} ^{+ \infty} dx  \frac{ e^{ i \pi k (x^2 - y^2 ) } }{ 4 \cosh \pi (x - y +m) \cosh \pi (x - y -m) } 
		 \label{eq:MassDefABJMZ}
		 \end{equation}
		 where the variables $x$ and $y$ parametrize the two $\mathfrak{u} (1)$'s. From \eqref{defIk} we rewrite it as:
		 \begin{equation*}
		 	\mz_{\mathrm{ABJ}(1 \vert 1)} (k, m)  = \frac{1}{2 \sinh (2 \pi m ) } \int_{- \infty} ^{+ \infty} dy  e^{- i \pi k y^2 }  \left[ \mI_{k} (y-m, 1) - \mI_{k} (y+m, 1)  \right].
		 \end{equation*}
		 As one may expect, the contribution from a single node coincides with the partition function of $U(1)_k$ theory with two massive hypermultiplets with masses $ y \pm m$. Without loss of generality, we take $k>0$ and, from \eqref{IkMord} together with \eqref{MordellG+} we get
		 \begin{align*}
		 	\mz_{\mathrm{ABJ}(1 \vert 1)} (k, m)  & = \frac{1}{2 \sinh (2 \pi m ) }  \left\{  i  e^{i \pi k \left( m^2 - \frac{1}{4} \right) } \left[ \int_{- \infty} ^{+ \infty} dy \frac{ e^{- i 2 \pi k m y} }{ 2 \sinh (\pi k  (y-m)) } -  \int_{- \infty} ^{+ \infty} dy \frac{ e^{i 2 \pi k m y} }{ 2\sinh (\pi k  (y+m)) } \right] \right. \\
		 	& + \left. \sqrt{ \frac{i}{k} }  \sum_{\alpha=0} ^{k-1}  (-1)^{\alpha} q^{ - \frac{ \alpha^2}{2}}  \int_{- \infty} ^{+ \infty} dy  e^{- i \pi k y^2 } \left[ \frac{ e^{2 \pi \alpha (y+m)} }{ e^{2 \pi k (y+m) }  - 1 }  -  \frac{ e^{ 2 \pi \alpha (y-m)} }{ e^{2 \pi k (y-m) }  - 1 }  \right]  \right\}  .
		 \end{align*}
		 The two integrals in the first line are the Fourier transform of $\sinh(\pi x)$ and are immediately solved. The two integrals in the second line, after a change of variables $y^{\prime} = k(y \pm m)$ are reduced again to Mordell integrals:
		 \begin{align*}
		 	\mz_{\mathrm{ABJ}(1 \vert 1)} (k, m)  & = \frac{1}{2 k \sinh (2 \pi m ) }  \left\{  e^{ - i \pi k  \left( m^2+ \frac{1 }{4} \right) } \tanh ( \pi m )  \right. \\
		 		& \left. + e^{- i \pi k m^2} \sqrt{ \frac{i}{k} } \sum_{\alpha =0} ^{k-1} (-1)^{\alpha} q^{- \frac{\alpha^2}{2} } \left[ \Psi_{-} \left( - \frac{ \alpha }{k} - i m, 0; 1, k \right) - \Psi_{-} \left( - \frac{ \alpha}{k} + i m, 0; 1, k \right) \right]  \right\} .
		 \end{align*}
		 Plugging the solution \eqref{MordellG-} and after some simplification,
		  \begin{align*}
		 	\mz_{\mathrm{ABJ}(1 \vert 1)} (k, m)  & = \frac{e^{ - i \pi k m^2}}{2 k \sinh (2 \pi m ) }  \left\{ e^{ - i \pi \frac{k}{4}  } \tanh ( \pi m )  \right. \\
		 		& + \sum_{\alpha =0} ^{k-1} (-1)^{\alpha} \left[ e^{i \pi k m^2}  \left( \frac{ e^{2 \pi m \alpha} }{ (-1)^{k} e^{2 \pi k m} -1}  - \frac{e^{- 2 \pi m \alpha} }{ (-1)^{k} e^{- 2 \pi k m} -1} \right) \right. \\
		 		& \left. \left. +  i  \sqrt{ \frac{i}{k} } \sum_{\beta=1} ^{k} q^{ - \frac{(\alpha - \beta)^2}{2} } \left(  \frac{ e^{2 \pi m \beta} }{ (-1)^{k} e^{2 \pi k m} -1} - \frac{e^{- 2 \pi m \beta} }{ (-1)^{k} e^{-2 \pi k m} -1}  \right)  \right] \right\} .
		\end{align*}
		The second line is a geometric sum, with a prefactor $e^{i \pi k m^2}$. Using the Gauss sum identity \eqref{Gaussid} to sum over $\alpha$ in the third line, we find another geometric sum, over $\beta$ this time, which cancels the contribution of the first line. After these simplifications we get:
		\begin{equation*}
		    	\mz_{\mathrm{ABJ}(1 \vert 1)} (k, m, \zeta=0 ) = \frac{1}{4 k \cosh (\pi m)^2  } . 
		\end{equation*}\par
		
		In general, unitary $\widehat{\mathsf A}_r$ quivers have topological symmetry $[ \prod_{p=0} ^{r} U(1)_{\mathrm{top},p} ] /U(1)$. This allows us to introduce an FI parameter $\zeta$ turning on a background twisted vector multiplet for the $U(1)_{\mathrm{top}}$ topological symmetry of ABJM. This can be reabsorbed in a simple change of variables, and the result is directly obtained from above replacing $\pm m \mapsto \frac{ 2\zeta}{k} \pm m$. We get 
		\begin{equation}
		    	\mz_{\mathrm{ABJ}(1 \vert 1)} (k, m, \zeta ) = \frac{1}{ 4 k \cosh \pi  \left( m - \frac{2 \zeta}{k} \right) \cosh \pi  \left( m + \frac{2 \zeta}{k} \right)   } . 
		  \label{ZABJM1simplified}
		\end{equation}
	    The partition function, as written in \eqref{eq:MassDefABJMZ}, is invariant under $k \leftrightarrow -k$ but the final expression \eqref{ZABJM1simplified} is not because, without loss of generality, we have assumed $k>0$ in the intermediate steps.
		The result agrees with \cite{RussoSchaposnik}, where the answer was obtained in a straightforward way using a change of variables $x^{\prime} =x-y$ in \eqref{eq:MassDefABJMZ}. Nevertheless, with our approach we can consider the more general case with arbitrary rational $k_1$ and $k_2$, which corresponds to deform the gravity dual by a Romans mass $F_0=k_1+k_2 $, cfr. Subsection \ref{sec:GTRomans}. Letting $k_2 \ne - k_1$ and also allowing generic masses $m_1,m_2$ and a FI parameter $\zeta$, the partition function is
		 \begin{align*}
		 	\mz_{\mathrm{ABJ}(1 \vert 1)} (k_1, k_2 , \vec{m}, \zeta ) & = \int_{- \infty} ^{+ \infty} dx \int_{- \infty} ^{+ \infty} dy \frac{ e^{i \pi k_1 x^2 + i \pi k_2 y^2 + i 2 \pi \zeta (x+y)} }{ 2 \cosh \pi (x-y+m_1) 2 \cosh \pi (y-x + m_2)  } \\
		 	& = e^{- i \pi \frac{\zeta^2}{k_{\mathrm{eff}}}}  \int_{- \infty} ^{+ \infty} dv \int_{- \infty} ^{+ \infty} dy \frac{ e^{i \pi k_1 v^2 + i \pi (k_1+k_2) y^2 - i 2 \pi k_1 v y } }{ 2 \cosh \pi (v+m_{+}) 2 \cosh \pi (v + m_{-})  } \\
		 	& = e^{- i \pi \frac{\zeta^2}{k_{\mathrm{eff}}}} \sqrt{\frac{i}{k_1+k_2}} \int_{- \infty} ^{+ \infty} dv \frac{ e^{i \pi k_{\mathrm{eff}} v^2 }}{  2 \cosh \pi (v+m_{+}) 2 \cosh \pi (v + m_{-}) } .
		 \end{align*}
		 To pass from the first to the second line we have used the change of variables \cite{RussoSchaposnik} 
		 \begin{equation*}
		    v = \left( x + \frac{\zeta}{k_1} \right)  - \left( y+ \frac{\zeta}{k_2} \right)  
		 \end{equation*}
		 together with a redefinition of the parameters 
		 \begin{equation*}
		     m_{+} : = m_1 - \frac{\zeta}{k_1} + \frac{\zeta}{k_2} , \quad   m_{-} : = - m_2 - \frac{\zeta}{k_1} + \frac{\zeta}{k_2} , \quad k_{\mathrm{eff}} = \left( \frac{1}{k_1} + \frac{1}{k_2} \right)^{-1} .
		 \end{equation*}
		 In the last line, we recognize the partition function of the $U(1)$ CS theory with two fundamentals at level $k_{\mathrm{eff}}$, studied in Subsection \ref{sec:A1AbelianNf2}. For generic $k_1$ and $k_2 \ne k_1$ the effective CS level $k_{\mathrm{eff}}$ is rational, and we assume $k_{\mathrm{eff}} = \frac{\kappa}{\varrho}>0$. The partition function is 
		 \begin{align*}
		     \mz_{\mathrm{ABJ}(1 \vert 1)} (k_1, k_2 , \vec{m}, \zeta ) = & \frac{ e^{- i \pi \frac{\zeta^2}{k_{\mathrm{eff}}}} }{ 2 \sinh \pi (m_1 + m_2) }  \left\{   \frac{1}{(-1)^{\kappa(-1\varrho)} e^{- 2 \pi \kappa m_+ }-1}\left[  \frac{1}{\sqrt{\kappa}} \sum_{\alpha=0} ^{\kappa -1} (-e^{- 2 \pi m_+})^{\alpha} e^{i \pi \frac{\varrho}{\kappa} \alpha^2} \right. \right. \\
		     & \left. \left. + \frac{1}{\sqrt{i \varrho}} \sum_{\beta=1} ^{\varrho} e^{- i \pi \frac{\kappa}{\varrho} \beta (\beta -1) - 2 \pi \frac{\kappa}{\varrho} m_+ \beta  } \right]   \ - ( \text{ replace $m_+$ with $m_-$ }) \right\} .
		 \end{align*}\par
	   The theory has a well defined $m \to 0$ limit. For $k_2=-k_1$ equation \eqref{ZABJM1simplified} gives directly $\frac{1}{4k}$, while the limit $m \to 0$ for generic $k_1$ and $k_2$ is given in Subsection \eqref{ZABJM1simplified} making use of the Gauss sum identity \eqref{Gaussid}, and follows straightforwardly from \cite{TRS}.

		\subsubsection{Non-Abelian $\mathsf{A}_2$ theory}
		\label{app:NonAbelianMordell}
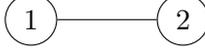
\begin{figure}[tbh]
				\centering
			\begin{tikzpicture}[auto,node distance=1cm]
				\node[circle,draw] (gauge1) at (1,0) {$2$};
				\node[circle,draw] (gauge2) at (-1,0) {$1$};
				\draw[-](gauge1)--(gauge2);
			\end{tikzpicture}
			\caption{The simplest non-Abelian $\mathsf{A}_2$ quiver.}
			\label{fig:A2NA}
			\end{figure}

		 	This Subsection contains an example of the application of the ideas of this Section to a non-Abelian quiver. We consider the simplest such theory, the $\mathsf{A}_2$ quiver with gauge group $U(1)_{k_1} \times U(2)_{k_2}$ and without any additional insertion, as in Figure \ref{fig:A2NA}. The bi-fundamental hypermultiplet has a real mass $m$.
		
			\par
			
			The partition function is 
			\begin{equation*}
			    \mathcal{Z}_{U(2)\times U(1) } ( k_1,k_2,m) = \frac{1}{2!} \int_{\R^2} dx_1 dx_2 \int_{- \infty} ^{+ \infty} dy \frac{ e^{i \pi k_1 (x_1 ^2 + x_2 ^2 ) + i \pi k_2 y } ~  \left( 2 \sinh \pi (x_1-x_2)  \right)^2 }{ 2 \cosh \pi (x_1 -y +m) 2 \cosh \pi (x_2 -y +m)  } .
			\end{equation*}
			A change of variables 
			\begin{equation*}
			    u = x_1 - y , \quad v= x_2 -y , \quad y^{\prime} = y + \frac{k_1}{2k_1 + k_2} (u+v)
			\end{equation*}
			allows to directly integrate out $y^{\prime}$, leaving 
			\begin{equation*}
			    \mathcal{Z}_{U(2)\times U(1) } ( k_1,k_2,m) = \sqrt{\frac{i}{2k_1 + k_2}}\int_{\R^2} \frac{du dv}{2} \frac{e^{i \pi \left( k_1 - \frac{k_1^2}{2 k_1 + k_2} \right) (u^2 + v^2) - i 2 \pi \frac{k_1 ^2 }{2k_1 + k_2} uv } ~\left( 2 \sinh \pi (x_1-x_2)  \right)^2}{ 2 \cosh \pi (u+m) 2 \cosh \pi (v+m) } .
			 \end{equation*}
		We discuss the two cases $k_1+k_2 =0$ and $k_1 +k_2 \ne 0 $ separately.\par
		When $k_1 = -k_2 \equiv k $, the CS coupling disappears after integrating over $y^{\prime}$. Expanding $\sinh \pi (u-v)^2$ and integrating over $v$ we get 
		\begin{align*}
		     \mathcal{Z}_{U(2)\times U(1) } ( k,-k,m) & = - \sqrt{\frac{i}{k}}\int_{- \infty} ^{+ \infty} du \frac{e^{i 2 \pi k m u } (e^{2 \pi (u+m)} +1 )}{  2 \cosh \pi (u+m) 2 \cosh (\pi ku) }  \\
		     & = - \sqrt{ \frac{i}{k} } \frac{ e^{ \pi m } }{2 k  \left[ \cosh (\pi m ) \cos \left( \frac{\pi}{2k} \right) + i  \sinh (\pi m ) \sin \left( \frac{\pi}{2k} \right) \right] } .
		\end{align*}
		When $k= \pm 1$ the partition function takes the specially simple form 
		\begin{equation*}
		    \mathcal{Z}_{U(2)\times U(1) } ( \pm 1 ,\mp 1 ,m) = \frac{ \sqrt{\mp i } }{ \vert k \vert (1 - e^{- 2 \pi m}) } .
		\end{equation*}
		\par
		When $k_1 \ne - k_2$ we have to invoke the Mordell integrals. It is convenient to slightly deform the denominator, replacing 
		\begin{equation*}
		    \prod_{a=1}^{2} 2 \cosh \pi (x_a -y +m) \mapsto 2 \cosh \pi (x_1 -y +m_1) 2 \cosh \pi (x_2 -y +m_2) ,
		\end{equation*}
		and eventually take the limit $m_1, m_2 \to m $ in the final expression. Besides, it is also more efficient to integrate first over $x_1$ and $x_2$ obtaining 
		 \begin{align*}
		     \mathcal{Z}_{U(2)\times U(1) } ( \vec{k}, \vec{m} ) = 2 e^{- \pi (m_1 + m_2)} \int_{- \infty} ^{+ \infty} dy~ e^{i \pi k_2 y^2 + 2 \pi y } & \left[  \mI_{k_1} \left( y-m_1, \frac{3}{2} \right) \mI_{k_1} \left( y-m_2, - \frac{1}{2} \right) \right. \\ 
		    & \left.  - \mI_{k_1} \left( y-m_1, \frac{1}{2} \right) \mI_{k_1} \left( y-m_2, \frac{1}{2} \right) \right] .
		 \end{align*}
		 The $\mI_{k_1}$ integrals give an overall denominator 
		 \begin{equation*}
		     \frac{1}{[ e^{2 \pi k_1 (y-m_1) } +1 ] [ e^{2 \pi k_1 (y-m_2) } +1 ]  } ,
		 \end{equation*}
		 whence we see that, thanks to the splitting of the masses, the last integral over $y$ can be solved again using the formula \eqref{IkMord}, this time with a rational effective CS level $\frac{k_2}{k_1}$. The resulting expression is a long multiple sum, which however admits a well-defined limit $m_1, m_2 \to m$, despite an overall factor $[2\sinh \pi (m_2-m_1)]^{-1}$, which can be dealt with in exactly the same manner as we have done in Subsection \ref{sec:A1AbelianNf2}. We conclude mentioning that the argument presented here is easily extended to ABJ theory with ranks $1$ and $2$ and arbitrary, possibly rational CS levels $k_1,k_2$, although it requires a convenient rewriting of the denominator and produces twice the number of terms than the theory with a single bi-fundamental that we have just discussed.

		\subsection{Abelian quivers at $k = \pm 1$}
		Beyond selected example that can be analyzed with the methods of this paper for a whole family of CS levels $\vec{k}$, the iterative application of Mordell's formula gives the quiver partition function when the CS levels are an alternating string of $+1$ and $-1$,
		\begin{equation}
		\label{eq:kCSpm1}
			\vec{k} = (1,-1,1 , \dots, -1) .
		\end{equation}
		In particular, when the rank is even, say $2r$, then $\vec{k}$ consists of alternating $+ 1$ and $- 1$, with exactly $r$ of each sign. When the rank is odd, say $2r+1$, then we take the middle node without CS couplings, in order to ensure 
		\begin{equation*}
			\sum_{p=1} ^{2r+1}  k_p =0 
		\end{equation*}
		for every rank. With such choice, the quiver is invariant under $\vec{k} \leftrightarrow - \vec{k}$. This symmetry is the diagonal action of the S-duality in the space of couplings. Let us stress that the restrictive choice of $\vec{k}$ is a sufficient condition that ensures the solvability through iterative application of Mordell's formula, but not necessary, as proved explicitly in the previous Subsections.\par
\begin{figure}[tbh]
	\centering
	\begin{tikzpicture}[auto,node distance=1cm]
		\node[circle,draw] (gauge1) {$1$};
		\node[circle,draw] (gauge2) [left = of gauge1]{$1$};
		\node[draw=none] (gaugemid) [left = of gauge2]{$\cdots$};
		\node[circle,draw] (gauge3) [left = of gaugemid]{$1$};
		\node[circle,draw] (gauge4) [left = of gauge3]{$1$};
		\draw[-](gauge1)--(gauge2);
		\draw[-](gaugemid)--(gauge2);
		\draw[-](gaugemid)--(gauge3);
		\draw[-](gauge4)--(gauge3);
	\end{tikzpicture}
	\caption{Abelian $\mathsf{A}_r$ quiver.}
	\label{fig:AbAr0}
	\end{figure}
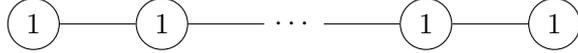

		With this condition, an example of theory solvable with the methods presented in the present work is the linear $\mathsf{A}_r$ quiver, with gauge group $U(1)^{r}$, represented in Figure \ref{fig:AbAr0}.
		 
		 \par
	
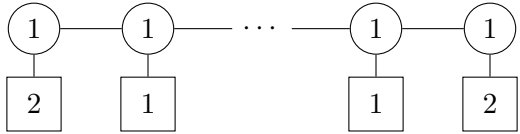
\begin{figure}[htb]
	\centering
	\begin{tikzpicture}[auto,square/.style={regular polygon,regular polygon sides=4}]
		\node[circle,draw] (gauge1) at (3,0) {$1$};
		\node[circle,draw] (gauge2) at (1.5,0) {$1$};
		\node[draw=none] (gaugemid) at (0,0) {$\cdots$};
		\node[circle,draw] (gauge3) at (-1.5,0) {$1$};
		\node[circle,draw] (gauge4) at (-3,0) {$1$};
		\node[square,draw] (fl1) at (3,-1) {$2$};
		\node[square,draw] (fl2) at (1.5,-1) {$1$};
		\node[square,draw] (fl3) at (-1.5,-1) {$1$};
		\node[square,draw] (fl4) at (-3,-1) {$2$};
		\draw[-](gauge1)--(gauge2);
		\draw[-](gaugemid)--(gauge2);
		\draw[-](gaugemid)--(gauge3);
		\draw[-](gauge4)--(gauge3);
		\draw[-](gauge1)--(fl1);
		\draw[-](gauge2)--(fl2);
		\draw[-](gauge3)--(fl3);
		\draw[-](gauge4)--(fl4);
	\end{tikzpicture}
	\caption{Abelian $\mathsf{A}_{r-1}$ quiver with one fundamental at each interior node and two fundamentals at the outermost nodes. In this picture the CS levels are all set to zero.}
	\label{fig:newArminus1quiver}
	\end{figure}

	On one hand, inspection of formula \eqref{defIk} has led us to a sufficient condition for the partition function of a linear quiver CS theory to be solved by iterative application of Mordell integrals. On the other hand, these theories are simple enough to be studied from a different angle. Let us focus on the even rank case. The partition function of the $\mathsf{A}_{2r}$ quiver in Figure \ref{fig:AbAr0} with CS levels \eqref{eq:kCSpm1} is 
	\begin{equation*}
	    \mz_{U(1)^{2r}} (\vec{m}) = \int_{- \infty} ^{+ \infty} dx_1 e^{i \pi x_1 ^2} \prod_{p=2} ^{2r} \int_{- \infty} ^{+ \infty} dx_p \frac{ e^{i \pi (-1)^{p-1} x_p ^2  } }{ 2 \cosh \pi (x_p - x_{p-1} + m_{p-1})  } .
	\end{equation*}
	We change variables 
	\begin{align*}
	    v_1 & = x_1 \\
	    v_2 & = x_2 - v_1 \\
	    v_3 & = x_3 - (v_2+v_1) \\
	    & \vdots \\
	    v_{2r} & = x_{2r} - \sum_{p=1}^{2r} v_p 
	\end{align*}
	and get 
		\begin{equation*}
	    \mz_{U(1)^{2r}} (\vec{m}) = \int_{- \infty} ^{+ \infty} dv_1 e^{-i 2 \pi v_1 (v_2 + v_4 + \cdots + v_{2r} ) } \prod_{p=2} ^{2r} \int_{- \infty} ^{+ \infty} dv_p \frac{ ( \text{ CS couplings } ) }{2 \cosh \pi (v_p + m_{p-1})} ,
	\end{equation*}
	with the bracket containing the CS couplings and mixed CS couplings in terms of the new variables $(v_2, \dots, v_{2r})$. The denominator, which carries the matter dependence, is completely factorized. The integral over the first variable yields a constraint on the variables at even nodes, 
	\begin{equation}
	    \delta \left( \sum_{p^{\prime} =1} ^{r} v_{2p^{\prime}} \right) .
	\label{deltaconstrevenvp}
	\end{equation}
	Besides, one can check that, thanks to the choice \eqref{eq:kCSpm1}, there is no CS level at the odd nodes, except for mixed CS couplings
	\begin{equation*}
	    \exp \left( - i 2 \pi v_p \sum_{p^{\prime} = (p-1)/2} ^{r} v_{2 p^{\prime}} \right) , \qquad \text{$p$ odd}. 
	\end{equation*}
	Thus the integral over $v_p$ can be solved straightforwardly for all odd $p$, yielding 
	\begin{equation*}
	    \int_{- \infty} ^{+ \infty} dv_p \frac{ e^{- i 2 \pi v_p \sum_{p^{\prime} = (p-1)/2} ^{r} v_{p^{\prime}} } }{ 2 \cosh \pi (v_p + m_{p-1}) } = \frac{1}{2 \cosh \pi \left( \sum_{p^{\prime} = (p-1)/2} ^{r} v_{2 p^{\prime}} + m_{p-1} \right)} , \qquad \text{$p$ odd}. 
	\end{equation*}
	We are left with the integral over the variables $v_{2 p^{\prime}}$, $p^{\prime}=1, \dots, r$, but we have the delta function \eqref{deltaconstrevenvp} to get rid of one of the variables, for example $v_2$. The advantage is that all the CS couplings are cancelled by \eqref{deltaconstrevenvp}, and we find:
	\begin{equation*}
	   \mz_{U(1)^{2r}} (\vec{m}) = \int_{\R^{r-1}} \frac{1}{2 \cosh \pi (\sum_{p^{\prime} =1}^{r} v_{2 p^{\prime}} - m_1 )} \prod_{p^{\prime}=2} ^{r} \frac{ dv_{2 p^{\prime}} }{ 2 \cosh \pi \left( \sum_{s=p^{\prime}} ^{r} v_{2s} + m_{2p^{\prime} -2}  \right) 2 \cosh \pi \left( v_{2p^{\prime}}  + m_{2p^{\prime}-1}  \right)  } 
	\end{equation*}
	Therefore the CS interactions can be removed from the computations, which are now reduced to $r-1$ integrals. We change again variables 
		\begin{align*}
	    y_1 & = v_{2r} \\
	    y_2 & = v_{2r-2} +y_1  \\
	    y_3 & = v_{2r-4} + y_2 \\
	    & \vdots \\
	    y_{r-1} & = v_4 + y_{r-2} 
	\end{align*}
	and arrive at 
	\begin{align*}
	   \mz_{U(1)^{2r}} (\vec{m}) & = \int_{\R^{r-1}} \frac{1}{2 \cosh \pi (y_1+ m_1 ^{\prime \prime} ) } \prod_{  p=1} ^{r-1} \frac{d y_p}{ 2 \cosh \pi ( y_p + m_p ^{\prime} ) 2 \cosh \pi ( y_p - y_{p+1} + m_p ^{\prime \prime} ) }  
	 \end{align*}
	 where in the formula $y_r \equiv 0$ and we have renamed the masses 
	 \begin{equation*}
	     m_p ^{\prime} = m_{2r - 2p } , \quad m_p ^{\prime \prime} =  m_{2r - 2p +1 } .
	 \end{equation*}
	In the latter form, we recognize the partition function of a linear quiver gauge theory of type $\mathsf{A}_{r-1}$, without CS term and with additional fundamental matter insertions, one at each node except for the first and last node, that yield two fundamentals. This is represented in Figure \ref{fig:newArminus1quiver}. This last theory has manifest $\mN=4$ theory, which was expected from the choice \eqref{eq:kCSpm1}.
	The partition function in this new form can be evaluated introducing FI terms $\zeta_{p^{\prime}}$ \cite{SS}, which can either be related to FI couplings in the original theory or we can take the limit $\zeta_{p^{\prime}} \to 0$ at the end. Note also that both the iterative application of Mordell formula and the method of \cite{SS} require the masses to be generic, but the limit of equal masses can be safely taken at the end of the calculations.\par
	
	 \par
	
	For the special case $r=2$ our formula states the equality of the partition function of the $\mathsf{A}_4$ quiver with alternating CS levels $+1$ and $-1$ with that of SQED with three fundamental flavour ($N_f=3$ can be seen by direct computations, or starting with $r=2$ and ungauging the second node in the $\mathsf{A}_2$ quiver), which are known to be dual \cite{Jafferis:2008em}.
	\par
	\medskip
	
\begin{figure}[tbh]
	\centering
	\begin{tikzpicture}[auto,node distance=1cm,square/.style={regular polygon,regular polygon sides=4}]
		\node[circle,draw] (gauge1) at (1,1) {$1$};
		\node[square,draw] (mat1) at (3,1) {\hspace{7pt} };
		\node[draw=none] (aux1) at (3,1) {$n_{\scriptscriptstyle{1}}$};
		\node[circle,draw] (gauge2) at (1,0) {$1$};
		\node[square,draw] (mat2) at (3,0) {\hspace{7pt} };
		\node[draw=none] (aux2) at (3,0) {$n_{\scriptscriptstyle{2}}$};
		\node[draw=none] (gaugemid) at (0,-1) {$\cdots$};
		\node[circle,draw] (gauge3) at (-1,0) {$1$};
		\node[square,draw] (mat3) at (-3,0) {\hspace{7pt} };
		\node[draw=none] (aux3) at (-3,0) {$n_{\scriptscriptstyle{r-1}}$};
		\node[circle,draw] (gauge4)  at (-1,1)  {$1$};
		\node[square,draw] (mat4) at (-3,1) {\hspace{7pt} };
		\node[draw=none] (aux4) at (-3,1) {$n_{\scriptscriptstyle{r}}$};
		\node[circle,draw] (gext) at (0,2) {$1$};
		\node[square,draw] (matext) at (2,2) {\hspace{7pt} };
		\node[draw=none] (auxext) at (2,2) {$n_{\scriptscriptstyle{0}}$};
		\draw[-](gauge1)--(gauge2);
		\draw[-](gaugemid)--(gauge2);
		\draw[-](gaugemid)--(gauge3);
		\draw[-](gauge4)--(gauge3);
		\path (gauge1) edge [] node {} (gext);
		\path (gext) edge [] node {} (gauge4);
		\draw[-](gauge1)--(mat1);
		\draw[-](gauge2)--(mat2);
		\draw[-](gauge3)--(mat3);
		\draw[-](gauge4)--(mat4);
		\draw[-](gext)--(matext);
	\end{tikzpicture}
	\caption{Abelian $\widehat{\mathsf{A}}_r$ extended quiver.}
	\label{fig:extendedAr}
	\end{figure}
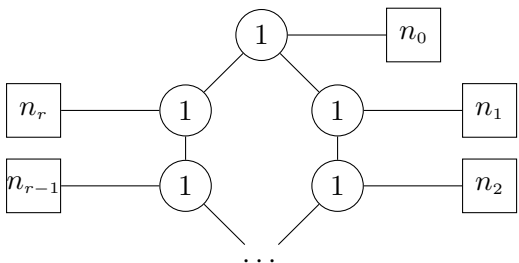

		The particularly suitable choice of CS levels \eqref{eq:kCSpm1} allows us to study a much wider class of quivers, such as extended $\widehat{\mathsf{A}}_r$ quivers with insertion of fundamental matter at any node, as in Figure \ref{fig:extendedAr}. Specializing to $n_p=0$ for all $p$, the resulting theories are Abelian sub-cases of \cite{Kimura:moduli,JafferisTom}. The Abelian $\widehat{\mathsf{A}}_r$ quiver with CS levels \eqref{eq:kCSpm1} corresponds to the gauge theoretical realization of the M-crystal model \cite{LeeCrystal} derived in \cite{Kim:2007ic,Hosomichi:2008CS}. Although a complete analytical solution seems hard to find, it should be possible to obtain explicit solutions for every $r$ through an algorithmic iteration of formula \eqref{IkMord}.\par
		The building blocks in the solution are the integrals $\mI_{k} (y, \check{\xi})$ defined in \eqref{defIk} at $k = \pm 1$ and $\check{\xi}=1$, or $\check{\xi}= \frac{1}{2}$ for boundary nodes of a linear quiver without additional matter insertion. They are evaluated as:
		\begin{equation}
			\label{mIepshalf}
			\mI_{k} \left( y, \frac{1}{2} \right) =  \frac{ i k }{e^{2 \pi y} +1} \left[ 1- e^{k i \pi \left( y^2 + \frac{1}{4} \right) } \right], \qquad k \in \left\{ \pm 1 \right\} 
		\end{equation}
		and 
		\begin{equation}
		\label{mIeps1}
			\mI_{k} \left( y, 1 \right) = \frac{ e^{\frac{i \pi}{4} k } }{ e^{2 \pi y } -1 } \left[ - 1 + e^{k  i \pi y^2 + \pi y  } \right] , \qquad k \in \left\{ \pm 1 \right\} .
		\end{equation}\par

		 \par

		\section{Wilson loops in ABJ theory}
		\label{sec:WLABJ}
	
		This Section is dedicated to the study of vacuum expectation values of $\frac{1}{2}$-BPS Wilson loops in ABJ(M) theories \cite{DrukTranca,Lee:2010hk}, whenever the Wilson loop is in a type of representation of $U(N_1 \vert N_2)$ called \emph{typical} representation (also known as \emph{long} representation in more physical settings). This distinction between types of representations emerges when considering Lie supergroups and superalgebras and has not been discussed in the context of Wilson loops of ABJ(M) theories before. Hence, we explain this first.
		
		\subsection{On Lie superalgebras representations}
		\label{sec:WLsuperalgebras}
		
While every finite-dimensional $\mathfrak{g}$-module of a semi-simple Lie
algebra $\mathfrak{g}$ is completely reducible (that is, every
representation decomposes into a direct sum of irreducible representations), this no longer holds for Lie superalgebras. 
A consequence of the classical Djokovic--Hochschild theorem \cite{djokovic1976semisimplicity} states that all simple Lie superalgebras, with the exception of the family $\left\{ \mathfrak{osp}(1,2n), n \ge 1 \right\}$ of ortho-symplectic Lie superalgebras, have indecomposable (that is, not completely reducible) representations.\par

This leads to the definition of two types of irreducible representations for
a Lie superalgebra $\mathfrak{g}$. Let $\mu $ be a highest weight for a finite
dimensional irreducible representation $\mathcal{R} \left( \mu \right)$ of $%
\mathfrak{g}.$ If the representation cannot be extended to an indecomposable representation of $\mathfrak{g}$, then it is called a typical representation. These are the ones that satisfy the usual properties of the irreducible representations of a Lie algebra. More involved are the atypical representations, which can be extended, with another $\mathfrak{g}$-module, in a manner that the new representation is an indecomposable representation of $ \mathfrak{g}$. Atypical representations appear, for example, in the decomposition of the tensor product of two typical representations.\par
By focusing on Wilson loops with typical representations we will be able to exploit a powerful mathematical factorization property for the characters of such representations \cite{Berele}.

	\subsection{Wilson loops in typical representations}
	\label{sec:WLdeftypical}
		
		$\frac{1}{2}$-BPS Wilson loops in ABJ(M) theories can be constructed as the trace of the holonomy of a $\mathfrak{u}(N_1 \vert N_2)$-valued superconnection \cite{DrukTranca}. We therefore consider an irreducible representation $\mathcal{R}(\mu)$ of the supergroup $U(N_1 \vert N_2)$ with highest weight labelled by a partition $\mu$. We henceforth identify $\mathcal{R} (\mu) \simeq \mu$, further identified with the Young diagram representing the partition $\mu$.\par
The vev of the $\frac{1}{2}$-BPS Wilson loop in the representation $\mu$ is \cite{MaPutExact}: 
		\begin{align*}
			\langle W_{\mu} \rangle_{N_1, N_2;k} = \frac{1}{ \mz_{\mathrm{ABJ} (N_1 \vert N_2)} (k) } & \int_{\R^{N_1} } d^{N_1} x  \int_{\R^{N_2} } d^{N_2} y  \ \mathfrak{s}_{\mu} \  ( e^{2 \pi x} \vert e^{2 \pi y} )  e^{i \pi k \left( \sum_{a=1} ^{N_1} x_a ^2 - \sum_{\dot{a}=1} ^{N_2} y_{\dot{a}} ^2  \right) }  \\
			& \times  \frac{ \prod_{1 \le a < b \le N_1}  \left( 2 \sinh  \pi (x_b - x_a ) \right)^2  \prod_{1 \le \dot{a} < \dot{b} \le N_2} \left( 2 \sinh \pi (y_{\dot{b}} - y_{\dot{a}} ) \right)^2  }{ \prod_{a=1}^{N_1} \prod_{\dot{a}=1}^{N_2}\left( 2 \cosh \pi ( x_a - y_{\dot{a}} ) \right)^2  } .
		\end{align*}
		$ \mz_{\mathrm{ABJ} (N_1 \vert N_2)} (k)$ is the ABJ partition function, and we are denoting $\langle \cdots \rangle_{N_1,N_2;k}$ the vevs taken in $U(N_1)_k \times U(N_2)_{-k}$ ABJ theory. Indices associated to the first node are labelled $a,b, \dots$ while indices corresponding to the second node are labelled by $\dot{a}, \dot{b}, \dots$, hence undotted indices are always meant to run from $1$ to $N_1$ and dotted indices run from $1$ to $N_2$. Moreover, $\mathfrak{s}_{\mu} ( \cdot \vert \cdot )$ is the supersymmetric Schur polynomial \cite{Macdonaldbook,BumpGam} (also known as hook Schur polynomial) associated to the partition $\mu$, and $e^{2 \pi x}$ and $e^{2 \pi y}$ stand for $(e^{2 \pi x_1} , \dots, e^{2 \pi x_{N_1}} )$ and $(e^{2 \pi y_1} , \dots, e^{2 \pi y_{N_2}} )$ respectively. Vevs of correlators of Wilson loops are taken inserting additional supersymmetric Schur polynomials in the matrix model. Notice that if $N_1=0$ or $N_2=0$ the supersymmetric Schur polynomial degenerates in the usual Schur polynomial, and the vev of a Wilson loop in $U(N_1)_{k}$ or $U(N_2)_{-k}$ pure CS theory is recovered.\par
\begin{figure}[htb]
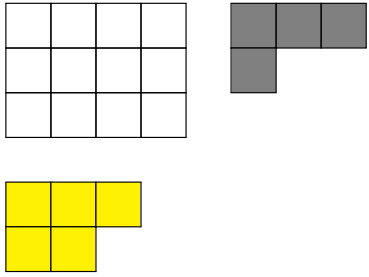

		\centering
		\ytableausetup{centertableaux}
		\begin{ytableau}
		 \ & \ & \ & \ & \none & *(gray) \ & *(gray) \ & *(gray) \ \\
		 \ & \ & \ & \ & \none & *(gray) \ \\
		 \ & \ & \ & \ \\
		 \none & \none \\
		 *(yellow) \ & *(yellow) \ & *(yellow) \  \\
		 *(yellow) \ & *(yellow) 
		\end{ytableau}
		\caption{Decomposition of a typical (\textit{i.e.} long) representation $\mu$. In this example, $N_1= 4,N_2=3$, the representation $\mu \simeq (7,5,4,3,2)$ is decomposed into $\kappa \simeq (4,4,4)$ (white), $\gamma \simeq (3,1)$ (gray) and $\eta \simeq (3,2)$ (yellow). Note that in the decomposition of $\mathfrak{s}_{\mu}$ it appears $\eta^{\prime} \simeq (2,2,1)$, and not $\eta$.}
		\label{fig:longrep}
		\end{figure}

		We now assume $\mu $ to be a typical representation of $U(N_1 \vert N_2)$, which implies that its associated Young diagram fills the upper-left $N_1 \times N_2$ rectangle. These representations have the remarkable factorization property \cite[Thm. 6.20]{Berele}
		\begin{equation}
		\label{factlongrep}
			\mathfrak{s}_{\mu} (X \vert Y ) = \mathfrak{s}_{\gamma} (X )  \mathfrak{s}_{\eta^{\prime}} ( Y ) \prod_{a=1} ^{N_1} \prod_{\dot{a} =1}^{N_2} \left( X_a + Y_{\dot{a}} \right) ,
		\end{equation}
		with $\mu = \left( \kappa + \gamma \right) \sqcup \eta $, with $\kappa$ the $N_1 \times N_2$ rectangular Young diagram, $\gamma$ the Young diagram consisting of the boxes of $\mu$ on the right of $\kappa$ and $\eta$ the Young diagram consisting of the boxes below $\kappa$, as in Figure \ref{fig:longrep}. The representation $\eta^{\prime}$ appearing in the factorization formula \eqref{factlongrep} is the conjugate representation of $\eta$, corresponding to the conjugate Young diagram. 
		\par

		\subsection{Two Wilson loops}
			One can foresee from \eqref{factlongrep} that part of the contribution from a long representation $\mu$ will cancel against the contribution from a bi-fundamental hypermultiplet. When the correlator of two Wilson loops is considered, one gets rid of the denominator in the two-matrix model, simplifying the computations. Taking the vev $\langle W_{\mu } W_{\tilde{\mu}} \rangle_{N_1, N_2;k} $, with $\vec{\mu}:=(\mu, \tilde{\mu})$ a pair of long representations, and using \eqref{factlongrep} we obtain:
			\begin{align}
				\langle W_{\mu} W_{\tilde{\mu}} \rangle_{N_1, N_2;k} & = \frac{1}{\mz_{\mathrm{ABJ}} }  \int_{\mathbb{R}^{N_1}} \mathfrak{s}_{\gamma} (e^{2 \pi x}) \mathfrak{s}_{\tilde{\gamma}} (e^{2 \pi x}) \prod_{1 \le a < b \le N_1} \left( 2 \sinh \pi (x_b - x_a) \right)^2 ~ \prod_{a=1} ^{N_1} e^{i \pi k x_a ^2 + 2 \pi N_2 x_a} ~dx_a \notag \\
					& \times \int_{\mathbb{R}^{N_2}} \mathfrak{s}_{\eta ^{\prime}} (e^{2 \pi y}) \mathfrak{s}_{\tilde{\eta} ^{\prime}} (e^{2 \pi y}) \prod_{1 \le \dot{a} < \dot{b} \le N_2} \left( 2 \sinh \pi (y_{\dot{b}}- y_{\dot{a}}) \right)^2 ~ \prod_{\dot{a}=1} ^{N_2} e^{- i \pi k y_{\dot{a}} ^2 + 2 \pi N_1  y_{\dot{a}}} ~dy_{\dot{a}} . \label{2WLintegral}
			\end{align}
			The correlator of two such Wilson loops in ABJ theory is therefore factorized into two pairs of Wilson loops, one pair for each node. Shifting variables and using basic properties of the Schur polynomials \cite{Macdonaldbook} we obtain:
			\begin{equation}
			\label{eq:fact2W}
				\langle W_{\mu} W_{\tilde{\mu}} \rangle_{N_1, N_2;k}  = C_{N_1, N_2; k} ^{\vec{\mu}} ~ \frac{\mz_{N_1;k} \mz_{N_2;-k} }{\mz_{\mathrm{ABJ} (N_1 \vert N_2)_{k}} }  \langle W_{\gamma} W_{\tilde \gamma} \rangle_{N_1; k}  \langle W_{\eta ^{\prime}} W_{\tilde \eta ^{\prime}} \rangle_{N_2; -k} .
			\end{equation}
			Here $\mz_{N_p,k} $ is the partition function of pure $U(N_p)$ bosonic Chern--Simons theory at renormalized level $k$, and $\langle \cdots \rangle_{N_p, k}$ is the average in the pure CS theory at node $p=1,2$. The shift of variables moves the integration cycle away from the real axis, but it can be translated back without changing the answer. The overall coefficient arising from the shift of variables is 
			\begin{equation*}
				C_{N_1, N_2;k} ^{\vec{\mu}} = \exp \left[ \frac{i \pi }{k} \left( N_2 \left( N_1 N_2  + 2 \vert \vec{\gamma} \vert \right) - N_1 \left( N_1 N_2 +2  \vert \vec{\eta}^{\prime} \vert  \right) \right) \right] ,
			\end{equation*}
			where $\vert \vec{\gamma} \vert $ is a shorthand for $\vert \gamma \vert + \vert \tilde{\gamma} \vert$, and the same for $\vert \vec{\eta}^{\prime} \vert$. Recall that $\vert \gamma \vert$ is the number of boxes in the Young diagram $\gamma$. Closely related results have been obtained in \cite{Kimura}, where the operator formalism was used to prove the factorization of the Hopf link invariant.\par
				The factorization property \eqref{eq:fact2W} is stable under deformation of the gravity dual theory by a Romans mass, taking different levels $k_1, k_2$. The procedure goes identically as above and gives
			\begin{equation}
				\langle W_{\mu} W_{\tilde{\mu}} \rangle_{N_1, N_2; k_1, k_2}  =  \frac{\mz_{N_1;k_1} \mz_{N_2;k_2} }{\mz_{\mathrm{ABJ} (N_1 \vert N_2)_{k_1,k_2}}  } ~C_{N_1, N_2; k_1 ,k_2} ^{\vec{\mu}} ~ \langle W_{\gamma} W_{\tilde \gamma} \rangle_{N_1; k_1}  \langle W_{\eta ^{\prime}} W_{\tilde \eta ^{\prime}} \rangle_{ N_2; k_2} ,   
				\label{fact2WLABJRomans}
			\end{equation}
			with refined coefficient
			\begin{equation*}
				C_{N_1, N_2;k_1, k_2} ^{\vec{\mu}} = \exp \left[ \frac{i \pi N_2 }{k_1} \left(  N_1 N_2 + 2 \vert \vec{\gamma} \vert  \right) + \frac{i \pi N_1 }{k_2} \left(  N_1 N_2 +2  \vert \vec{\eta}^{\prime} \vert  \right) \right] .
			\end{equation*}\par
		    The expression \eqref{fact2WLABJRomans} can be further reduced using a character expansion:
			\begin{equation*}
				\langle W_{\gamma} W_{\tilde \gamma} \rangle_{N_1; k_1}  = \sum_{\nu} {\mathsf{N}_{\gamma \tilde{\gamma}}}^{\nu} \langle W_{\nu} \rangle_{N_1;k_1}  ,
			\end{equation*}
			with ${\mathsf{N}_{\gamma \tilde{\gamma}}}^{\nu}$ the Littlewood--Richardson coefficients, and analogously for $\langle W_{\eta ^{\prime}} W_{\tilde \eta ^{\prime}} \rangle_{N_2; k_2}$. The vev of a Wilson loop in Chern--Simons theory along an unknot wrapping a great circle is known \cite{TDoli}, see \eqref{eq:WLCSk}, and the final form of \eqref{fact2WLABJRomans} is:
			\begin{equation*}
				\langle W_{\mu} W_{\tilde{\mu}} \rangle_{N_1, N_2; k_1, k_2}  = \frac{\mz_{N_1;k_1} \mz_{N_2;k_2} }{\mz_{\mathrm{ABJ} (N_1 \vert N_2)_{k_1,k_2}}  }  C_{N_1, N_2;k_1, k_2} ^{\vec{\mu}}  \sum_{\nu, \tilde{\nu}}  {\mathsf{N}_{\gamma \tilde{\gamma}}}^{\nu}  {\mathsf{N}_{\eta \tilde{\eta}}}^{\tilde{\nu}} (\dim_{q_1} \nu ) (\dim_{q_2} \tilde{\nu} ) e^{i \pi \left[ \frac{ \mathsf{C}_{2;N_1} (\nu)}{k_1} + \frac{ \mathsf{C}_{2;N_2} (\tilde{\nu})}{k_2} \right] }   . 
			\end{equation*}
			\par
			\medskip
			There exists an equivalent derivation, which consists in inverting the variables of one of the two Schur polynomials in each integrals in \eqref{2WLintegral}, using the identity
			\begin{equation}
			\label{invertschur}
				\mathfrak{s}_{\nu} (X_1 ^{-1}, \dots, X_{N} ^{-1}) = \prod_{a=1} ^{N} X_{a} ^{- \nu_1 } \mathfrak{s}_{\nu^{\ast}} ( X_1, \dots, X_{N}) ,
			\end{equation}
			with the starred partition defined as 
			\begin{equation}
			\label{defstarrednu}
				\nu^{\ast} = \left(  \nu_1 - \nu_N , \nu_1 - \nu_{N-1}, \dots, \nu_1 - \nu_2 \right) .
			\end{equation}
			We work directly with generic $k_1, k_2$ as the computations are identical. Exploiting \eqref{invertschur} we recognize in each factorized integral the vev of a Wilson loop wrapping a Hopf link in pure CS theory \cite{Kimura}: 
			\begin{equation}
			    \langle W_{\mu} W_{\tilde{\mu}} \rangle_{N_1, N_2; k_1, k_2}  = \frac{\mz_{N_1;k_1} \mz_{N_2;k_2} }{\mz_{\mathrm{ABJ} (N_1 \vert N_2)_{k_1,k_2}}  } ~ e^{ - i \pi \left[  \frac{N_1}{k_1} \tilde{\gamma}_1 ^2+ \frac{N_2}{k_2} (\tilde{\eta}^{\prime} _1) ^2 \right] } C_{N_1, N_2;k_1, k_2} ^{\vec{\mu}}  ~ \langle W_{\gamma \tilde{\gamma}^{\ast} } \rangle_{N_1; k_1}  \langle W_{\eta ^{\prime} (\tilde{\eta} ^{\prime})^{\ast}} \rangle_{ N_2; k_2} .
			\label{eq:2WLfactwithstarred}
			\end{equation}

			\subsubsection{Inverting one of the two Wilson loops}
			    A different correlator of two $\frac{1}{2}$-BPS Wilson loops than \eqref{2WLintegral} was considered in \cite{MoriyamaWL2p}, with one loop carrying inverted variables, mimicking the Hopf link invariant of \cite{Kimura}. This correlator has the integral representation 
			    \begin{align*}
				\langle W_{\mu} \overline{W}_{\tilde{\mu}} \rangle = \frac{1}{\mz_{\mathrm{ABJ}} } &  \int_{\mathbb{R}^{N_1}} \int_{\mathbb{R}^{N_2}}  \mathfrak{s}_{\mu} (e^{2 \pi x} \vert e^{2 \pi y} ) \mathfrak{s}_{\tilde{\mu}} (e^{-2 \pi x} \vert e^{-2 \pi y} ) ~  \prod_{a=1} ^{N_1} e^{i \pi k_1 x_a ^2 } ~dx_a ~\prod_{\dot{a}=1} ^{N_2} e^{i \pi k_2 y_{\dot{a}} ^2 } ~dy_{\dot{a}} \\
				& \frac{ \prod_{1 \le a < b \le N_1} \left( 2 \sinh \pi (x_b - x_a) \right)^2  \prod_{1 \le \dot{a} < \dot{b} \le N_2} \left( 2 \sinh \pi (y_{\dot{b}}- y_{\dot{a}}) \right)^2 }{ \prod_{a=1}
			^{N_1} \prod_{\dot{a}=1}^{N_2} \left( 2 \cosh \pi (x_a - y_{\dot{a}}) \right)^2 } 
			\end{align*}
			where in the left-hand side we have omitted the subscript, $\langle W_{\mu} \overline{W}_{\tilde{\mu}} \rangle \equiv \langle W_{\mu} \overline{W}_{\tilde{\mu}} \rangle_{N_1, N_2;k_1, k_2}$, to avoid clutter. We have also considered generic CS levels $k_1,k_2$ as we have seen that the argument holds with no difference. Using \eqref{factlongrep} on both supersymmetric Schur polynomial, with 
			\begin{equation*}
			    \mathfrak{s}_{\tilde{\mu}} (e^{- 2 \pi x}  \vert e^{- 2 \pi y}  ) 
			     = \mathfrak{s}_{\tilde{\gamma}} (e^{- 2 \pi x} ) \mathfrak{s}_{\tilde{\eta}^{\prime}} (e^{- 2 \pi y})   \prod_{a=1}^{N_1} \prod_{\dot{a}=1}^{N_2} \left( 2 \cosh \pi (x_a -y_{\dot{a}} ) \right) e^{- \pi x_a - \pi y_{\dot{a}}}  ,
			\end{equation*}
			we get 
			\begin{align*}
				\langle W_{\mu} \overline{W}_{\tilde{\mu}} \rangle & = \frac{1}{\mz_{\mathrm{ABJ}} }  \int_{\mathbb{R}^{N_1}} \mathfrak{s}_{\gamma} (e^{2 \pi x}) \mathfrak{s}_{\tilde{\gamma}} (e^{-2 \pi x}) \prod_{1 \le a < b \le N_1} \left( 2 \sinh \pi (x_b - x_a) \right)^2 ~ \prod_{a=1} ^{N_1} e^{i \pi k_1 x_a ^2} ~dx_a  \\
					& \times \int_{\mathbb{R}^{N_2}} \mathfrak{s}_{\eta ^{\prime}} (e^{2 \pi y}) \mathfrak{s}_{\tilde{\eta} ^{\prime}} (e^{-2 \pi y}) \prod_{1 \le \dot{a} < \dot{b} \le N_2} \left( 2 \sinh \pi (y_{\dot{b}}- y_{\dot{a}}) \right)^2 ~ \prod_{\dot{a}=1} ^{N_2} e^{i \pi k_2 y_{\dot{a}} ^2 } ~dy_{\dot{a}} .
			\end{align*}
			We find that the factorization persists, but the observables we get now are Hopf link invariants in $U(N_1)_{k_1}$ and $U(N_2)_{k_2}$ pure CS theory, instead of the correlator of two unlinked unknots:
			\begin{equation*}
			    	\langle W_{\mu} \overline{W}_{\tilde{\mu}} \rangle_{N_1, N_2 ; k_1, k_2} = \frac{\mz_{N_1;k_1} \mz_{N_2;k_2} }{\mz_{\mathrm{ABJ} (N_1 \vert N_2)_{k_1,k_2}}  } ~ \langle W_{\gamma \tilde{\gamma}} \rangle_{N_1;k_1} \langle W_{\eta^{\prime} \tilde{\eta}^{\prime}} \rangle_{N_2;k_2}  .
			\end{equation*}\par
			We could as well run the argument that led to \eqref{eq:2WLfactwithstarred} backwards. Inverting the variables in one of the two (ordinary) Schur polynomials in each integral using \eqref{invertschur} disentangles the Hopf link and gives the correlator of two circular Wilson loops, 
            \begin{equation*}
            	\langle W_{\mu} \overline{W}_{\tilde{\mu}} \rangle = \frac{\mz_{N_1;k_1} \mz_{N_2;k_2} }{\mz_{\mathrm{ABJ} (N_1 \vert N_2)_{k_1,k_2}}  } ~ e^{ i \pi \left[  \frac{N_1}{k_1} \tilde{\gamma}_1 ^2 +\frac{N_2}{k_2} (\tilde{\eta}^{\prime} _1) ^2 \right] } ~  \langle W_{\gamma} W_{\tilde{\gamma}^{\ast}} \rangle_{N_1;k_1} \langle W_{\eta^{\prime}} W_{(\tilde{\eta}^{\prime})^{\ast}} \rangle_{N_2;k_2}  .
            \end{equation*}
            The upshot is that having the variables of one of the two supersymmetric Schur polynomials inverted has the effect to switch the role of the partitions $\tilde{\gamma}$ and $\tilde{\eta}^{\prime}$ with that of the starred ones $\tilde{\gamma}^{\ast}$ and $(\tilde{\eta}^{\prime})^{\ast}$.

		\subsection{Three or more Wilson loops}
			Consider three long $U(N_1 \vert N_2)$ representations $\vec{\mu} = \left( \mu^{(1)}, \mu^{(2)}, \mu^{(3)} \right)$, and let 
			\begin{equation}
			    \left\langle W_{\vec{\mu}} \right\rangle \equiv 	\left\langle \prod_{j=1} ^{3} W_{\mu^{(3)}} \right\rangle_{N_1,N_2; k_1,k_2}
			\label{shorthandmultiWL}
			\end{equation}
			denote the correlator of three $\frac{1}{2}$-BPS Wilson loops carrying the representations $\vec{\mu}$ in ABJ theory with ranks $N_1$ and $N_2$, and we have allowed generic CS levels $k_1$ and $k_2$. We also denote for shortness $\mathfrak{s}_{\vec{\gamma}} (e^{2\pi x}) = \prod_{j=1} ^{3} \mathfrak{s}_{\gamma^{(j)}} (e^{2\pi x})  $, and likewise for $\mathfrak{s}_{\vec{\eta} ^{\prime}} (e^{2\pi y}) $. The correlator of the three Wilson loops, using \eqref{factlongrep}, is
			\begin{align*}
			 \left\langle W_{\vec{\mu}} \right\rangle & = \frac{1 }{\mz_{\mathrm{ABJ}} }  \int_{\mathbb{R}^{N_1}}  \int_{\mathbb{R}^{N_2}}   \prod_{1 \le a < b \le N_1} \left( 2 \sinh \pi (x_b - x_a) \right)^2   \prod_{1 \le \dot{a} < \dot{b} \le N_2} \left( 2 \sinh \pi (y_{\dot{b}} - y_{\dot{a}}) \right)^2   \\
				& \times \mathfrak{s}_{\vec{\gamma}} (e^{2 \pi x}) \mathfrak{s}_{\vec{\eta} ^{\prime}} (e^{2 \pi y})  ~ \left[ \prod_{a=1} ^{N_1} \prod_{\dot{a}=1} ^{N_2} \left( e^{2 \pi x_a} + e^{2 \pi y_{\dot{a}}} \right) e^{2 \pi ( x_a + y_{\dot{a}})} \right] ~ \prod_{a=1} ^{N_1} e^{i \pi k x_a ^2  } ~dx_a \prod_{\dot{a}=1} ^{N_2} e^{-i \pi k y_{\dot{a}} ^2  } ~d y_{\dot{a}} .
			\end{align*}
			The term in square bracket on the second line is a symmetric polynomials both in the variables $e^{2 \pi x_a}$ and $e^{2 \pi y_{\dot{a}}}$, and we can expand it in the Schur basis using the dual Cauchy identity \eqref{eq:dualCauchy}:
			\begin{equation}
			    \left[ \prod_{a=1} ^{N_1} \prod_{\dot{a}=1} ^{N_2} \left( e^{2 \pi x_a} + e^{2 \pi y_{\dot{a}}} \right) e^{2 \pi ( x_a + y_{\dot{a}})} \right] = \left( \prod_{a=1} ^{N_1} e^{2 \pi N_2 x_a} \right)\left( \prod_{\dot{a}=1} ^{N_2} e^{3 \pi N_1 y_{\dot{a}}} \right) \sum_{\nu} \mathfrak{s}_{\nu^{\prime}} (e^{2 \pi x} )  \mathfrak{s}_{\nu} (e^{-2 \pi y} ) ,
			\label{applydualcauchyto3WL}
			\end{equation}
			with the sum running over all partition of length at most $\min \left\{ N_1, N_2 \right\}$. At this point, the correlator is given by a finite sum of terms, each one completely factorized between the two nodes. We can exploit \eqref{invertschur} to invert the variables in the second Schur polynomial in \eqref{applydualcauchyto3WL}, and get the partition $\nu^{\ast}$ instead of $\nu$. We find 
			\begin{align}
			    \left\langle W_{\vec{\mu}} \right\rangle  = \frac{\mathcal{Z}_{N_1;k_1} \mathcal{Z}_{N_2;k_2} }{\mz_{\mathrm{ABJ}} } ~  \sum_{\nu } \tilde{C}_0 (\nu_1)  \tilde{C}_1 \left( \vec{\gamma}, \nu \right) \tilde{C}_2 \left( \vec{\eta} ^{\prime}, \nu^{\ast} \right) ~ \left\langle W_{\nu^{\prime}} W_{\vec{\gamma}}  \right\rangle_{N_1; k_1} \left\langle W_{\nu^{\ast}} W_{\vec{\eta}^{\prime}} \right\rangle_{N_2; k_2}  ,
			\label{factmultiWLcorrel}
			\end{align}
			where the coefficients are defined as 
			\begin{align*}
			    \tilde{C}_0 (\nu_1) & = \exp \left[ i \pi \frac{ N_1 N_2 ^2  }{k_1} + i \pi \frac{ N_2 }{k_2} \left( \frac{3}{2} N_1 - \nu_1  \right)^2  \right] , \\
			     \tilde{C}_1 \left( \vec{\gamma}, \nu \right) & =  \exp \left[  i 2 \pi \frac{N_2}{k_1} \left( \vert \vec{\gamma} \vert + \vert \nu \vert \right)   \right] , \\
			     \tilde{C}_2 \left( \vec{\eta} ^{\prime}, \nu^{\ast} \right) & =  \exp \left[  i 2 \pi \frac{N_1}{k_2} \left( \vert \vec{\eta} \vert + \vert \nu^{\ast} \vert \right)   \right].
			\end{align*}
			We are using, as in \eqref{shorthandmultiWL}, the shorthand notation $W_{\vec{\gamma}} \equiv \prod_{j} W_{ \gamma ^{(j)} }$, $\vert \vec{\gamma} \vert = \sum_j \vert \gamma ^{(j)} \vert$ and so on. We have also used $\vert \eta ^{\prime} \vert = \vert \eta \vert$, but note that $\vert \nu^{\ast} \vert \ne \vert \nu \vert$.\par
			Formula \eqref{factmultiWLcorrel} is factorized into two correlators of four ordinary Wilson loops is two pure CS theories, disconnected and without matter. Each correlator can be further simplified expanding pairwise the products of two Schur polynomials in the Schur basis, using the Littlewood--Richardson rule. Repeating this step twice reduces completely the vev $\langle W_{\vec{\mu}} \rangle$ to a finite sum of products of two ordinary Wilson loop vevs is two pure CS theories: 
			\begin{align*}
			    \left\langle W_{\vec{\mu}} \right\rangle  = \frac{\mathcal{Z}_{N_1;k_1} \mathcal{Z}_{N_2;k_2} }{\mz_{\mathrm{ABJ}} }  ~ \sum_{\nu } & \tilde{C}_0 (\nu_1)  \tilde{C}_1 \left( \vec{\gamma}, \nu \right)  ~ 
			    \left[ \sum_{\tilde{\nu}, \hat{\nu}, \check{\nu}} {\mathsf{N}_{\nu^{\prime} \gamma ^{(1)}} }^{\tilde{\nu}} {\mathsf{N}_{\gamma ^{(2)} \gamma ^{(3)}} }^{\hat{\nu}} {\mathsf{N}_{\tilde{\nu} \hat{\nu}} }^{\check{\nu}} ~ \langle W_{\check{\nu}} \rangle_{N_1;k_1}  \right] \\
			    & \times \tilde{C}_2 \left( \vec{\eta} ^{\prime}, \nu^{\ast} \right) \left[  \sum_{\tilde{\sigma}, \hat{\sigma}, \check{\sigma}}   {\mathsf{N}_{\nu^{\ast} \eta ^{(1) \prime}} }^{\tilde{\sigma}} {\mathsf{N}_{\eta ^{(2) \prime} \eta ^{(3) \prime}} }^{\hat{\sigma}} {\mathsf{N}_{\tilde{\sigma} \hat{\sigma}} }^{\check{\sigma}} ~  \langle W_{\check{\sigma}} \rangle_{N_2;k_2} \right] . 
			\end{align*}
			The Wilson loop vevs are known, cfr. \eqref{eq:WLCSk}, and the coefficients ${\mathsf{N}_{\mu \nu}}^{\tilde{\nu}}$ are the Littlewood--Richardson coefficients, and recall that the sum over $\nu$ only includes a finite number of terms.\par
			\medskip
			From the derivation, it is clear that the method applies to the correlator of any number of Wilson loops greater than two. Consider ABJ theory with ranks $N_1$ and $N_2$ and CS levels $k_1$ and $k_2$. Let $\vec{\mu}$ be a set of $n_{\mathrm{W}} \ge 2$ irreducible typical $U(N_1 \vert N_2)$ representations, and take the correlator of the $n_{\mathrm{W}}$ $\frac{1}{2}$-BPS Wilson loops in the representations $\vec{\mu}$. The recipe to compute the correlator is:
			\begin{itemize}
			    \item apply the factorization \eqref{factlongrep} to all the $n_{\mathrm{W}}$ supersymmetric Schur polynomials, and 
			    \item simplify two of the products arising from \eqref{factlongrep} with the denominator coming from the bi-fundamental hypermultiplets.
			    \item Apply $n_{\mathrm{W}} - 2$ times the dual Cauchy identity \eqref{eq:dualCauchy} to expand all the remaining products in the numerator in the Schur basis.
			    \item Use \eqref{invertschur} to bring all the Schur polynomials with variables $e^{- 2 \pi x}$ or $e^{- 2 \pi y}$ into functions of $e^{2 \pi x}$ and $e^{2 \pi y}$.
			    \item Expand the product of ordinary Schur polynomials pairwise using the Littlewood--Richardson rule. Repeat this step until the products are completely reduced.
			    \item The final result is a \emph{finite} sum of Wilson loop vevs in pure CS theory, wrapping an unknotted great circle in $\mathbb{S}^3$.
			\end{itemize}
			Besides, we notice that if some of the supersymmetric Schur polynomials have inverted variables \cite{MoriyamaWL2p}, the recipe does not change and they are taken care of in the fourth step.\par
			As a sample application, consider the particular case of four rectangular $N_1 \times N_2$ Young diagrams, $\vec{\mu} = (\kappa, \kappa, \kappa, \kappa )$. From \eqref{factlongrep} we obtain 
			\begin{align*}
			    \langle \left( W_{\kappa} \right)^4 \rangle =  \frac{1 }{\mz_{\mathrm{ABJ}} } & \int_{\mathbb{R}^{N_1}}  \int_{\mathbb{R}^{N_2}}   \prod_{1 \le a < b \le N_1} \left( 2 \sinh \pi (x_b - x_a) \right)^2   \prod_{1 \le \dot{a} < \dot{b} \le N_2} \left( 2 \sinh \pi (y_{\dot{b}} - y_{\dot{a}}) \right)^2   \\
				& \times  \left[ \prod_{a=1} ^{N_1} \prod_{\dot{a}=1} ^{N_2} 2 \cosh \pi (x_a - y_{\dot{a}} ) \right]^2 ~ \prod_{a=1} ^{N_1} e^{i \pi k x_a ^2 + 4 \pi N_2 x_a } ~dx_a \prod_{\dot{a}=1} ^{N_2} e^{-i \pi k y_{\dot{a}} ^2  + 4 \pi N_1 y_{\dot{a}} } ~d y_{\dot{a}} ,
			\end{align*}
			which, except for the normalization by $\mathcal{Z}_{\text{ABJ}}$, is the partition function of pure $U(N_1 + N_2)$ CS theory on the lens space $L(2,1) \simeq \mathbb{S}^3/\mathbb{Z}_2$, evaluated in the background of a fixed, generic flat connection that breaks the gauge symmetry 
			\begin{equation*}
			    U(N_1 + N_2) \longrightarrow U(N_1) \times U(N_2) .
			\end{equation*}
			Following the steps listed above, we get 
			\begin{align*}
			    \langle \left( W_{\kappa} \right)^4 \rangle =   \frac{\mathcal{Z}_{N_1;k_1} \mathcal{Z}_{N_2;k_2} }{\mz_{\mathrm{ABJ}} }  ~ \sum_{\nu , \tilde{\nu} }  \hat{C} (\nu, \tilde{\nu}) & \left[ \sum_{\hat{\nu}} {\mathsf{N}_{\nu^{\prime} \tilde{\nu}^{\prime} } }^{ \hat{\nu} } (\dim_{q_1} \hat{\nu}) q_1 ^{- \frac{1}{2} \mathsf{C}_{2;N_1} (\hat{\nu})} \right] \\ & \times  \left[ \sum_{\hat{\sigma}} {\mathsf{N}_{\nu^{\ast} \tilde{\nu}^{\ast} } }^{ \hat{\sigma} } (\dim_{q_1} \hat{\sigma}) q_1 ^{- \frac{1}{2} \mathsf{C}_{2;N_1} (\hat{\sigma})} \right] ,
			\end{align*}
			with coefficient 
			\begin{equation*}
			    \hat{C} (\nu, \tilde{\nu}) = \exp \left[  i \pi \frac{N_1}{k_1}  \left( N_2 ^2 + 2 \vert \nu \vert + 2 \vert \tilde{\nu} \vert \right) + i \pi \frac{N_2}{k_2}  \left(  \left( 2 N_1 - \nu_1 - \tilde{\nu}_1 \right)^2  + \vert \nu^{\ast} \vert + \vert \tilde{\nu}^{\ast} \vert  \right) \right] .
			\end{equation*}
			The complete partition function of pure $U(N)$ CS theory on $L(2,1)$ is obtained from this expression, dropping the overall normalization and summing over all $N_1$ and $N_2$ with $N_1+N_2=N$ fixed.

		\subsection{Necklace quivers}
		\label{sec:WLnecklace}
		    We now discuss the insertion of supersymmetric Schur polynomials in the matrix model describing quiver CS theories 
		    \begin{equation*}
		        U(N_0)_{k_0} \times U(N_1)_{k_1} \times \cdots \times U(N_r)_{k_r} .
		    \end{equation*}
		    We focus for clarity on an extended $\widehat{\mathsf{A}}_r$-type quiver, periodically identifying the nodes $r+1 \equiv 0$, being the discussion for linear quivers completely analogous. Let us fix $p\in \left\{ 0, \dots, r \right\}$ and consider a typical $U(N_p \vert N_{p+1}) $ representation $\mu$. The average of the supersymmetric Schur polynomial $\mathfrak{s}_{\mu}$ is 
		    \begin{align*}
		        \langle \mathfrak{s}_{\mu} \rangle = \int_{\R^{r+1}} & \mathfrak{s}_{\mu} \left( e^{2 \pi x_p} \vert e^{2 \pi x_{p+1}}  \right) ~ \prod_{p=1} ^{r} \prod_{a=1}
		       ^{N_p} e^{i \pi k_p x_a ^2 } dx_a \\
		       & \frac{ \prod_{1 \le a < b \le N_p} \left( 2 \sinh \pi (x_{p,a} - x_{p,b}) \right)^2 \prod_{1 \le \dot{a} < \dot{b} \le N_{p+1} } \left( 2 \sinh \pi (x_{p+1,a} - x_{p+1,b}) \right)^2 }{ \prod_{a=1}^{N_p} \prod_{\dot{a}=1} ^{N_{p+1}} 2 \cosh \pi (x_{p,a} - x_{p+1, \dot{a}}) }  .
		    \end{align*}
		    The identity \eqref{factlongrep} has the net effect to cut the edge joining the $p^{\text{th}}$ node to the $(p+1)^{\text{th}}$, leaving behind the correlator of two Wilson loops, one in the $U(N_p)$ representation $\gamma$ and the other in the $U(N_{p+1})$ representation $\eta^{\prime}$, computed in a $\mathsf{A}_{r+1}$ linear quiver gauge theory.\par
		    The correlator of more than one supersymmetric Schur polynomial, taken in typical representations of different supergroups $U(N_p \vert N_{p+1})$, cuts the edges joining each pair of nodes involved in the definition of the supersymmetric Schur polynomials. The final expression is factorized into the correlators of Wilson loops in disconnected linear quivers, with the loop operator inserted at the first or last node of each sub-quiver.\par
\begin{figure}[tbh]
	\centering
	\begin{tikzpicture}[auto]
		\node[circle,draw] (gauge0) at (-1,1) { \hspace{10pt} };
		\node[draw=none] (aux0) at (-1,1) {$N_0$};
    	\node[circle,draw] (gauge1) at (1,1) { \hspace{10pt} };
		\node[draw=none] (aux1) at (1,1) {$N_1$};
		\node[circle,draw] (gauge2) at (1,-1) { \hspace{10pt} };
		\node[draw=none] (aux2) at (1,-1) {$N_2$};
    	\node[circle,draw] (gauge3) at (-1,-1) { \hspace{10pt} };
		\node[draw=none] (aux3) at (-1,-1) {$N_3$};
		\draw[-,draw=blue] (gauge0.north east)--(gauge1.north west);
	    \draw[-,draw=red] (gauge1.south east)--(gauge2.north east);
	    \draw[-] (gauge0)--(gauge1);
	    \draw[-] (gauge1)--(gauge2);
	    \draw[-] (gauge2)--(gauge3);
	    \draw[-] (gauge3)--(gauge0);
	\end{tikzpicture}\hspace{0.1\textwidth}
	\begin{tikzpicture}[auto]
		\node[circle,draw] (gauge0) at (-3,-1) { \hspace{10pt} };
		\node[draw=none] (aux0) at (-3,-1) {$N_0$};
    	\node[circle,draw] (gauge1) at (1,1) { \hspace{10pt} };
		\node[draw=none] (aux1) at (1,1) {$N_1$};
		\node[circle,draw] (gauge2) at (1,-1) { \hspace{10pt} };
		\node[draw=none] (aux2) at (1,-1) {$N_2$};
    	\node[circle,draw] (gauge3) at (-1,-1) { \hspace{10pt} };
		\node[draw=none] (aux3) at (-1,-1) {$N_3$};
		\node[draw=none] (wl1) at (0.7,1.1) {{\color{red}${\scriptstyle \bullet}$}};
		\node[draw=none] (wl2) at (0.7,0.9) {{\color{blue}${\scriptstyle \bullet}$}};
			\node[draw=none] (wl3) at (1,-0.7) {{\color{red}${\scriptstyle \bullet}$}};
		\node[draw=none] (wl4) at (-3,-0.7) {{\color{blue}${\scriptstyle \bullet}$}};
	    \draw[-] (gauge2)--(gauge3);
	    \draw[-] (gauge3)--(gauge0);
	\end{tikzpicture}
	\caption{Left: $\widehat{\mathsf{A}}_3$ quiver with two supersymmetric Schur polynomial insertions, represented as a blue and a red line respectively. Right: the same quantity is factorized into two disjoint sub-quivers, with blue and red dots denoting ordinary Schur polynomial insertions.}
	\label{fig:necklaceBerelefact}
	\end{figure}
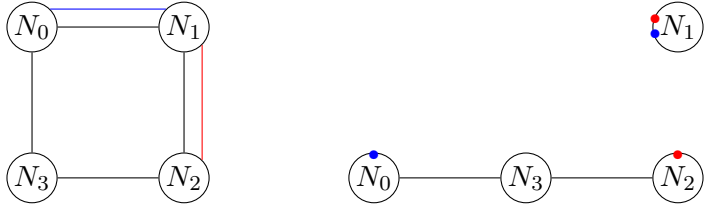

		    Consider, for example, a necklace quiver with four nodes, and take a typical $U(N_0 \vert N_1)$ representations $\mu$ and a typical $U(N_1 \vert N_2)$ representations $\tilde{\mu}$, as in Figure \ref{fig:necklaceBerelefact}. We find 
		    \begin{equation*}
		        \langle \mathfrak{s}_{\mu} \mathfrak{s}_{\tilde{\mu}} \rangle_{\widehat{\mathsf{A}}_3} = \frac{\mathcal{Z}_{N_1;k_1} \mathcal{Z}_{\mathsf{A}_3} }{\mathcal{Z}_{\widehat{\mathsf{A}}_3}} ~ \langle W_{\gamma} W_{\tilde{\eta}^{\prime}} \rangle_{\mathsf{A}_3}   \langle W_{\eta^{\prime}} W_{\tilde{\gamma}}\rangle_{N_1;k_1} .
		    \end{equation*}

		     \par
	The special case 
	\begin{equation*}
	    \left\langle \prod_{p=0}^{r} \mathfrak{s}_{\mu^{(p)}} \right\rangle  \qquad \text{with $ \mu^{(p)}$ a typical $U(N_p \vert N_{p+1})$ representation}
	\end{equation*}
	is completely factorized into correlators of pairs of Wilson loops in ordinary, bosonic $U(N_p)$ CS theory with renormalized CS level $ k_p $, for all $p=0,1,\dots, r$.


	\section{Schur expansion and its perturbative meaning}
	\label{sec:Schurexpansion}
	
		In this Section, we exploit the Cauchy identity \eqref{eq:Cauchyid} in different classes of Chern--Simons-matter theories and uncover a relation between the partition function of such theories and formal power series encoding topological invariants of simple links and (un)knots. As we will see, the series that appear are coarse-grained versions of generating functions. The invariants we obtain on the right-hand side are associated to either the unknot, a collection of unlinked unknots, or the Hopf link, coloured by $U(N)$ or $SU(N)$ representations. If we denote by $t$ the variable in the generating-like series of such link invariants, we find that it is related to the physical quantities of the gauge theory we started with through 
		\begin{equation}
		\label{eq:deftfugacity}
			t = - e^{2 \pi m }
		\end{equation}
		where $m$ is a real mass parameter. If there are more mass parameters, associated to the Cartan subalgebra of the flavour symmetry, we get a collection $\left\{ t_j \right\} = \left\{ - e^{2 \pi m_j} \right\}$.\par
		The simple rewriting 
		\begin{equation*}
			t = e^{- i 2 \pi \left( \frac{1}{2} +i m \right)} \equiv e^{- i 2 \pi \lambda } 
		\end{equation*}
		shows that $t$ is a fugacity for the variable $\lambda= \frac{1}{2} +i m$, which is ubiquitous in the calculations of Section \ref{sec:exactZ}. More accurately stated, in Sections \ref{sec:momentsSW} and \ref{sec:exactZ} we have found that the results are functions of the fugacity $e^{i 2 \pi \lambda ~\text{sign} \Re k }$, but it is in fact a matter of conventions whether we choose to expand in positive or negative powers of $t$, as will be clear from the examples below.\par
		As we already pointed out in Subsection \ref{sec:Lessons}, the partition function is holomorphic in $\lambda$, which is precisely the holomorphic variables of Jafferis \cite{JafferisLoc}, but further constrained by the $\mN \ge 3$ supersymmetry in all the theories considered in the present work. Besides, we have found holomorphy in the vertical strip $\left\{  0 < \Re \lambda <1 , - \infty < \Im \lambda < + \infty  \right\}$ \cite{Mordell}, thus the partition functions are holomorphic functions of $t \in \C \setminus \R_\ge 0 $.\par
		The fact that turning off background values for the flavour symmetry corresponds to ``take the Euler characteristic'', $t \to -1$, may point toward an interpretation in terms of categorification of link invariants \cite{Gukov:2004,Gukov:2007}, although not in the direction of the Khovanov--Rozansky homology. However, as we will see explicitly in the examples below, the quantities we obtain with our prescription have a too simple structure to capture homological data. In conclusion, there are obstructions in embedding the results presented in this Section into some homological theory of knots, as we spell in more detail in Appendix \ref{app:noknot}.\par
		\medskip
		Before diving into the detailed analysis, a remark is in order. It is important to bear in mind that the Cauchy identity \eqref{eq:Cauchyid} is algebraic, and is meant as an equality of the coefficients of the book-keeping variables $\left\{ t_j \right\}$ order by order in a (possibly formal) series expansion.\footnote{The dual Cauchy identity \eqref{eq:dualCauchy}, instead, is a finite sum and this issue does not show up.} We will use the symbol ``$ \stackrel{\text{pert.}}{=} $'' to signify that the equality between the left- and the right-hand side will be understood as equating the coefficients of each variable $t_j$ order by order. Note that the distinction between perturbative and non-perturbative in all the formulas in this Section is meant as functions of the fugacities $\left\{ t_j \right\}$ of the global symmetries, and not as functions of the gauge or CS couplings.
		
		\subsubsection{A toy example: Dawson's integral}
		To set the ground for the Schur expansion of physically sensible theories in the forthcoming Subsections, we firstly present our argument in a toy model. Consider the integral 
		\begin{equation}
		\label{eq:Dawson}
		    F_{\text{Dawson}} (t^{-1}) = \int_{- \infty} ^{+ \infty} \frac{dx}{\sqrt{\pi}} \frac{e^{- x^2}}{x+t^{-1}} = \underbrace{\sqrt{\pi} e^{- i \frac{\pi}{2} \mathrm{sign}(t) - \frac{1}{t^2}}  }_{\text{non-pert.}} + t + \frac{1}{2} t^3 + \frac{3}{4} t^5 + \frac{15}{8} t^7 + \frac{105}{16} t^9 + \dots 
		\end{equation}
		known as Dawson's integral \cite{DLMF7}. The Gaussian damping term plays the role of the CS coupling in this toy example, and moreover we ave chosen to write $t^{-1}$ instead of $t$ to mimic what we get from massive hypermultiplets in the physical theories. In the right-hand side we have identified a non-perturbative part in $t$ and a formal power series in $t$. The customary expansion of a Stieltjes transform such as \eqref{eq:Dawson} consists in considering the denominator as a geometric series, giving:
		\begin{equation*}
		    F_{\text{Dawson}} (t^{-1}) \stackrel{\text{pert.}}{=} t + \frac{1}{2} t^3 + \frac{3}{4} t^5 + \frac{15}{8} t^7 + \frac{105}{16} t^9 + \dots 
		\end{equation*}
		The agreement of this solution with \eqref{eq:Dawson} can be checked to arbitrarily high order in $t$, once the non-perturbative term is discarded.\par

			\subsection{Schur expansion of $\mathsf{A}_1$ theories with adjoint matter}
			\label{sec:SchurSUNadj}
		
			\subsubsection{Schur expansion: $SU(2)$ Chern--Simons theory with one adjoint hypermultiplet}
			
				Let us consider the partition function of $SU(2)_k$ CS theory with one adjoint hypermultiplet. We turn on a real mass $m$ associated to the $U(1)$ flavour symmetry rotating the adjoint, and define the fugacity $t= - e^{2 \pi m}$, as in \eqref{eq:deftfugacity}. The partition function is 
				\begin{align*}
					\mathcal{Z}_{SU(2), 1_{\mathrm{adj}}} (m) & = \int_{- \infty } ^{+ \infty} dx_1  \int_{- \infty } ^{+ \infty} dx_2  ~ \delta (x_1 + x_2) \frac{ \left( 2 \sinh  \pi (x_1 - x_2) \right)^2 ~e^{ i \pi k (x_1 ^2 + x_2 ^2)} }{ ( 2 \cosh \pi (x_1 - x_2 + m ))( 2 \cosh \pi (x_2 - x_1 + m )) } \\
						& = \int_{- \infty } ^{+ \infty} dx \frac{ 4 \sinh (2 \pi x )^2 ~ e^{i 2 \pi k x^2 }}{ (2 \cosh \pi (2x +m )) (2 \cosh \pi (2x -m ))   }  .
				\end{align*}
				Rewriting the denominator and using the Cauchy identity \eqref{eq:Cauchyid} we arrive at 
				\begin{equation}
				\label{eq:ZSU2Schurexp}
					\frac{\mathcal{Z}_{SU(2)_k, 1_{\mathrm{adj}}} (m) }{ \mathcal{Z}_{SU(2)_k} } \stackrel{\text{pert.}}{=} \sum_{\nu=0}^{\infty} t^{\nu+1} \langle W_{\nu \nu} \rangle_{SU(2)} ,
				\end{equation}
				where we have used the definition \eqref{eq:deftfugacity}. The sum runs over isomorphism classes of irreducible $SU(2)$ representations, in one-to-one correspondence with non-negative integers $\nu$. We recognize the generating function of the vevs of a Wilson loops running along a Hopf link in $\mathbb{S}^3$, computed in $SU(2)$ CS theory with renormalized coupling $k= k_{\text{bare}} +2$. These vevs in turn are given by coloured Jones polynomials \cite{WittenJones}.\par
				In the spirit of knot homology theory and its physical interpretation \cite{Gukov:2004}, we may try to interpret \eqref{eq:ZSU2Schurexp} as the Poincar\'{e} polynomial of some knot homology, up to some simple overall factor. The fugacity $t$ corresponds on the physical side to a fugacity for the $U(1)$ symmetry rotating the adjoint hypermultiplet, as occurs for example in \cite{Chung:2014}, and turning off the mass parameters sends $t \to -1$, giving the Jones polynomials as the Euler characteristic of the would-be homological theory. While this may be seen as a hint toward the interpretation of the result as a categorification of the Hopf link invariant, a closer look at \eqref{eq:ZSU2Schurexp} suggests that such an interpretation is not correct, at least in the present form. Further discussion on this point is presented in Appendix \ref{app:noknot}.\par

			\subsubsection{Schur expansion: $SU(N)$ Chern--Simons theory with one adjoint hypermultiplet}
				We now generalize the discussion above to higher rank, considering $SU(N)$ theory. We may consider $U(N)$ theory as well, and the procedure goes through in precisely the same way.\par
				The partition function of $SU(N)$ Chern--Simons theory coupled to one adjoint is 
				\begin{equation*}
					\mathcal{Z}_{SU(N)_k, 1_{\mathrm{adj}}} (m)  = \int_{\R^N} d^{N} x ~ \delta \left( \sum_{a=1} ^{N} x_a \right) ~ \frac{ \prod_{1 \le a \ne b \le N} 2 \sinh \pi (x_a - x_b)  }{ \prod_{a,b=1} ^N 2 \cosh (x_a - x_b + m)} ~ e^{i \pi k \sum_{a=1} ^{N} x_a ^2 } .
				\end{equation*}
				The usual manipulations on the denominator, taking advantage of the $\delta$ function in the integrand to simplify the expression, and the application of the Cauchy identity \eqref{eq:Cauchyid} lead us to 
				\begin{equation*}
					\prod_{a=1}^{N} \prod_{b=1}^{N} \left( 1+e^{2 \pi x_a} e^{-2 \pi x_b + 2 \pi m} \right)^{-1} = \sum_{\nu} \mathfrak{s}_{\nu} (e^{2 \pi x})  \mathfrak{s}_{\nu} (- e^{- 2 \pi x + 2 \pi m})  ,
				\end{equation*}
				where the sum is over irreducible representation of $SU(N)$, which are equivalently represented by Young diagrams with at most $N-1$ rows. We also adopted a shorthand notation $ \mathfrak{s}_{\nu} (e^{2 \pi x}) :=  \mathfrak{s}_{\nu} (e^{2 \pi x_1}, e^{2 \pi x_2}, \dots, e^{2 \pi x_N} ) $. We obtain the expansion of the partition function 
				\begin{equation}
				\label{eq:SUNandlinks}
					\frac{\mathcal{Z}_{SU(N)_k, 1_{\mathrm{adj}}} (m) }{ \mathcal{Z}_{SU(N)_k} }\stackrel{\text{pert.}}{=} t^{\frac{N(N-1)}{2}} \sum_{\nu } t^{\vert \nu \vert } \left\langle W_{\nu \nu} \right\rangle_{SU(N)} .
				\end{equation}
				We have used the definition \eqref{eq:deftfugacity} of the fugacity $t$, and $\vert \nu \vert$ is the number of boxes in the Young diagram $\nu$.\par
				We find a (formal) polynomial in two variables $(q,t)$, which as a function of the variable $t$, looks similar to a generating function of HOMFLY-PT polynomials of the Hopf link coloured by $SU(N)$ representation. Note that the Hopf link is non-generic, since it is yields equal representations on the two components. Note also that it is not truly a generating function, because each summand is weighted by $t^{\vert \nu \vert}$, which does not distinguish between representation with the same value of $\vert \nu \vert$.\par

			\subsubsection{Schur expansion: $SU(N)$ Chern--Simons theory with $N_{\mathrm{adj}}$ adjoint hypermultiplets}
			\label{sec:SuNNadj}
				The computations can be extended to an arbitrary number $N_{\mathrm{adj}}=n$ of adjoint hypermultiplets with generic masses. The $SU(N)$ partition function is 
				\begin{equation*}
					\mathcal{Z}_{SU(N)_k, 1_{\mathrm{adj}}} (m)  = \int_{\R^N} d^{N} x ~ \delta \left( \sum_{a=1} ^{N} x_a \right) ~ \frac{ \prod_{1 \le a \ne b \le N} 2 \sinh \pi (x_a - x_b)  }{ \prod_{j=1} ^{N_{\mathrm{adj}}}\prod_{a,b=1} ^N 2 \cosh (x_a - x_b + m_j)} ~ e^{i \pi k \sum_{a=1} ^{N} x_a ^2 } .
				\end{equation*}
				We mimic the steps above and apply the Cauchy identity $n$ times, arriving at 
				\begin{equation*}
					\frac{\mathcal{Z}_{SU(N)_k, N_{\mathrm{adj}}} (m) }{ \mathcal{Z}_{SU(N)_k} } \stackrel{\text{pert.}}{=} \left( \prod_{j=1} ^{n} t_j ^{\frac{N(N-1)}{2}} \right) ~ \sum_{\nu^{(1)} } t_1 ^{\vert \nu^{(1)} \vert }  \cdots \sum_{\nu^{(n)} } t_{n} ^{\vert \nu^{(n)}  \vert }   \left\langle \prod_{j=1} ^{n} W_{\nu^{(j)} \nu^{(j)} } \right\rangle_{SU(N)} .
				\end{equation*}
				The average computes the correlator of $n=N_{\mathrm{adj}}$ pairwise unlinked Hopf links, each one with equally coloured components.

		
		\subsection{Schur expansion of $\mathsf{A}_1$ theories with fundamental matter}
		
		\subsubsection{Schur expansion: $U(1)$ theory with $N_f$ hypermultiplets}
		
		We now go back to the Abelian $\mathsf{A}_1$ CS theory with $N_f$ massive hypermultiplets, discussed in Subsection \ref{sec:AbelianA1Nf}. We assume an even number of hypermultiplets $N_f = 2n$ and write the partition function in the form
		\begin{equation*}
			\mz_{U(1),2n} (k, \vec{m}) = \int_{- \infty} ^{+ \infty} dx \frac{ e^{i \pi k x^2  + 2 \pi n x }}{ \prod_{j=1} ^{2n} \left(  1 -t_j e^{2 \pi x} \right) } ,
		\end{equation*}
		where $t_j = - e^{2 \pi m_j}$, as defined in \eqref{eq:deftfugacity}, and we used $\sum_{j=1} ^{N_f} m_j =0$ to drop an overall factor. We now exploit the Cauchy identity \eqref{eq:Cauchyid}. We thus write 
		\begin{equation}
		\label{eq:GenFhh}
			\prod_{j=1} ^{2n} \left( 1-t_j e^{2 \pi x} \right)^{-1} = \sum_{\nu=0} ^{\infty} \mathfrak{s}_{\nu} (e^{2 \pi x}) \mathfrak{s}_{\nu} (t_1, \dots, t_{2n})  
		\end{equation}
		and obtain:
		\begin{equation*}
			\frac{\mz_{U(1),2n} (k, \vec{m}) }{ \mathcal{Z}_{\mathrm{CS} (1)_k} } \stackrel{\text{pert.}}{=} \sum_{\nu=0} ^{\infty} \mathfrak{s}_{\nu} (t_1, \dots, t_{2n}) \langle W_{\nu +n} \rangle_{U(1)_{k}}
		\end{equation*}
		where $\langle W_{\nu + n} \rangle_{U(1)_{k}}$ stands for the vev of the Wilson loop in the $U(1)$ representation corresponding to $\nu + n \in \mathbb{Z}_{>0}$ computed in pure Chern--Simons theory on $\mathbb{S}^3$ at level $k$. Recall that the fugacities $t_j$ are defined in \eqref{eq:deftfugacity} as minus the fugacities for the maximal torus of the flavour symmetry.\par
		This Abelian case is particularly simple: recall from \eqref{eq:genhomo} that equation \eqref{eq:GenFhh} gives in fact the generating function of the homogeneous symmetric polynomials $\mathfrak{h}_{\nu} (t_1, \dots, t_{2n})$ \cite{Macdonaldbook}, and besides the Wilson loop is captured by a simple Gaussian integral. We get: 
		\begin{align}
			\mz_{U(1),2n} (k, \vec{m}) & \stackrel{\text{pert.}}{=} \sqrt{ \frac{i}{k} }  e^{ \frac{i \pi}{k} n^2 } \sum_{\nu =0} ^{\infty}  e^{ \frac{i \pi}{k} (\nu^2 + 2 \nu n ) } \mathfrak{h}_{\nu} (t_1, \dots, t_{2n}) \label{eq:AbelianSchurexp} \\
			& = \sqrt{\frac{i}{k}} \left\{ e^{ \frac{i \pi}{k} n^2 } -   e^{ \frac{i \pi}{k} (n+1)^2 } \mathfrak{h}_{1} (t_1, \dots, t_{2n})  +   e^{ \frac{i \pi}{k} (n+2)^2 } \mathfrak{h}_{2} (t_1, \dots, t_{2n}) + \dots  \right\} , \notag
		\end{align}
		where $\mathfrak{h}_{1} (t_1, \dots, t_{2n}) = \sum_{j=1}^{2n} t_j $, $\mathfrak{h}_{2} (t_1, \dots, t_{2n}) = \sum_{1 \le j \le l \le 2n} t_j t_l $ and so on, and recall that the number of flavours is $N_f=2n$. The result is a symmetric polynomial in the fugacities $t_j$.\par

		We can compare the result \eqref{eq:AbelianSchurexp} with the exact one obtained in Section \ref{sec:AbelianA1Nf}, but in doing so we have to bear in mind a few caveats:
		\begin{itemize}
			\item While  the physical parameters satisfy $\prod_{j=1} ^{2n} t_j = 1$, we should treat these as formal indeterminates, thus expanding for each $t_j$ independently.
			\item The formal expansion is in positive powers of $t_j$, hence it will be compared with $k<0$ in Subsection \ref{sec:AbelianA1Nf}. We could as well have begun with the expansion in negative powers of $t_j$, to be compared with $k>0$ in \ref{sec:AbelianA1Nf}. In each case, the choice must be made at the beginning, through the manipulations of the denominator before plugging the identity \eqref{eq:Cauchyid}. Nevertheless, the summation variable $\nu$ plays the role of a real irreducible $U(1)$ representation, and through the isomorphism with its conjugate representation we could extract the expansion for $k>0$.
			\item The Schur expansion will miss non-perturbative terms in $t_j$, namely those $\propto e^{- i \pi k \left( \frac{1}{2} + i m_j\right) ^2}$. 
		\end{itemize}
		After elementary manipulations of the result in Subsection \ref{sec:AbelianA1Nf} and dropping non-perturbative terms, the expansion relative to a single $t_j$ is 
		\begin{equation*}
			\frac{1}{1- (- t_j) ^{\vert k \vert }} \sum_{\alpha=0} ^{\vert k \vert -1 } t_j ^{\alpha} e^{ \frac{i \pi }{k} (\alpha+n)^2 } \prod_{s \ne j } \sum_{\beta_s=0} ^{\infty} \left( \frac{t_s}{t_j} \right)^{\beta_s}.
		\end{equation*}
		The prefactor should be expanded as a geometric series to compare with the Schur expansion. In this way, the terms $t_j ^{\beta \vert k \vert}$ kick in extending the summation range beyond $\alpha = \vert k \vert -1$,
		\begin{equation}
			\label{eq:MardellU1Nfexp}
			\sum_{\beta=0}^{\infty} \sum_{\alpha=0} ^{\vert k \vert -1}  (-1)^{\beta \vert k \vert } e^{ \frac{i \pi }{k} (\alpha+n)^2 }  t_j ^{\alpha+ \beta \vert k \vert } \prod_{s \ne j } \sum_{\beta_s=0} ^{\infty}  \left( \frac{t_s}{t_j} \right)^{\beta_s} .
		\end{equation}
		To check the agreement of the two expressions requires care in the power counting. So, for example, tho compare at order $t_j ^{1}$, one should take into account all the combinations, which in particular include a term from all the homogeneous polynomials in the Schur expansion, which contribute 
		\begin{equation*}
			 e^{ \frac{i \pi }{k} (n+1)^2} t_j + \sum_{s \ne j }  e^{ \frac{i \pi }{k} (n+2)^2} t_j t_s +  \sum_{s_1, s_2 \ne j }  e^{ \frac{i \pi }{k} (n+3)^2} t_j t_{s_1} t_{s_2} + \dots  
		\end{equation*}
		For $\mathfrak{h}_{\nu}$ with $\nu> \vert k \vert -1$ write $\nu = \alpha + \beta \vert k \vert $ and use 
		\begin{equation*}
		    e^{\frac{i \pi }{k} (n + \nu )^2 } = e^{\frac{i \pi }{k} (n + \alpha )^2  + i \pi \beta^2 \vert k \vert } = (-1)^{\beta \vert k \vert } e^{\frac{i \pi }{k} (n + \alpha )^2} .
		\end{equation*}
		On the side of the exact evaluation \eqref{eq:MardellU1Nfexp} in turn we see that all $\alpha$ and $\beta$ contribute, as they are partially cancelled by the $t_j ^{- \beta_s}$. Term by term comparison shows that the Schur expansion correctly reproduces the exact answer, with the non-perturbative contributions already discarded. 
		Let us stress once again that the agreement is understood in an algebraic sense, reading off the coefficients of the multiple expansion in $\left\{ t_j \right\}$.\par
		\medskip
		To conclude the analysis of the present theory, we note that the same expressions have been analyzed in \cite{Conifold:2006}, in the context of topological strings theory with non-compact branes. To make contact with that setting we specialize the masses 
		\begin{equation*}
			m_j = m \left( n - j + \frac{1}{2} \right)
		\end{equation*}
		(recall that $N_f=2n$) and define $t= - e^{2 \pi m} $. The homogeneous polynomials become a $q$-binomial, with $q$-parameter $t$: 
		\begin{equation*}
			\mathfrak{h}_{\nu} \left( t^{  n -  \frac{1}{2} } , \dots,  t^{  -n +  \frac{1}{2} } \right) = \left[ \begin{matrix} n \\ \nu \end{matrix} \right]_{t} .
		\end{equation*}
		Then, our expressions differ from \cite{Conifold:2006} only in the Gaussian term in the sum. This mismatch is exactly the factor due to the difference in the framing, as the Wilson loop vev on $\mathbb{S}^3$ in \cite{Conifold:2006} is computed in the natural framing instead of the matrix model framing.

		\subsubsection{Schur expansion: $U(N)$ and $SU(N)$ theory with $N_f$ hypermultiplets}
		\label{sec:SchurexpUNNf}
		
		The manipulations above have been presented in the Abelian theory for clarity, but are straightforwardly generalized to the non-Abelian setting. The partition function of $U(N)_k$ CS theory with $N_f$ fundamental hypermultiplets, studied in Section \ref{sec:A1NonAb-Nf}, is more suitably written for our purposes in the form 
		\begin{equation*}
			\mathcal{Z}_{U(N),N_f} = \int_{\R^N} \frac{ \prod_{1 \le a < b \le N }  \left(  2 \sinh \pi (x_b - x_a) \right)^2 }{ \prod_{a=1}^{N} \prod_{j=1} ^{N_f} \left( 1-t_j e^{2 \pi x_a} \right)}  \prod_{a=1}^{N}  e^{i \pi k x_a ^2 + \pi N_f x_a}  ~dx_a .
		\end{equation*}
		When the gauge group is $SU(N)$ the partition function includes a $\delta$-function $\delta \left( \sum_{a=1} ^N x_a \right)$ in the measure.\par
		Using the Cauchy identity \eqref{eq:Cauchyid}, we identify the average of a Schur polynomial in the Chern--Simons random matrix ensemble, which computes the vev of a Wilson loop. Note however that in principle we cannot reabsorb the $\pi N_f x_a$ in the exponential because it would move the integration contour away from the real axis, and the integrand has poles in the complex plane. Equivalently, the problem can be seen reabsorbing the shift into a redefinition of the masses, which would acquire half-integer imaginary part, rendering the integrand singular. To handle this, we pass from $q=e^{- i 2 \pi/k}$ to $q=e^{- g}$, $g>0$. Doing so, we can safely complete the square in the matrix model, and the change of variables shifts $2 \pi m_j \mapsto 2 \pi m_j + \frac{g}{2} N_f $.\par
		This problem does not arise in the $SU(N)$ theory, since the $\delta$-constraint on the eigenvalues would cancel the linear shift, and we are allowed to work directly with $q$ root of unity.\par
		With this distinction in mind, we get for the $U(N)$ case 
		\begin{equation}
		\label{eq:ZNfSchurW}
			\frac{\mathcal{Z}_{U(N),N_f} }{ \mathcal{Z}_{\mathrm{CS} (N)} } \stackrel{\text{pert.}}{=} q^{ -\frac{N_f^2}{8} } \sum_{\nu} \mathfrak{s}_{\nu} \left( q^{- \frac{N_f}{2} } t_1, \dots, q^{- \frac{N_f}{2} } t_{N_f} \right) \langle W_{\nu} \rangle_{\mathrm{CS}(N)} .
		\end{equation}
		The sum runs over Young diagrams associated to irreducible $U(N)$ representations, and the basic properties of the symmetric polynomials imply that all contributions with 
		\begin{equation*}
			\mathrm{length} (\nu) > \min \left\{ N, N_f \right\}
		\end{equation*}
		vanish. The average $\langle \cdots \rangle_{\mathrm{CS} (N)}$ means the vev in $U(N)$ CS theory with real $q=e^{-g}$.\par
		The overall factor in \eqref{eq:ZNfSchurW} is reminiscent of the effective CS coupling associated to a mixed flavour-R contact term \cite{Closset:2012vp}. Besides, we again notice how the result is more naturally written in terms of fugacities for the holomorphic variables $\lambda_j= \frac{1}{2} + i m_j$ rather than for the masses $m_j$ alone. The $q$-shift of the mass parameters seem likewise to originate from an effective coupling for the background fields. This $q$-shift can be brought out of the Schur polynomials and contributes a factor $q^{- \frac{N_f}{2} \vert \nu \vert }$ to each summand.\par
		The Wilson loop vev is known \cite{TDoli} and has been presented in equation \eqref{eq:WLCSk}, which we report here for clarity: 
		\begin{equation*}
			 \langle W_{\nu} \rangle_{\mathrm{CS} (N)}  = (\dim_q \nu ) ~q^{- \frac{1}{2} \mathsf{C}_{2;N} (\nu)  } .
		\end{equation*}\par
		In the $SU(N)$ theory instead we obtain 
		\begin{equation}
		\label{eq:ZSUNfSchurW}
			\frac{\mathcal{Z}_{SU(N),N_f}}{\mathcal{Z}_{SU(N)_k}} \stackrel{\text{pert.}}{=} \sum_{\nu} \mathfrak{s}_{\nu} \left( t_1, \dots, t_{N_f} \right) \langle W_{\nu} \rangle_{SU (N)_k} .
		\end{equation}
		The difference, besides the overall factor $q^{- \frac{N_f^2}{8}}$, is the specialization of the variables in the argument of the Schur polynomial, which are not renormalized by a $q$-shift.\par
		We have therefore written the partition function of Chern--Simons theory with $N_f$ fundamental hypermultiplets as a generating-like function of unknot invariants. From \eqref{eq:ZNfSchurW} we can also obtain the Schur expansion of the $\mathsf{A}_2$ quiver theory, simply dropping the constraint $\prod_{j=1} ^{N_f} t_j =(-1)^{N_f}$ (this would introduce a factor $\prod_{j=1}^{N_f} (-t_j) ^N $ in the matrix model, which we have set to 1) and gauging the $U(N_f)$ symmetry. Adding a CS term to the newly gauge node and using \eqref{eq:WLCSk} we find for the $\mathsf{A}_2$ quiver $U(N_1) \times U(N_2)$ CS theory 
		\begin{equation*}
			\frac{\mathcal{Z}_{\mathsf{A}_2} }{ \mathcal{Z}_{\mathrm{CS} (N)_1}  \mathcal{Z}_{\mathrm{CS} (N_2)}} \stackrel{\text{pert.}}{=} q_1 ^{ -\frac{N_f^2}{8} } \sum_{\nu}  \left( - q_1 ^{- \frac{N_2}{2}} \right)^{\vert \nu \vert} ( \dim_{q_1} \nu)  ( \dim_{q_2} \nu)  ~q_1 ^{- \frac{1}{2} \mathsf{C}_{2;N_1} (\nu)}  q_2 ^{- \frac{1}{2} \mathsf{C}_{2;N_2} (\nu)}
		\end{equation*}
		where $q_1$ and $q_2$ are the $q$-parameters of the two pure CS theories obtained removing the edge joining the two nodes of the $\mathsf{A}_2$ quiver.\par
		\medskip

		\subsubsection{Schur expansion: $SU(N)$ theory with fundamental and adjoint hypermultiplets}
			We can consider a theory with both $N_f$ fundamental and $N_{\mathrm{adj}}$ adjoint hypermultiplets. We will limit ourselves to $N_{\mathrm{adj}}=1$, being the effect of adding more adjoint matter studied in Subsection \ref{sec:SuNNadj}. We work with gauge group $SU(N)$ for concreteness, being the $U(N)$ theory completely analogous, up to a change of variables which generates a $q$-shift of the fugacities $t_j$.\par
			\begin{equation*}
				\mathcal{Z}_{SU(N),N_f,1_{\mathrm{adj}}} (\vec{m}, m_0 ) = \int_{\R^N} \delta \left( \sum_{a=1}^{N} x_a \right)   \frac{  \prod_{1 \le a < b \le N }  \left(  2 \sinh \pi (x_b - x_a) \right)^2 }{  \prod_{a,b=1} ^{N} \left( 1- t_0 e^{2\pi x_a - 2 \pi x_b} \right) } \prod_{a=1} ^{N} \frac{  e^{i \pi k x_a ^2 }  ~dx_a  }{  \prod_{j=1} ^{N_f} \left( 1- t_j e^{2 \pi x_a} \right)  }   .
			\end{equation*}
			Here the variables $\left\{ t_j \right\}$ are as in \eqref{eq:deftfugacity}, and we have denoted $m_0$ the mass of the adjoint and $t_0= -e^{2 \pi m_0}$ the corresponding fugacity. Combining the manipulations of Subsection \ref{sec:SchurSUNadj} with those of \ref{sec:SchurexpUNNf} we arrive at 
			\begin{equation*}
				\frac{\mathcal{Z}_{SU(N),N_f,1_{\mathrm{adj}}} (\vec{m}, m_0 ) }{ \mathcal{Z}_{SU(N)_k} } \stackrel{\text{pert.}}{=}  \sum_{\mu, \nu} t_0 ^{\vert \mu \vert} \mathfrak{s}_{\nu} ( t_1, \dots, t_{N_f} ) \langle W_{\mu} W_{\nu \mu} \rangle_{SU(N)_k} .
			\end{equation*}
			From the matrix model description we see that, adding fundamental matter to the theory with one adjoint, we have produced more interesting observables, which are correlators of two Wilson loops, one along an unknot and one along a Hopf link, with the latter not necessarily coloured by two equal representations.

		\subsubsection{Schur expansion: $4d$ $\mN=4$ SYM with defects}
			We now apply the ideas presented in this Section to a special case of four-dimensional gauge theory, namely  $\mN=4$ $U(N)$ super-Yang--Mills (SYM) on $\mathbb{S}^4$ with co-dimension 1 matter defects placed at the equatorial $\mathbb{S}^3 \subset \mathbb{S}^4$ \cite{Drukker:2010jp}. The partition function of such theory, as obtained from localization, is \cite{Drukker:2010jp,Robinson:2017sup}
			\begin{equation*}
				\mathcal{Z}_{U(N), N_f} ^{4d+\text{defect}} = \int_{\R^N} \prod_{1 \le a < b \le N } ( x_a - x_b)^2 ~ \prod_{a=1}^{N} \frac{ e^{- \frac{8 \pi^2}{g_{4d}} x_a ^2} ~ dx_a }{  \prod_{j=1} ^{N_f}  2 \cosh \pi (x_a + m_j) }  .
			\end{equation*}
			Applying identical manipulations as in Subsection \ref{sec:SchurexpUNNf}, we arrive at a perturbative expansion in the parameters $t_j$, exactly as in the purely $3d$ framework, but now the summands are vevs of Wilson loops computed in $4d, \mN=4$ SYM (with $q_{4d}=e^{- g_{4d}/16 \pi^2 }$): 
			\begin{equation*}
				\frac{ \mathcal{Z}_{U(N), N_f} ^{4d+\text{defect}} }{ \mathcal{Z}_{U(N), N_f} ^{4d, \mN=4} } \stackrel{\text{pert.}}{=} q_{4d} ^{ -\frac{N_f^2}{8} } \sum_{\nu} \mathfrak{s}_{\nu} \left( q_{4d} ^{- \frac{N_f}{2} } t_1, \dots, q_{4d} ^{- \frac{N_f}{2} } t_{N_f} \right) \langle W_{\nu} \rangle_{U(N)} ^{4d, \mN=4} .
			\end{equation*}

		\subsection{Schur expansion of necklace quiver theories}

        The focus of this Subsection is on quiver gauge theories of type $\widehat{\mathsf{A}}_r$. 
		\subsubsection{Schur expansion: ABJ}
		\label{sec:SchurexpABJ}
		We now consider the mass-deformed ABJ theory, whose partition function reads:
		\begin{align*}
			\mz_{\mathrm{ABJ}(N_1 \vert N_2)} (k,m) &= \int_{\R^{N_1}} d^{N_1} \vec{x} e^{i \pi k \sum_{a=1} ^{N_1} x_a ^2 }\int_{\R^{N_2}} d^{N_2} \vec{y} e^{- i \pi k \sum_{\dot{a}=1} ^{N_2} y_{\dot{a}} ^2 } \\
			& \times \frac{\prod_{1 \le a < b \le N_1} \left( 2 \sinh \pi (x_b - x_a) \right)^2 ~ \prod_{1 \le \dot{a} < \dot{b} \le N_1} \left( 2 \sinh \pi (y_{\dot{b}} - y_{\dot{a}}) \right)^2}{ \prod_{a=1}^{N_1} \prod_{\dot{a}=1}^{N_2} 2 \cosh \pi (x_a - y_{\dot{a}} +m_{-}) ~ 2 \cosh \pi (x_a - y_{\dot{a}} +m_{+} ) }
		\end{align*}
		where the physical values of the masses are $m_{\pm} = \pm m$, but here we treat them as independent. This can be achieved turning on a FI parameter which, upon changing variables, shifts the real masses. The second line is more conveniently written as
		\begin{equation*}
			e^{\pi (N_1 + N_2) (m_+ - m_-) } \left[ \prod_{a=1} ^{N_1} \prod_{\dot{a}=1} ^{N_2} \left( 1+ e^{2 \pi x_a} e^{- 2 \pi (y_{\dot{a}} +m_{-}) }  \right) ~ \left( 1+ e^{-2 \pi x_a} e^{2 \pi (y_{\dot{a}} +m_{+} ) }  \right) \right]^{-1} .
		\end{equation*}
		We now apply the Cauchy identity \eqref{eq:Cauchyid} 
		\begin{equation*}
			\left[ \prod_{a=1} ^{N_1} \prod_{\dot{a}=1} ^{N_2} \left( 1+ e^{\pm 2 \pi x_a} e^{\mp 2 \pi (y_{\dot{a}} +m_{\mp }) }  \right) \right]^{-1} = \sum_{\nu} \mathfrak{s}_{\nu} (- e^{\pm 2 \pi x}) \mathfrak{s}_{\nu} (e^{\mp 2 \pi (y+  m_{\mp} )}) ,
		\end{equation*}
	    adopting the usual shorthand $e^{2 \pi x}$ for $(e^{2 \pi x_1}, \dots, e^{2 \pi x_{N_1}})$ and likewise for $e^{2\pi y}$, and the sum runs over all partitions $\nu$ with
		\begin{equation}
		\label{eq:ABJmunulemin}
			\mathrm{length} (\nu) \le \min (N_1, N_2) .
		\end{equation}
		Therefore, bringing the common factors $e^{\mp 2 \pi m_{\mp}}$ out of the Schur polynomials in $e^{2 \pi y}$ we get 
		\begin{align*}
			\mz_{\mathrm{ABJ}(N_1 \vert N_2)} (k,m) & \stackrel{\text{pert.}}{=} e^{\pi (N_1 + N_2) (m_+ - m_-) } \sum_{\mu} \sum_{\nu} (- e^{ 2 \pi m_{+}  })^{\vert \mu \vert } (- e^{ - 2 \pi m_{-} } ) ^{ \vert \nu \vert } \\
			& \times \int_{\R^{N_1}} \mathfrak{s_{\mu}} (e^{2 \pi x}) \mathfrak{s_{\nu}} (e^{-2 \pi x})  ~ \prod_{1 \le a < b \le N_1} \left( 2 \sinh \pi (x_b - x_a) \right)^2  ~\prod_{a=1} ^{N_1} e^{i \pi k x_a ^2 } dx_a \\
			& \times  \int_{\R^{N_2}} \mathfrak{s_{\mu}} (e^{-2 \pi y}) \mathfrak{s_{\nu}} (e^{2 \pi y})  ~ \prod_{1 \le \dot{a} < \dot{b} \le N_2} \left( 2 \sinh \pi (y_{\dot{b}}- y_{\dot{a}}) \right)^2  ~\prod_{\dot{a}=1} ^{N_2} e^{-i \pi k y_{\dot{a}} ^2 } dy_{\dot{a}} .
		\end{align*}
		We find that the integrals are factorized into vevs of Wilson loops in pure Chern--Simons theory \cite{Kimura} at each node: 
		\begin{equation*}
			\mz_{\mathrm{ABJ}(N_1 \vert N_2)} (k,m)  \stackrel{\text{pert.}}{=} e^{\pi (N_1 + N_2) (m_+ - m_-) }  \sum_{\mu} \sum_{\nu} (- e^{2 \pi m_+})^{\vert \mu \vert }  (- e^{2 \pi m_- })^{- \vert \nu \vert } \langle W_{\mu \nu} \rangle_{N_1;k} \langle W_{\nu \mu } \rangle_{N_2;-k} 
		\end{equation*}
		where the equality is understood order by order in the Laurent expansion in the parameters $t_{\pm} = - e^{2 \pi m_{\pm} }$, and the two vevs compute Hopf link invariants respectively in $ U(N_1)$ and $U(N_2)$ pure Chern--Simons theory on $\mathbb{S}^3$ with renormalized levels $k$ and $-k$. Note also how the roles of the two representations $\mu, \nu$ are swapped between the two nodes. The restriction \eqref{eq:ABJmunulemin}, which arises here from an elementary property of the symmetric polynomials, matches with the analysis of the quiver variety of $\widehat{\mathsf{A}}_1$, which only includes $U(N)$ representations for $N=\min \left\{ N_1, N_2 \right\}$.\par
		The result does not rely on the specific choice of CS levels and immediately extends to generic $(k_1, k_2)$.

		\subsubsection{Schur expansion: necklace quivers}
		\label{sec:NecklaceSchurExp}
		
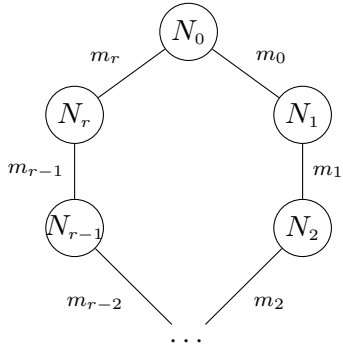
\begin{figure}[tbh]
			\centering
	\begin{tikzpicture}[auto]
		\node[circle,draw] (gext) at (0,2.6) {\hspace{13pt} };
		\node[draw=none] (auxext) at (0,2.6) {$N_{\scriptscriptstyle 0}$};
		\node[circle,draw] (g1) at (1.5,1.5) {\hspace{13pt} };
		\node[draw=none] (aux1) at (1.5,1.5) {$N_{\scriptscriptstyle 1}$};
		\node[circle,draw] (g2) at (1.5,0) {\hspace{13pt} };
		\node[draw=none] (aux2) at (1.5,0) {$N_{\scriptscriptstyle 2}$};
		\node[circle,draw] (gr) at (-1.5,1.5) {\hspace{13pt} };
		\node[draw=none] (auxr) at (-1.5,1.5) {$N_{\scriptscriptstyle r}$};
		\node[circle,draw] (g3) at (-1.5,0) {\hspace{13pt} };
		\node[draw=none] (aux3) at (-1.5,0) {$N_{\scriptscriptstyle r-1}$};
		\node[draw=none] (gaugemid) at (0,-1.5) {$\cdots$};
		\path (g1) edge [] node {$\scriptstyle m_1$} (g2);
		\path (g2) edge [] node {$\scriptstyle m_2$} (gaugemid);
		\path (gaugemid) edge [] node {$\scriptstyle m_{r-2}$} (g3);
		\path (g3) edge [] node {$\scriptstyle m_{r-1}$} (gr);
		\path (gr) edge [] node {$\scriptstyle m_{r}$} (gext);
		\path (gext) edge [] node {$\scriptstyle m_{0}$} (g1);
	\end{tikzpicture}
	\caption{Mass-deformed non-Abelian $\widehat{\mathsf{A}}_r$ extended quiver.}
	\label{fig:NAbExtAr}
	\end{figure}

		ABJ theory belongs to the class of extended $\widehat{\mathsf{A}}_r$ quivers. We now show how the Schur expansion holds for the whole $\widehat{\mathsf{A}}_r$ family of theories, with mass deformation and without any additional matter content beyond the bi-fundamental hypermultiplets linking the gauge nodes, as depicted in quiver notation in Figure \ref{fig:NAbExtAr}. These $\mN=3$ CS theories have been constructed in \cite{Kimura:moduli,JafferisTom}. The result we find is a series expansion in the parameters $t_p =-e^{2 \pi m_p}$, with the coefficients being vevs of a Wilson loop in pure Chern--Simons theory with gauge group $U(N_p)$ and level $k_p$.\par
	
			 \par
	
	The partition function of the theory is: 
		\begin{align*}
			\mz_{\widehat{\mathsf{A}}_r} (\vec{k}, \vec{m}) &= \int_{\mathbb{R}^{N_0}} d^{N_0} \vec{x}_0  \int_{\mathbb{R}^{N_1}} d^{N_1} \vec{x}_1 \cdots \int_{\mathbb{R}^{N_r}} d^{N_r} \vec{x}_r \prod_{p=0} ^{r} e^{i \pi k_p \sum_{a=1} ^{N_p} x_{p, a} ^2 }  \\
				& \times  \prod_{p=0}^{r}  \frac{ \prod_{1 \le a < b \le N_p } \left( 2 \sinh \pi (x_{p, b} - x_{p, a}) \right)^2 }{ \prod_{a=1} ^{N_p} \prod_{\dot{b}=1} ^{N_{p+1}} 2 \cosh \pi (x_{p+1, \dot{b}} - x_{p,a} + m_p )   } ,
		\end{align*}
		with periodic identification of the labels, $r+1 \equiv 0$.	At the level of the matrix model, the eigenvalues associated to each gauge node interact among themselves as in pure $U(N_p)_{k_p}$ Chern--Simons theory, and also interact with the nearest neighbours through the denominator.\par
		We apply the Cauchy identity \eqref{eq:Cauchyid} at each edge of the quiver in Figure \ref{fig:NAbExtAr}, expanding in the fugacities associated to the masses of the bi-fundamental hypermultiplets. We obtain the expressions 
		\begin{align*}
			\frac{ 1 }{ \prod_{a=1} ^{N_p} \prod_{\dot{b}=1} ^{N_{p+1}}  2 \cosh \pi ( x_{p,a} - x_{p+1,\dot{b}} + m_p ) } & = \prod_{a=1} ^{N_p} \prod_{\dot{b}=1} ^{N_{p+1}} \frac{ e^{\pi (x_{p,a} + m_p) - \pi x_{p+1,\dot{b}}}  }{ 1 + e^{ 2 \pi (x_{p,a} +m_p) } e^{- 2 \pi x_{p+1,\dot{b}}}  } \\
			&= e^{\pi ( N_{p+1} + N_p ) m_p - \pi N_{p+1} \sum_{a=1} ^{N_p} x_{p,a} -  \pi N_{p} \sum_{\dot{b}=1} ^{N_{p+1}} x_{p+1,\dot{b}} } \\
			& \times  \sum_{\nu^{(p)} } (-1)^{\vert \nu^{(p)} \vert } \mathfrak{s}_{\nu^{(p)}} (e^{ 2 \pi (x_p +m_p) } )   \mathfrak{s}_{\nu^{(p)}} (e^{-2 \pi x_{p+1} } )   ,
		\end{align*}
		where the sum runs over partitions $\nu^{(p)}$ satisfying 
		\begin{equation}
		\label{eq:constainnup}
			\mathrm{length} ( \nu^{(p)} ) \le \min (N_p, N_{p+1}) .
		\end{equation}
		Each set of variables $e^{2 \pi x_p}$ appears with plus sign in the exponent in the Schur $\mathfrak{s}_{\nu^{(p)}}$ and with minus sign in $\mathfrak{s}_{\nu^{(p-1)}}$. Besides, as above, we have written $\mathfrak{s}_{\nu^{(p)}} (e^{2 \pi x_p})$ as a shorthand for the Schur polynomial in the $N_p$ variables $(e^{2 \pi x_{p,1}}, \dots, e^{2 \pi x_{p,N_p}})$. 
		Putting all such contributions together we get 
		\begin{align*}
			\mz_{\widehat{\mathsf{A}}_r} (\vec{k}, \vec{m}) & \stackrel{\text{pert.}}{=}   \int_{\mathbb{R}^{N_0}}  \int_{\mathbb{R}^{N_1}} \cdots \int_{\mathbb{R}^{N_r}} \prod_{p=0} ^{r} \prod_{a=1} ^{N_p} e^{ \left[  i \pi k_p  x_{p,a} ^2 + \pi \left( N_{p+1} - N_{p-1} \right) x_{p,a}  \right]} ~dx_{p,a} \\
			& \times \prod_{p=0} ^{r} ~ \prod_{1 \le a < b \le N_p } \left( 2 \sinh \pi (x_{p, b} - x_{p, a}) \right)^2  \\
			& \times e^{\pi \sum_{p=0}^{r} (-1)^p m_p (N_{p+1}+ N_p)}   \sum_{\vec{\nu}}  (-1)^{\vert \vec{\nu} \vert }  \prod_{p=0} ^{r} e^{ (-1)^p 2 \pi m_p \vert \nu^{(p)}  \vert } ~ \mathfrak{s}_{\nu^{(p-1)}} (e^{-2 \pi x_p})  \mathfrak{s}_{\nu^{(p)}} (e^{2 \pi x_p } ) ,
 		\end{align*}
 		with the sum running over $(r+1)$-tuples of partitions 
 		\begin{equation*}
 			\vec{\nu} = (\nu^{(0)}, \dots, \nu^{(r)}) ,
 		\end{equation*}
 		with all partitions $\nu^{(p)}$ constrained according to \eqref{eq:constainnup}. The integrals are now suitably factorized in each summand. Completing the squares in the Gaussian term at each node and comparing with \cite{Kimura}, we obtain
 		\begin{align}
			\frac{ \mz_{\widehat{\mathsf{A}}_r} (\vec{k}, \vec{m})}{ \prod_{p=0} ^r \mathcal{Z}_{\mathrm{CS} (N_p);k_p}} & \stackrel{\text{pert.}}{=}  e^{\pi \sum_{p=0}^{r} \left[ m_p (N_{p+1}+ N_p) + \frac{i}{2 k_p} N_p (N_{p+1} - N_{p-1})^2 \right]} \label{eq:SchurExpnecklace}\\
			& \times  \sum_{\vec{\nu}}   \prod_{p=0} ^{r} t_p \vert \nu^{(p)} \vert  ~ e^{ \frac{i}{2 k_p} (N_{p+1} - N_{p-1}) \left( \vert \nu^{(p)} \vert - \vert \nu^{(p-1)} \vert \right) } ~ \langle W_{\nu^{(p)} \nu^{(p-1)}} \rangle_{N_p;k_p} \notag ,
		\end{align}
		where the average in each summand is the vev of a Wilson loop in $U(N_p)_{k_p}$ Chern--Simons theory, computing the Hopf link invariant in the representations $(\nu^{(p)}, \nu^{(p-1)})$. As always, the two sides of the equality are understood as formal series expansions in the parameters $t_p = - e^{2 \pi m_p}$. These global symmetry fugacities serve as book-keeping variables in the expansion, while all other variables are integrated. Furthermore, if we think of each $\langle W_{\bullet \bullet} \rangle_{N_p;k_p}$ as a ring homomorphism from the ring of $U(N_p)$ representations to $\C [q_p, q_p^{-1} ]$, we notice the emergence of a trace of the product of $r+1$ such maps as a direct consequence of the quiver being necklace-shaped. This trace is taken on the ring of $U(N_{\text{max}})$ representations, with $N_{\text{max}}= \max_p N_p$ and $\langle W_{\mu \nu} \rangle_{N_p;k_p}$ understood to vanish if either $\mu$ or $\nu$ is not a $U(N_p)$ representation. The trace structure appears more clearly when $N_p=N$ and $k_p= \pm k$ for all $p=0,1, \dots, r$.\par
		From the properties of pure CS theory and its relation with the level $k$ WZW model \cite{WittenJones}, only \emph{integrable} representations contribute to each Hopf link invariant. This introduces an effective ``mod $k_p$'' periodicity \cite{Naculich:2007nc} of the coefficients of each $t_p$.

		\subsubsection{Schur expansion: the M-crystal model}
		\label{sec:SchurMCrystal}
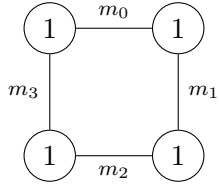
\begin{figure}[tbh]
	\centering
	\begin{tikzpicture}[auto,node distance=1cm]
		\node[circle,draw] (gauge1) {$ 1 $};
		\node[circle,draw] (gauge0) [left = of gauge1]{$ 1 $};
		\node[circle,draw] (gauge2) [below = of gauge1]{$ 1 $};
		\node[circle,draw] (gauge3) [below = of gauge0]{$ 1 $};
		\path (gauge0) edge [] node {$\scriptstyle m_0$} (gauge1);
		\path (gauge1) edge [] node {$\scriptstyle m_1$} (gauge2);
		\path (gauge2) edge [] node {$\scriptstyle m_2$} (gauge3);
		\path (gauge3) edge [] node {$\scriptstyle m_3$} (gauge0);
	\end{tikzpicture}
	\caption{Mass-deformed $\widehat{\mathsf{A}}_3$ extended quiver. For $\vec{k}= (1,-1,1,-1)$ the associated Chern--Simons theory is the gauge theoretical realization of the M-crystal model with four vertices.}
	\label{fig:Mcrystal4}
	\end{figure}

		A simple yet interesting example of the above setting corresponds to the Abelian model with alternating $\pm 1$ CS levels, $\vec{k}=(+1,-1, \dots, -1)$. This quiver gauge theory describes the M-crystal model \cite{Kim:2007ic,Hosomichi:2008CS}, see for example Figure \ref{fig:Mcrystal4}. Specializing the computations above and after a few simplifications we get 
		\begin{equation*}
			\frac{ \mz_{\widehat{\mathsf{A}}_r} (\pm 1 , \vec{m}) }{ \prod_{p=0} ^{r} \mz_{\mathrm{CS} (1); (-1)^p } } \stackrel{\text{pert.}}{=} e^{2 \pi \vert \vec{m} \vert } \sum_{\vec{\nu} \in \mathbb{Z}_{\ge 0} ^{r+1}} \prod_{p=0}^{r}  \left( -e^{2 \pi m_p \nu^{(p)}  + i 2 \pi  (-1)^{p} \nu^{(p-1)} \nu^{(p)}  } \right) .
		\end{equation*}
		The series is clearly not convergent, but this was expected as the right-hand side has the meaning of an algebraic expansion in multiple variables. In conclusion, a perturbative expansion of the partition function of the M-crystal model has all the terms $t_0 ^{\nu^{(0)}} \cdots t_r ^{\nu ^{(r)}}$ with coefficient $1$.\par

		 \par
	
	A simple generalization of the above formula to the Abelian necklace quiver with arbitrary $\vec{k}$ gives 
	\begin{equation}
			\frac{\mz_{\widehat{\mathsf{A}}_r} (\vec{k} , \vec{m})}{{ \prod_{p=0} ^{r} \mz_{\mathrm{CS} (1); k_p } }}  \stackrel{\text{pert.}}{=} (-1)^r \left( \prod_{p=0} ^r t_p \right) \sum_{\vec{\nu} \in \mathbb{Z}_{\ge 0} ^{r+1}} \prod_{p=0}^{r} t_p ^{\nu^{(p)}}  e^{- i \pi \nu^{(p)} \left( \frac{1}{k_p} +  \frac{1}{k_{p-1}} \right) + i \frac{2 \pi}{k_p} \nu^{(p)} \nu^{(p-1)} } .
    \label{AbelianneckScurformula}
	\end{equation}
	When $r=1$ and $k_2=-k_1$ we get the Abelian ABJM theory, exactly solved in Subsection \ref{sec:AbelianABJM}. As a consistency check, we expand the geometric series 
	\begin{equation*}
		\sum_{\alpha=0}^{k-1} \frac{ t_j ^{\alpha} }{  t_j ^{k}  -1 } = - \sum_{\alpha=0} ^{k-1} \sum_{\beta =0} ^{\infty} t_j ^{\alpha + \beta k } = -  \sum_{\alpha=0} ^{\infty} t_j ^{\alpha} 
	\end{equation*}
	in the answer from Subsection \ref{sec:AbelianABJM}, and confirm that the Schur expansion reproduces the correct coefficients to all orders in $t_1, t_2$, although, as expected, it misses all the terms proportional to $e^{i \pi m_j ^2}$.\par
	Explicit Schur expansions for higher rank quivers are given in Appendix \ref{app:AbelianquiverSchurexp}.

    \subsection{Schur expansion for Wilson loops}	
    \label{sec:1WLSchurExp}
		
		It is possible to combine the ideas used in this Section with those of Section \ref{sec:WLABJ} to study Wilson loops.\par
		We come back to the setting of Section \ref{sec:WLABJ} and consider the vev of a single $\frac{1}{2}$-BPS Wilson loop in ABJ theory, with ranks $N_1$ and $N_2$. We assume the Wilson loop carries a typical representation $\mu$ of the supergroup $U(N_1 \vert N_2)$ \cite{DrukTranca}. We write
		\begin{align*}
			\frac{ \mathfrak{s}_{\mu} (e^{2 \pi x} \vert e^{2 \pi y}) }{  \prod_{a=1}^{N_1} \prod_{\dot{a}=1}^{N_2}\left( 2 \cosh \pi ( x_a - y_{\dot{a}} ) \right)^2 } & =  \mathfrak{s}_{\gamma} (e^{2 \pi x}) \mathfrak{s}_{\eta^{\prime}} (e^{2 \pi y}) \prod_{a=1}^{N_1} \prod_{\dot{a}=1}^{N_2} \frac{ e^{2 \pi x_a + 2 \pi y_{\dot{a} } } }{  e^{2 \pi x_a} + e^{2 \pi  y_{\dot{a} } } } \\
				& = \mathfrak{s}_{\gamma} (e^{2 \pi x}) \mathfrak{s}_{\eta^{\prime}} (e^{2 \pi y}) \left( \prod_{a=1}^{N_1}  e^{2 \pi x_a} \right) \sum_{\nu }  (-1)^{ \vert \nu \vert } \mathfrak{s}_{\nu} (e^{2 \pi x}) \mathfrak{s}_{\nu} (e^{- 2 \pi y}) ,
		\end{align*}
		where the first equality follows from the factorization property \eqref{factlongrep}, while to pass from the first to the second line we have used the Cauchy identity \eqref{eq:Cauchyid} and brought out the factor $(-1)$ from $\mathfrak{s}_{\nu} (- e^{- 2 \pi y})$. The sum over $\nu$ runs over all partitions
		\begin{equation*}
			\left\{ \nu \ : \  \mathrm{length} (\nu) \le \min (N_1, N_2) \right\} .
		\end{equation*}
		It is important to stress the difference between the results we present in this Subsection and the ones we have found in Section \ref{sec:WLABJ}. There, the correlator of two or more Wilson loops in ABJ has been taken into account, and the factorization of the final result into vevs of Wilson loops in CS theories without matter is exact. Here, instead, we consider a single Wilson loop in ABJ, and we use the Cauchy identity to expand the interaction between the two nodes. In turn, the latter is the Schur expansion of the $\mathsf{A}_2$ quiver of Subsection \ref{sec:SchurexpUNNf}.\par
		With the Schur expansion, the expectation value of the Wilson loop decomposes into a sum of contributions, indexed by the partition $\nu$, factorized into two multiple integrals, one for each node:
		\begin{equation}
			\langle W_{\mu} \rangle_{N_1, N_2 ;k} \stackrel{\text{pert.}}{=} \frac{1}{ \mz_{\mathrm{ABJ} } }  \sum_{\nu }  (-1)^{ \vert \nu \vert } \mathcal{W}^{(1)} _{\gamma \nu } \mathcal{W}^{(2)} _{\nu \eta^{\prime }} ,
	    \label{eq:1WLfact2}
		\end{equation}
		\begin{align*}
			\mathcal{W}^{(1)} _{\gamma \nu }  &:= \int_{\R^{N_1}} \prod_{1 \le a < b \le N_1}  \left( 2 \sinh \pi (x_b - x_a) \right)^2  ~ \mathfrak{s}_{\gamma} (e^{2 \pi x})  \mathfrak{s}_{\nu} (e^{2 \pi x})  ~\prod_{a=1}^{N_1} e^{ i \pi k  x_a ^2 + 2 \pi x_a  }  ~ dx_a, \\
			\mathcal{W}^{(2)} _{\nu \eta^{\prime }} & := \int_{\R^{N_2}} \prod_{1 \le \dot{a} < \dot{b} \le N_1}  \left( 2 \sinh \pi (y_{\dot{b}} - y_{\dot{a}} ) \right)^2 ~ \mathfrak{s}_{\nu} (e^{2 \pi y})  \mathfrak{s}_{\eta^{\prime}} (e^{- 2 \pi y}) ~ \prod_{\dot{a}=1} ^{N_2} e^{ - i \pi k  y_{\dot{a}} ^2  } ~ dy_{\dot{a}}  .
		\end{align*}
		The first function corresponds to integration over the Cartan subalgebra of $\mathfrak{u} (N_1)$ and the second to integration over the Cartan subalgebra of $\mathfrak{u} (N_2)$. In the integral over the second node, we have reflected variables $y_{\dot{a}} \mapsto - y_{\dot{a}}$. The term $\prod_a e^{2 \pi x_a}$ in $\mathcal{W}_{\gamma \nu } ^{(1)}$ can be removed with a shift of variables and translating back the integration cycle onto $\R$, obtaining 
		\begin{equation*}
			\mathcal{W}^{(1)} _{\gamma \nu } = e^{ \frac{i \pi}{k} ( N_1 + 2 \vert \gamma \vert + 2 \vert \nu \vert )} \int_{\R^{N_1}}  \mathfrak{s}_{\gamma} (e^{2 \pi x})  \mathfrak{s}_{\nu} (e^{2 \pi x}) ~ \prod_{1 \le a < b \le N_1}  \left( 2 \sinh \pi (x_b - x_a) \right)^2 ~ \prod_{a=1}^{N_1}  e^{i \pi k x_a ^2} dx_a .
		\end{equation*}\par
		\medskip

		\subsubsection{Rectangular partition}
			The simplest case is the expectation value of a Wilson loop in a rectangular representation $\mu = \kappa$, so $\gamma = \emptyset = \eta$. We get:
			\begin{align*}
				\mathcal{W}^{(1)} _{\emptyset \nu } & = e^{ \frac{i \pi}{k} ( N_1 + 2 \vert \nu \vert )} \int_{\R^{N_1}}  \mathfrak{s}_{\nu} (e^{2 \pi x})  ~ \prod_{1 \le a < b \le N_1}  \left( 2 \sinh \pi (x_b - x_a) \right)^2 ~ \prod_{a=1}^{N_1}  e^{i \pi k x_a ^2} dx_a \\
					& =  \mathcal{Z}_{N_1;k } ~ q^{- \frac{ \mathsf{C}_{2; N_1} (\nu) }{2} - \frac{N_1}{2} -  \vert \nu \vert }  \dim_q \nu 
			\end{align*}
			and
			\begin{align*}
				\mathcal{W}^{(2)} _{ \nu \emptyset} & =  \int_{\R^{N_2}} d^{N_2} y \ \mathfrak{s}_{\nu} (e^{2 \pi x})  \ e^{-  \sum_{\dot{a}=1} ^{N_2} i \pi k  y_{\dot{a}} ^2  }  \prod_{1 \le \dot{a} < \dot{b} \le N_2}  \left( 2 \sinh \pi (y_{\dot{b}} - y_{\dot{a}} ) \right)^2  \\
					& =\mathcal{Z}_{N_2;-k}  \ q^{\frac{ \mathsf{C}_{2; N_2} (\nu) }{2} }  \dim_{q^{-1}} \nu 
			\end{align*}
			In both evaluations, the second line follows from the Wilson loop vev \eqref{eq:WLCSk} and $\mathcal{Z}_{N_1,k} $ and $\mathcal{Z}_{N_2, -k}$ are the corresponding normalizations. Noting that $\dim_{q^{-1}} \nu = \dim_q \nu$, the vev of the Wilson loop in a rectangular representation $\kappa$ of the supergroup $U(N_1 \vert N_2)$ is then 
			\begin{equation*}
				\langle W_{\kappa} \rangle_{N_1,N_2;k} \stackrel{\text{pert.}}{=} \frac{ \mathcal{Z}_{N_1;k} \mathcal{Z}_{N_2;-k} }{\mz_{\mathrm{ABJ} (N_1 \vert N_2)_{k}} } q^{- \frac{N}{2}} \ \sum_{\nu} (-q)^{- \vert \nu \vert } \left( \dim_q \nu \right)^2 q^{\frac{ \mathsf{C}_{2; N_2} (\nu) - \mathsf{C}_{2; N_1} (\nu) }{2} }  .
			\end{equation*}
			In particular, for ABJM theory, $N_1 = N = N_2$, the quadratic Casimir cancels and we arrive to the simpler formula
			\begin{equation*}
				\langle W_{\kappa} \rangle_{N,N;k} \stackrel{\text{pert.}}{=} \frac{ \mathcal{Z}_{N;k}  \mathcal{Z}_{N;-k} }{\mz_{\text{ABJM} (N)_k} } q^{- \frac{N}{2}} \ \sum_{\nu} (-q)^{- \vert \nu \vert } \left( \dim_q \nu \right)^2  .
			\end{equation*}

		\subsubsection{Arbitrary typical representation}

			We now tackle the general case of a typical (long) but otherwise arbitrary representation $\mu$, and give two equivalent, and in fact related, evaluations of the vev of the Wilson loop in ABJ theory.\par
			Both approaches require to invert the variables in a Schur polynomial, which can be done using the identity \eqref{invertschur}.\par
			The first procedure mimics \cite{Kimura}, and extends the result to the unknot Wilson loop. Inverting variables in $\mathcal{W}^{(1)}_{\gamma \nu}$ using \eqref{invertschur} we identify $\mathcal{W}^{(1)}_{\gamma \nu}$ and $\mathcal{W}^{(2)}_{\nu \eta^{\prime}}$ with Hopf link invariants computed in $U(N_1)_k$ and $U(N_2)_{-k}$ Chern--Simons theory on $\mathbb{S}^3$, respectively. Explicitly: 
			\begin{equation}
			\label{Wloopexpavg}
				\langle W_{\mu} \rangle_{N_1, N_2 ;k} = \frac{\mz_{N_1;k} \mz_{N_2;-k} }{\mz_{\mathrm{ABJ} (N_1 \vert N_2)_{k}} } ~ \sum_{\nu} C_{\gamma \nu} (q)~ \langle W_{ \gamma \nu^{\ast}} \rangle_{N_1;k} \langle W_{\nu \eta^{\prime}} \rangle_{N_2;-k}  ,
			\end{equation}
			where the averages in the sum are the Hopf link invariants and the summands are weighted by  
			\begin{equation}
			\label{WloopweightC}
				 C_{\gamma \nu} (q) = (-1)^{\vert \nu \vert} q^{ - (1+ \nu_1) \left( \vert \gamma \vert + \vert \nu \vert + \frac{1 + \nu_1}{2} N_1 \right)} .
			\end{equation}
			The partition $\nu^{\ast}$ has been defined in \eqref{defstarrednu}. In the operator formalism the expansion \eqref{Wloopexpavg} takes the form
			\begin{equation*}
				\langle 0 \vert T S T \vert \mu \rangle_{N_1, N_2 ;k} = \frac{\mz_{N_1;k} \mz_{N_2;-k} }{\mz_{\mathrm{ABJ} (N_1 \vert N_2)_{k}} } ~ \sum_{\nu} C_{\gamma \nu} (q)~ \langle \eta \vert T S T \vert \nu \rangle_{N_1, k}  \langle \nu^{\ast} \vert T S T \vert \gamma \rangle_{N_2, -k} ,
			\end{equation*}
			where $T,S$ are the $SL(2, \mathbb{Z})$ modular matrices. Note that the two $\nu$'s are treated differently: one is considered as a $U(N_1)$ representation and the other as a $U(N_2)$ representation, with the latter twisted by the starred partition. The appearance of the operator $TST$ rather than $S$ is because the matrix model presentation computes the observables in a special instance of the Seifert framing, rather than in the natural $\mathbb{S}^3$ framing.\par
			The alternative path consists in applying the inversion formula \eqref{invertschur} to $\mathfrak{s}_{\nu} (e^{2 \pi y})$. Similar manipulations lead to: 
			\begin{equation}
			\label{WlambdasimpleW}
				\langle W_{\mu} \rangle_{N_1, N_2 ;k} =  \frac{\mz_{N_1;k} \mz_{N_2;-k} }{\mz_{\mathrm{ABJ} (N_1 \vert N_2)_{k}} } ~ q^{- \frac{N_1}{2} - \vert \gamma \vert } \sum_{\nu} \tilde{C}_{\nu \eta^{\prime}}  (q) \langle W_{\gamma} W_{\nu} \rangle_{N_1;k} \langle W_{\nu^{\ast}}  W_{\eta^{\prime}}  \rangle_{N_2;-k} 
			\end{equation}
			with coefficient
			\begin{equation*}
				\widetilde{C}_{\nu \eta^{\prime}}  (q) = (-q)^{\vert \nu \vert } q^{- \nu_1 \left( \vert \eta^{\prime} \vert + \vert \nu \vert - \frac{\nu_1}{2} N_2 \right)} .
			\end{equation*}
			The expression \eqref{WlambdasimpleW} is expressed as a sum of correlators of two pairs of (unlinked) unknots, one pair in each CS theory. These correlators can be further reduced with a character expansion in the Schur basis:
			\begin{equation*}
				\langle W_{\mu} \rangle_{N_1, N_2 ;k} =  \frac{\mz_{N_1;k} \mz_{N_2;-k} }{\mz_{\mathrm{ABJ} (N_1 \vert N_2)_{k}} } ~ q^{- \frac{N_1}{2} - \vert \gamma \vert } \sum_{\nu} \widetilde{C}_{\nu \eta^{\prime}}  (q) \sum_{\tilde{\nu}, \hat{\nu}} {\mathsf{N}_{\gamma \nu}}^{\tilde{\nu}}  \langle W_{\tilde{\nu} } \rangle_{N_1;k}  {\mathsf{N}_{\eta^{\prime} \nu^{\ast} }}^{\hat{\nu}}  \langle W_{\hat{\nu}} \rangle_{N_2;-k} ,
			\end{equation*}
			where, as above, ${\mathsf{N}_{\gamma \nu}}^{\tilde{\nu}}$ are the Littlewood--Richardson coefficients.\par
			The solvability is again preserved if we turn on a Romans mass in the dual theory as prescribed in \cite{GaiTom}. The above computation is straightforwardly generalized to $k_2 \ne - k_1$ and gives:
			\begin{equation*}
				\langle W_{\mu} \rangle_{N_1, N_2 ;k_1, k_2} =  \frac{\mz_{N_1;k_1} \mz_{N_2;k_2} }{\mz_{\mathrm{ABJ} (N_1 \vert N_2)_{k_1, k_2}} } ~ \sum_{\nu} C_{\gamma \nu } (q_1) ~ \langle W_{ \gamma \nu^{\ast}} \rangle_{N_1;k_1} \langle W_{\nu \eta^{\prime}} \rangle_{N_2;k_2} 
			\end{equation*}
			with summands weighted by \eqref{WloopweightC} with $q= q_1 = e^{- \frac{i2 \pi}{k_1}}$, hence independent of $k_2$.\par
			Our derivation complements previous results \cite{MoriyamaWL1,NiiWLSeiberg} extending the analysis to a broader class of representations.

		\subsection{The Schur expansion does not probe dualities}
			The Schur expansions we developed are useful tools to read off the coefficients in a perturbative expansion in the parameters $t_j$, given in terms of simple topological invariants. Such expansions, however, are of no use in an attempt to test gauge theory dualities. We discuss here the reasons.\par
			First of all, although dual theories should have the same global symmetries, the flavour symmetry with respect to which we apply a Schur expansion in one theory, may not (and in general, shall not) be mapped into a symmetry useful for a Schur expansion in the dual picture. Therefore, the fugacities used for the Schur expansions in the dual theories, namely $\left\{ t_j ^{\text{electric}} \right\}$ and  $\left\{ t_j ^{\text{magnetic}} \right\}$, are not mapped into each other by the duality map in general.\par
			A second crucial aspect is that nothing guarantees that, given a theory suitable for the Schur expansion, its dual admits a meaningful expansion at all. When both theories have CS couplings, we are able to give an expansion on both sides of the duality, although using different fugacities. However, for theories whose dual is not a Chern--Simons theory, it is possible that the Schur expansion would lead to ill-defined quantities.\par
			To exemplify the problems, consider the duality between ABJM theory with $k=1$ and super-Yang--Mills with one adjoint and one fundamental hypermultiplet \cite{ABJM}. The former theory is a particular case of Subsection \ref{sec:SchurexpABJ}, and can be in expanded in the fugacities associated to a $U(1) \times U(1)$ symmetry rotating the bi-fundamentals. As in Subsection \ref{sec:SchurexpABJ}, we are identifying the Cartan subalgebra $\mathfrak{u}(1)_{\text{flavour}} \oplus \mathfrak{u}(1)_{\text{top}} $ of the global symmetry with the Cartan subalgebra of an enhanced $\mathfrak{u} (2)_{\text{flavour}}$, through a simple change of variables that shifts the masses, $( m, -m) \mapsto( m - \zeta, -m - \zeta )$. If, on the other hand, we try to expand SYM for the fugacity $t_0$ associated to the $U(1)_{\text{adj}}$ flavour symmetry rotating the adjoint, we obtain vevs of Wilson loops in $U(N)$ SYM with only one fundamental. The latter is a bad theory, in the sense that the localized integral in the UV does not capture the IR behaviour. Alternatively, we may attempt an expansion using the flavour symmetry $U(1)_{\text{fund}}$ rotating the fundamentals. This procedure gives as coefficients of the powers of the variable $t $ the vevs of Wilson loops wrapping a great circle in $\mathbb{S}^3$ in SYM with one adjoint, which give ill-defined answers if we naively try to compute them from the localized path integral.
	
	\subsection{Comments on the $U(N)$ theory with $N_f$ fundamental hypermultiplets}
				
				As a final observation, and departing from the previous use of Schur expansions, we discuss further the partition function $\mathcal{Z}_{U(N),N_f}$. The expression \eqref{eq:ZNfSchurW} appears in topological string theory \cite{Conifold:2006} in the study of non-compact branes on the resolved conifold. There, the fugacities $t_j$ correspond to diagonal holonomies of the gauge fields along a circle $\mathbb{S}^1$, determined as the locus where a non-compact brane intersects $\mathbb{S}^3$. Replacing a brane with an anti-brane in the framework of \cite{Conifold:2006} corresponds here to exchange the Cauchy identity \eqref{eq:Cauchyid} with the dual Cauchy identity \eqref{eq:dualCauchy}, which describes the Schur expansion of the matrix model 
		\begin{equation*}
			\mathcal{Z}_{U(N),N_f} ^{\text{ferm.}} = \int_{\R^N} \prod_{1 \le a < b \le N }  \left(  2 \sinh \pi (x_b - x_a) \right)^2  ~\prod_{a=1}^{N} \left[ \prod_{j=1} ^{N_f} 2 \cosh \pi (x_a + m_j ) \right] ~ e^{- \frac{1}{2g} ( 2 \pi x_a) ^2} ~ dx_a .
		\end{equation*}
		The choice of notation ``ferm.'' for this matrix model will be justified in Subsection \ref{sec:ZUNfanddual}.\par
		In the Abelian theory, in particular, replacing a brane with an anti-brane \cite{Conifold:2006} switches from the generating function of the complete homogeneous symmetric polynomials to that of the elementary symmetric polynomials $\mathfrak{e}_{\alpha}$ \cite{Macdonaldbook}, cfr. equation \eqref{eq:genele}.\par
		We compare now $\mathcal{Z}_{U(N),N_f}$ with $\mathcal{Z}_{U(N),N_f} ^{\text{ferm.}}$. 
				The former is the partition function of $U(N)$ CS theory at level $k$ on $\mathbb{S}^3$ coupled to $N_f$ fundamental hypermultiplets. We have introduced the latter to mimic the pair of identities \eqref{eq:Cauchyid}-\eqref{eq:dualCauchy} at the level of matrix integrals. Nevertheless, there are several physical motivations to study both $\mathcal{Z}_{U(N),N_f}$ and $\mathcal{Z}_{U(N),N_f} ^{\text{ferm.}}$.\par
				As we have already mentioned, in topological string theory on the conifold it is important to have both functions \cite{Conifold:2006,Okuyama:2006eb}. Moreover, the matrix model $\mathcal{Z}_{U(N),N_f} ^{\text{ferm.}}$ with all the masses vanishing, has been studied in \cite{Anninos:2016klf} in the context of fermionic quantum mechanics, and solved in \cite{TGrass} for any $\left\{ m_j \right\} \subset \R^{N_f}$. A third motivation for the introduction of the ``fermionic'' partition function comes from looking at each summand in the Schur expansions. 
				Consider a fixed $\nu$ in the sums over representations which corresponds to a symmetric $SU(N)$ representation. The associated reduced coloured knot invariants have been categorified in \cite{GukovStosic1,GukovStosic2}. The corresponding homologies posses a mirror symmetry which exchanges the symmetric representation $\nu$ with the totally antisymmetric representation $\nu^{\prime}$. Generalizing this operation to the present context, replacing one representation by its conjugate, switches from the Cauchy identity \eqref{eq:Cauchyid}, to the dual Cauchy identity \eqref{eq:dualCauchy}.\par
				A fourth, heuristic argument to consider the pair $\mathcal{Z}_{U(N),N_f}$ $\mathcal{Z}_{U(N),N_f} ^{\text{ferm.}}$ is presented in Subsection \ref{sec:ZUNfanddual}.

		\subsubsection{Averages of characteristic polynomials}
		\label{sec:charpol}
		
		It has been shown in \cite{Tchar} that $\mathcal{Z}_{U(N),N_f}$ computes the average of the inverse of the product of characteristic polynomials in the Stieltjes--Wigert random matrix ensemble, which describes the CS matrix model \cite{T0}. Explicitly: 
		\begin{equation*}
			\frac{ \mathcal{Z}_{U(N),N_f} }{ \mz_{\mathrm{CS} (N)} }  \propto  \left\langle \left[ \prod_{j=1} ^{N_f}  \det \left( \widetilde{t}_j ^{\vee} - \mathbf{X} \right) \right]^{-1}  \right\rangle_{\mathrm{SW} (N)} ,
		\end{equation*}
		with $\mathbf{X}$ a random Hermitian matrix whose eigenvalues are $(x_1, \dots, x_N)$. The spectral parameters $\widetilde{t}_j ^{\vee}$ are related to the physical quantities through 
		\begin{equation*}
			\widetilde{t}_j ^{\vee} = - q^{- N - \frac{N_f}{2}} e^{- 2 \pi m_j } = q^{- N - \frac{N_f}{2}} t_j^{-1} ,
		\end{equation*}
		with $q=e^{-g}$.\par
		The average of the inverse product of characteristic polynomials in the Hermitian random matrix ensemble with Stieltjes--Wigert weight is calculated exactly, and is a $N_f \times N_f$ determinant:
		\begin{equation}
		\label{eq:ZUNNfcharpol}
			\frac{ \mathcal{Z}_{U(N),N_f} }{ \mz_{\mathrm{CS} (N)} } =  \frac{ c_{N,N_f} }{ \prod_{1 \le j < l \le N_f}  \left( \widetilde{t}_j  ^{\vee} - \widetilde{t}_l  ^{\vee} \right) }  \det_{1 \le j,l \le N_f} \left[  \mathfrak{p} ^{\vee} _{N+l-1} \left( \widetilde{t}_j  ^{\vee} \right) \right] ,
		\end{equation}
		where $ \mathfrak{p}^{\vee} _n ( \widetilde{t} ^{\vee} )$ are the Cauchy transform of the Stieltjes--Wigert orthogonal polynomials, and the constant $c_{N,N_f}$ in \eqref{eq:ZUNNfcharpol} does not depend on the spectral parameters $\widetilde{t}_j  ^{\vee}$. We refer to \cite{Tchar} for more details, proofs and references.\par
		We obtain the analogous expression for the other matrix integral considered, in terms of averages of products of characteristic polynomials in the Stieltjes--Wigert ensemble \cite{TGrass}:
		\begin{equation*}
			\frac{ \mathcal{Z}_{U(N),N_f} ^{\text{ferm.}}}{ \mz_{\mathrm{CS} (N)} } \propto  \left\langle \prod_{j=1} ^{N_f}  \det \left( \widetilde{t}_j - \mathbf{X} \right) \right\rangle_{\mathrm{SW} (N)} ,
		\end{equation*}
		Here, the spectral parameters $\widetilde{t}_j $ are related to the parameters of the gauge theory as 
		\begin{equation*}
			\widetilde{t}_j = - q^{-N + \frac{N_f}{2}} e^{- 2 \pi m_j } = q^{-N} \left( q^{- \frac{N_f}{2}} t_j\right)^{-1} .
		\end{equation*}
		The average of the product of characteristic polynomials is explicitly given by a $N_f \times N_f$ determinant 
		\begin{equation*}
			\frac{ \mathcal{Z}_{U(N),N_f} ^{\text{ferm.}}}{ \mz_{\mathrm{CS} (N)} } =  \frac{ c_{N,N_f} ^{\text{ferm.}} }{ \prod_{1 \le j < l \le N_f}  (\widetilde{t}_j - \widetilde{t}_l) }  \det_{1 \le j,l \le N_f} \left[  \mathfrak{p} _{N+l-1} \left( \widetilde{t}_j \right) \right] ,
		\end{equation*}
		where $\mathfrak{p}_{n} (\widetilde{t} )$ are the Stieltjes--Wigert polynomials, and $c_{N,N_f} ^{\text{ferm.}}$ is a numerical constant. We refer to \cite{TGrass} for details and a detailed list of references. We remark that a closely related result was obtained in \cite{Okuyama:2006eb}, in the context of topological string theory on the conifold.

		\subsubsection{Bosonic versus fermionic matrix models}
		\label{sec:ZUNfanddual}
		We can recast the two expressions in a unified formalism, integrating over auxiliary variables: 
		\begin{equation}
		\label{eq:replicaUNNf}
			\mathcal{Z}_{U(N),N_f} ^{\epsilon}  =  \int d \mathbf{X} ~ e^{- \frac{1}{2g} \mathrm{Tr} \left( \log \mathbf{X} \right)^2 } ~ \prod_{j=1} ^{N_f}  \int  e^{ - \bar{\psi} ^{j} \left(  \mathbf{X} - \widetilde{t}_j ^{\epsilon}  \right) \psi ^{j} }  ~\prod_{a=1}^{N}  \frac{ d\bar{\psi}^{j} _a d\psi^{j} _a }{ 2 \pi } .
		\end{equation}
		In this expression, $\epsilon \in \left\{ \pm 1 \right\}$, with $\epsilon =-1$ giving $\mathcal{Z}_{U(N),N_f}$ and $\epsilon =+1$ giving $\mathcal{Z}_{U(N),N_f}^{\text{ferm.}}$. The spectral parameters are respectively $\widetilde{t}_j ^{\vee}$ and  $\widetilde{t}_j$ for $\epsilon=-1,+1$. The integration is over $N_f$ $N$-component vectors $\psi^{j}= (\psi^j _a)_{a=1, \dots,N}$, for $j=1, \dots, N_f$, and their conjugates $\bar{\psi}^j=(\bar{\psi}^j _a)_{a=1, \dots,N}$. These vectors have Grassmann-even entries when $\epsilon=-1$ and Grassmann-odd entries when $\epsilon=+1$. We recall that $\widetilde{t}_j ^{\vee}, \widetilde{t}_j <0$ from their definition in terms of the physical variables, which guarantees that the inner integral in \eqref{eq:replicaUNNf} is well posed. Besides, we have dropped an overall constant.\par
		Written in the form \eqref{eq:replicaUNNf}, we see that switching from the Cauchy identity \eqref{eq:Cauchyid} to the dual Cauchy identity \eqref{eq:dualCauchy} passes from the Schur expansion of the matrix model \eqref{eq:replicaUNNf} with bosonic fields to the Schur expansion of \eqref{eq:replicaUNNf} with fermionic fields.\footnote{The suggestive form \eqref{eq:replicaUNNf} does not seem to allow a unified treatment of fermionic and bosonic versions of the quantum mechanical model of \cite{Anninos:2016klf}, because the matrix model representation $\mathcal{Z}_{U(N),N_f}^{\text{ferm.}}$ has been originally derived using a Wick rotation that is forbidden in the bosonic counterpart.}\par
		\medskip
		The ideas of the present Subsection can be applied to $4d$ $\mN=4$ SYM with co-dimension 1 matter defects sitting on a great $\mathbb{S}^3$ inside $\mathbb{S}^4$. 
		The analogue of \eqref{eq:replicaUNNf} is 
		\begin{equation}
		\label{eq:replica4ddefect}
			\mathcal{Z}_{U(N),N_f} ^{4d+\text{defect}, \epsilon }  =  \int d \mathbf{X} \int\frac{ d\bar{\psi} d\psi }{ (2 \pi)^{N} }~ e^{- \mathrm{Tr} \left[ \frac{8 \pi^2}{g_{4d}}  \mathbf{X}^2 + \sum_{j=1}  \bar{\psi} ^{j} \left(  e^{\mathbf{X}} - \widetilde{t}_j ^{\epsilon}  \right) \psi ^{j} \right] } .
		\end{equation}
		Here $\epsilon=-1$ corresponds to the physical $4d$ $\mN=4$ theory and $\epsilon=+1$ is its counterpart using a fermionic matrix representation. Compared to the purely three-dimensional theory, we have removed the $\log^2$-interaction, at the cost of an exponential term in the action. Rescaling $\mathbf{X}_{ab} \to \sqrt{g_{4d}/16 \pi^2} \mathbf{X}_{ab} $ we can expand the interaction term in \eqref{eq:replica4ddefect} in a power series in 
		\begin{equation*}
			g_{\text{eff}} = \frac{ \sqrt{g_{4d}} }{4 \pi } .
		\end{equation*}
		The resulting effective action includes infinitely many vertices:
		\begin{equation*}
			 \bar{\psi} ^{j} \left(  e^{\mathbf{X}} - \widetilde{t}_j ^{\epsilon}  \right) \psi ^{j} =  \bar{\psi} ^{j} _a \left( 1 - \widetilde{t}_j ^{\epsilon}  \right)\delta_{ab} \psi ^{j} _b \ + \ g_{\text{eff}} \bar{\psi} ^{j} _a X_{ab}  \psi ^{j} _b \ + \ \frac{g_{\text{eff}} ^2}{2}  \bar{\psi} ^{j} _a X_{ac} X_{cb}  \psi ^{j} _b + \dots 
		\end{equation*}
		and can be analyzed by standard perturbative techniques in random matrix theory \cite{DiFrancesco:2004}. \par 
		The upshot of this digression is that we may as well describe four-dimensional $\mN=4$ SYM with defects using random matrix theory, and equivalently represent it as a theory of massive scalars $\bar{\psi}^j, \psi^j$ in the vector representation of $U(N)$ which interact with a (zero-dimensional) gluon $\mathbf{X}$ in the adjoint representation of $U(N)$. We also naturally get an associated theory in which the bosons are replaced by zero-dimensional fermions.\par

		\section{Outlook}
		
		The explicit evaluation of the Mordell integral has been a valuable analytical tool, giving explicit results, containing both perturbative and non-perturbative information. It naturally comes split in two terms, the first of which is a Gauss sum, refined by powers of $e^{i 2 \pi \lambda }$.\par
		The Gauss sum in itself may be worth of further consideration, taking into account its intricate behavior, especially when the parameter is irrational, and known ``renormalization'' features appear, leading for example to spiral patterns when plotting the function on the complex plane, while moving the value of the parameter, which in our case would be the Chern--Simons level \cite{coutsias1987disorder,berry1988renormalisation}. 
		Scaling theories of such patterns were studied in these, and posterior works, precisely using what can be considered as a real-space renormalization procedure, based on grouping the terms in the sum into blocks. The first consideration of a renormalization equation can be considered to be the functional equation given for Gauss sums by Hardy and Littlewood \cite{hardy1914some}. 
		
		Even for rational values of the parameter we have non-trivial behavior \cite{coutsias1987disorder,berry1988renormalisation}, and that case would be relevant to, say, Abelian Chern--Simons-matter theories.\footnote{Its relation with the irrational $k$ case has been commented in Subsection \ref{sec:Lessonkrational}.} The partition functions we studied are expressed in terms of such Gauss sums. However, the dependence on the mass and the second term in Mordell's solution introduce a new behaviour, cfr. the plots in Subsection \ref{sec:AbelianA1Nf1}. Therefore, a further look into deeper mathematical features, in particular interpretations of such sums from the point of view of dynamical systems and ergodic theory \cite{marklof1999limit} may be worthwhile. It has already been argued that localization results may be a useful playground to further understand renormalization behavior in a broader sense \cite{Gukov:2016tnp}.
		
		We have considered generic long representations when discussing Wilson loops in the ABJ and ABJM matrix models. This leads to the possibility of applying the Berele--Regev factorization \eqref{factlongrep}. As usual in the representation theory of Lie superalgebras, everything is more involved and less well-known if one wants to consider atypical representations. It would be interesting to consider the rather newer result, quoted in \cite[Sec. 5.7]{Feigin}, for atypical representations, where seemingly the supersymmetric Schur can be expressed as a sum of factorized terms, and try to carry out the same type of computations with it.
		
		There are further possible Schur expansions that we have not discussed. For example, we may consider ABJ(M) theories with orthogonal and symplectic gauge group, that describes orientifolds, or more general ortho-symplectic quiver theories, with gauge group the alternating product of orthogonal and unitary symplectic groups. It turns out that the different Haar measures involved in the resulting matrix model admit an expression in terms of the $U(N)$ Haar measure times a sum of Schur polynomials. This expressions can be found for example in Macdonald's book \cite[Ch. 5 Ex. 9]{Macdonaldbook} and they have been applied to matrix models in \cite{tierz2020matrix}, but have not been considered for orientifold ABJ(M) matrix models. Combining this property with the Cauchy identity to deal with the contribution from the bi-fundamental hypermultiplets, would lead again to Schur expansions given by sums of Wilson loop vevs in $U(N)$ Chern--Simons theory on the three-sphere.

\acknowledgments	

The work is partially supported by FCT Project PTDC/MAT-PUR/30234/2017. The work of LS is supported by the FCT through the doctoral grant SFRH/BD/129405/2017.

	\clearpage
	
\appendix
	
		\section{Mordell integrals at $\lambda \ne 0$}
		\label{app:mordellproof}
		In this Appendix we explain some subtlety related to the integrals \eqref{MordellG+}-\eqref{MordellG-}. We mainly review results from \cite{Mordell} and comment on how to properly combine them.\par
\begin{figure}[htb]
		\centering
			\includegraphics[width=0.4\textwidth]{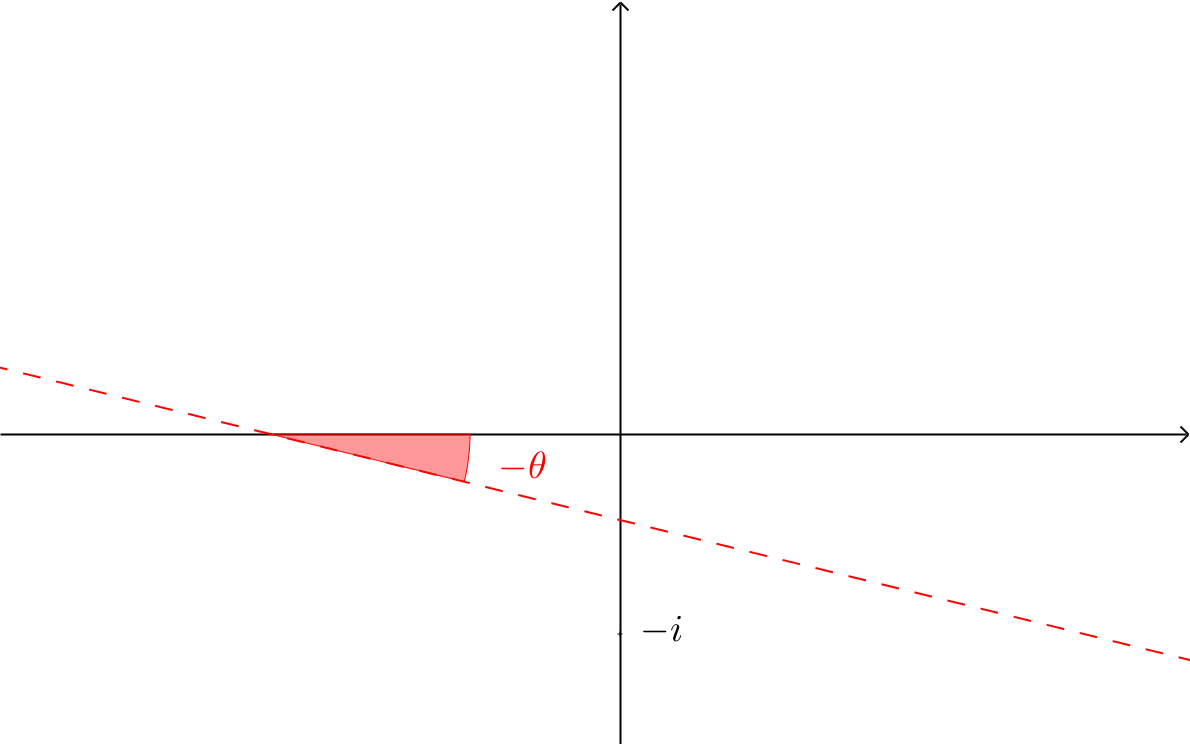}%
			\hspace{0.1\textwidth}%
			\includegraphics[width=0.4\textwidth]{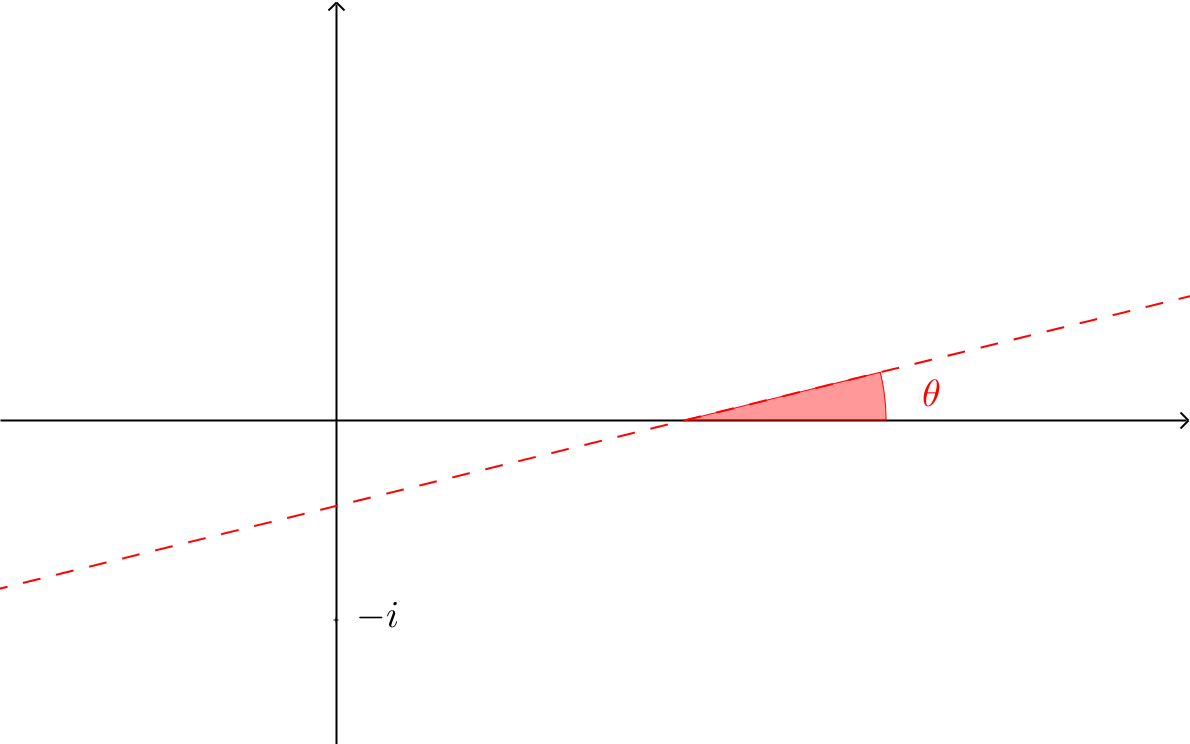}
			\caption{Choice of integration contour for the Mordell integrals, shifted and rotated by a small angle with respect to the real axis. Left: contour for $\Psi_{+}$, rotated by a negative angle $- \theta <0$. Right: contour for $\Psi_{-}$, rotated by a positive angle $\theta >0$.}
			\label{fig:contourMordell}
		\end{figure}

		In his original paper \cite{Mordell}, Mordell gave the formulas \cite[Eq. (8.1)-(8.2)]{Mordell}
		\begin{align}
			\Psi_{+} (\xi) & := \int_{- \infty } ^{+ \infty} dx \frac{ e^{i \pi \frac{\kappa}{\varrho} x^2 - 2 \pi x \xi }  }{ e^{2 \pi x} - 1 } = \frac{ 1 }{ e^{ i \pi \varrho \left(  2 \xi - \kappa \right) }  - 1  } \left[ - \sqrt{ \frac{ i \varrho}{\kappa} } \sum_{\alpha=1} ^{\kappa} e^{i \pi \frac{\varrho}{\kappa} \left( \xi + \alpha \right)^2   }  + i \sum_{\beta=1}^{\varrho} e^{ i \pi \beta \left( 2 \xi - \frac{\kappa}{\varrho} \beta  \right)  }  \right] , \label{MordellL0+} \\
			\Psi_{ -} (\xi) & :=\int_{- \infty } ^{+ \infty} dx \frac{ e^{- i \pi \frac{\kappa}{\varrho} x^2 - 2 \pi x \xi } }{ e^{ 2 \pi x}  - 1 } = \frac{ 1 }{ e^{i \pi \varrho \left( 2 \xi  - \kappa \right) } -1  } \left[ \sqrt{- \frac{ i \varrho}{ \kappa} } \sum_{\alpha= 0} ^{\kappa-1} e^{- i \pi \frac{\varrho}{\kappa} \left( \xi- \alpha \right)^2   }  + i \sum_{\beta=1}^{ \varrho} e^{ i \pi \beta \left( 2 \xi  + \frac{\kappa}{\varrho} \beta  \right)  }   \right] , \label{MordellL0-} 
		\end{align}
		valid for $\kappa, \varrho \in \mathbb{Z}_{>0}$. The integration contour can be taken either along the real axis avoiding $x=0$ by a small semicircle, or on a straight line inclined with respect to the real axis and intersecting the imaginary axis between $0$ and $-i$. The inclination should be a negative angle for $\Psi_{+}$ and a positive angle for $\Psi_{-}$. We follow this latter choice, and represent the inclined straight line in Figure \ref{fig:contourMordell}. The final result is independent of the angle $\theta$ between the integration axis and the real axis.\par
		\par
\begin{figure}[htb]
		\centering
			\includegraphics[width=0.5\textwidth]{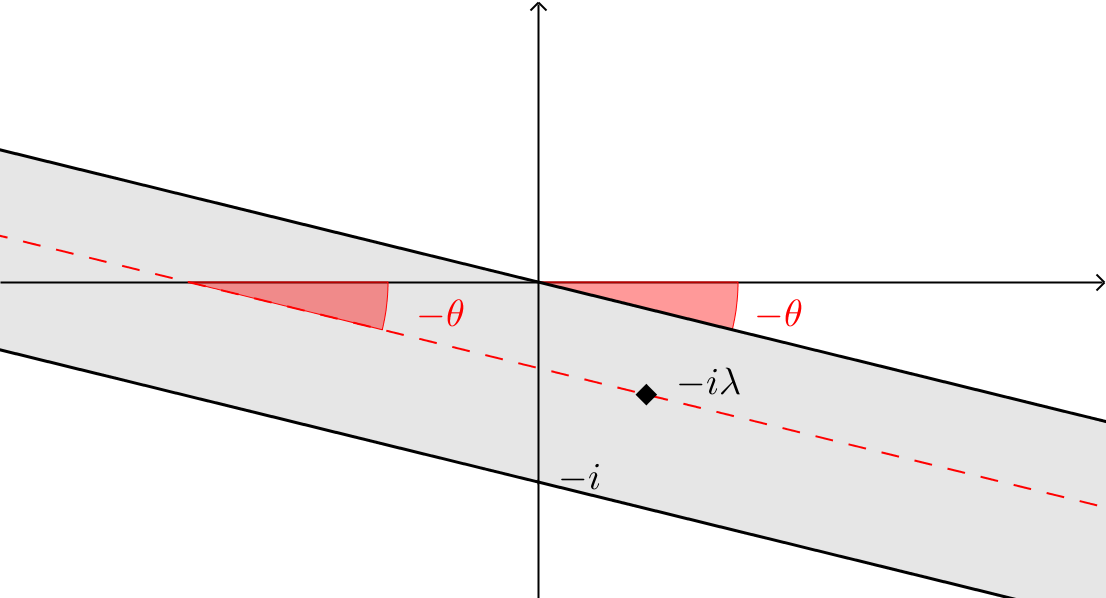}%
			\caption{Choice of integration contour for the Mordell integral $\Psi_{+}$. The angle $0 < \theta < \frac{\pi}{2}$ is arbitrary, and $-i \lambda$ must lie in the shaded region.}
			\label{fig:howtolambda}
		\end{figure}

		On the other hand, the integral 
		\begin{equation*}
			\tilde{\Psi} (\lambda, \xi ) := \int_{\R - i \lambda } dx \frac{ e^{i \pi \tilde{\kappa} x^2 - 2 \pi x (\xi + \tilde{k} \lambda ) }  }{ e^{2 \pi x} - 1 }
		\end{equation*}
		with 
		\begin{equation*}
			\Im (\tilde{\kappa}) >0 , \qquad 0 < \Re \lambda <1 
		\end{equation*}
		is equivalent to \cite[Eq. (3.8)]{Mordell}
		\begin{equation*}
			\tilde{\Psi} (\lambda, \xi ) = e^{i \pi \lambda \left( 2 + 2 \xi + \tilde{\kappa} \lambda \right) } \int_{- \infty} ^{+ \infty } dx \frac{ e^{i \pi \tilde{\kappa} x^2 - 2 \pi x \xi  }  }{ e^{2 \pi x} - e^{i 2 \pi \lambda } }
		\end{equation*}
		now with the integration cycle along the real axis. The proof of this formula \cite{Mordell} makes explicit that one can move the original integration contour in the region $0 < \Re \lambda < 1$ without changing the result. The same is true for the contour of $\Psi_{\pm}$, where we are free to chose where to intersect the imaginary axis. We can therefore introduce the parameter $\lambda$ also in $\Psi_{\pm}$, obtaining the integrals defined in \eqref{MordellG+}-\eqref{MordellG-}. However, in order not to get out of the proper region, we must impose $\theta$-dependent restrictions on $\lambda$, as can be seen from Figure \ref{fig:howtolambda}. So the formula \eqref{MordellG+} for $\Psi_{+}$ hold for 
		\begin{equation*}
			r \sin \theta < \Re \lambda < 1 + r \sin \theta , \qquad - r \cos \theta < \Im \lambda < r \cos \theta 
		\end{equation*}
		where $r \ge 0$ is arbitrary, and similarly for $\Psi_{-}$. Rotating $\theta \to 0^{+}$ we recover the constraints $0 < \Re \lambda < 1 $ and $\Im \lambda$ arbitrary for $\tilde{\Psi}$ defined above. In particular we can fix $0 < \Re \lambda < 1 $ and $0 < \theta \ll \frac{\pi}{2}$, so that we are free to chose $\Im \lambda$ arbitrarily large. With this choice, we change variables in \eqref{MordellL0+} and recover \eqref{MordellG+}, and likewise in \eqref{MordellL0-} to recover  \eqref{MordellG-}, with the integration contour now arbitrarily close to the real axis.\par
		\par
		\medskip
		Another subtle aspect is that the denominators in the right-hand side of \eqref{MordellG+}-\eqref{MordellG-} seem to have a sign ambiguity when $\kappa $ is a multiple of $ \varrho$ and $ \vert \Re \xi \vert =  \frac{1}{2}$ and $\Re \lambda = \frac{1}{2}$. This happens because the result for $\Re \left( \xi - \frac{\kappa}{\varrho} \lambda \right) \in \mathbb{Z}$ is obtained by analytic continuation, and this should be performed at the end of the computations. The result is unique and unambiguous if we move slightly away from such points, for example by a shift $\xi \mapsto \xi + \varepsilon $ for a small $\varepsilon$, and take the limit at the end. Stated differently, the apparent sign ambiguity would only be an artefact of the intermediate steps and will disappear after simplifications in the final answer. We have also checked it for the solutions in Section \ref{sec:exactZ}.\par
		
		\section{Cauchy identities, symmetries and knot homologies}
		\label{app:noknot}
				In this Appendix we discuss further similarities and differences between the expansions presented in Section \ref{sec:Schurexpansion} and generating functions of knot invariants.\par

		\subsubsection{$SU(N)$ Chern--Simons theory with one adjoint and the Hopf link}
		    We first look back at one of the simplest yet suggestive expressions we have found in Subsection \ref{sec:SchurSUNadj}, which is the Schur expansion of the $SU(2)$ Chern--Simons theory with one adjoint hypermultiplet. The Schur expansion of the partition function is given in \eqref{eq:ZSU2Schurexp}, which we report here for clarity: 
			\begin{equation*}
					\frac{\mathcal{Z}_{SU(2)_k, 1_{\mathrm{adj}}} (m) }{ \mathcal{Z}_{SU(2)_k} } \stackrel{\text{pert.}}{=} \sum_{\nu=0}^{\infty} t^{\nu+1} \langle W_{\nu \nu} \rangle_{SU(2)} ,
				\end{equation*}
				with $t=-e^{2 \pi m}$ and $m$ the mass of the adjoint.\par
				As we have already commented in Subsection \ref{sec:SchurSUNadj}, one may think of \eqref{eq:ZSU2Schurexp} as categorifying the Hopf link invariant when the two components are coloured by two equal representations, and $t \to -1$ gives the associated Jones polynomial as the Euler characteristic of the putative homology. There are, however, several obstructions for such an interpretation to hold, as we now discuss.\par
				Firstly, we must underline that, in contrast to Khovanov homology, the refinement parameter $t$ here is graded by the representation (\emph{i.e.} by the color), while the realization of coloured knot and link homologies (see \cite{Gukov:2007,GukovStosic1,GukovStosic2} and references therein) is usually in terms of a given representation. That is, typically the color is not a grading, but rather some fixed data.\par
				A more serious problem is the fact that knot and link homologies are constructed from the reduced knot invariants, normalized by the unknot invariant, so that the coefficients are integers and can be interpreted as the dimensions of certain vector spaces. In the present context, there is no natural way to normalize each term in the sum by the corresponding unknot invariant. Furthermore, even if we implemented such \emph{ad hoc} normalization, the simple structure of the link invariants we obtain does not seem to lead to a rich homological theory.\par
				Going to higher rank, an additional problem shows up: the Schur expansion of the $SU(N)$ theory with one adjoint, given in \eqref{eq:SUNandlinks}, is a function of the single variable $t$, and Hopf link invariants coloured by a $SU(N)$ representation $\nu$  are weighted by $t^{\vert \nu \vert}$. This quantity does not distinguish between different invariants with same $\vert \nu \vert$, and could at most be interpreted as an unrefined version of a generating function.

		\subsubsection{Higher rank theories}
				
				Expressions \eqref{eq:ZNfSchurW}-\eqref{eq:ZSUNfSchurW} assemble together coloured (unreduced) HOMFLY-PT polynomials for the unknot, with additional gradings related to the color $\nu$. Turning off the mass deformations associated to the torus $U(1)^{N_f}$ gives the ``Euler characteristic'', $t_j \to -1$.\par
				On a purely mathematical ground, the existence of a pair of Cauchy identities \eqref{eq:Cauchyid}-\eqref{eq:dualCauchy} invokes the definition of a new partition function $\mathcal{Z}^{\text{ferm.}} (\vec{m})$ for a given $\mathcal{Z} (\vec{m})$. Such a definition may not seem justified from the perspective of the original Chern--Simons-matter theory. Looking at the partition functions $\mathcal{Z}_{SU(N)}$ and $\mathcal{Z}^{\text{ferm.}} _{SU(N)}$ through the lenses of topological strings, though, it is in fact important to include $\mathcal{Z}^{\text{ferm.}} _{SU(N)}$, see for example \cite{Conifold:2006,Okuyama:2006eb}. Further aspects of the existence of such a pair of matrix models have been discussed in Subsection \ref{sec:ZUNfanddual}.\par
				The transposition map switching from one Cauchy identity to the other is lifted to the mirror map for knot homologies discussed in \cite{GukovStosic1,GukovStosic2}, with the two models being the generating functions of the mirror dual homologies. Note that the mirror symmetry under replacement $\nu \mapsto \nu^{\prime}$ holds for reduced knot invariants \cite{GukovStosic1}, while for a Hopf link the transposition map is not lifted to any symmetry. Moreover, the relation holds for knot invariants normalized by the unknot, hence such relations will not be grasped by a Schur expansion, in which only unknots and Hopf links appear.\par
				More concretely, the mirror symmetry discussed in \cite{GukovStosic1} categorifies the identity 
				\begin{equation*}
					\frac{\mathfrak{P}_{\nu} [\text{knot}] (a,q) }{\mathfrak{P}_{\nu} [\text{unknot}] (a,q) } = 	\frac{\mathfrak{P}_{\nu^{\prime}} [\text{knot}] (a, q^{-1} ) }{\mathfrak{P}_{\nu^{\prime}} [\text{unknot}] (a, q^{-1}) } 
				\end{equation*}
				satisfied by the reduced HOMFLY-PT polynomials. Here we have denoted $\mathfrak{P}_{\nu} [\text{knot}] (a,q)$ the (unreduced) HOMFLY-PT polynomial for a knot coloured by the representation $\nu$. The corresponding knot invariant is obtained setting $a=q^{N}$. Since this identity holds upon normalization by the unknot invariant, and in our description we can only obtain (products of) unknots or Hopf links, the picture provided by the Schur expansion does not retain any footprint of the mirror symmetry discovered in \cite{GukovStosic1}. That the map $\nu \mapsto \nu^{\prime}$ is not a symmetry was already clear from Subsection \ref{sec:ZUNfanddual}, because it switches from a theory of (zero-dimensional) bosons to a theory of (zero-dimensional) fermions.\par	
				\medskip
			
			A simple example with interesting features is the non-Abelian theory with a single flavour, $N_f=1$. In that case, the expansion would be 
			\begin{equation}
			\label{eq:ZNf1knot}
				\frac{\mathcal{Z} (q,t) }{ \mathcal{Z}_{\mathrm{CS}} (q) } \stackrel{\text{pert.}}{=} \sum_{\nu \in S_{N}} t^{\vert \nu \vert } \left\langle W_{\nu} \right\rangle_{\mathrm{CS}} ,
			\end{equation}
			where the sum runs over symmetric representations. We have not specified whether the theory has gauge group $U(N)$ or $SU(N)$ since the result is analogous, up to the minor modifications mentioned in Subsection \ref{sec:SchurexpUNNf} above. We have also slightly changed the notation, compared to Subsection \ref{sec:SchurexpUNNf}, to stress the dependence on the two parameters $(q,t)$. Trading the Cauchy identity \eqref{eq:Cauchyid} for the dual one \eqref{eq:dualCauchy} replaces 
			\begin{equation*}
				\left\langle W_{\nu} \right\rangle_{\mathrm{CS}}  \mapsto \left\langle W_{\nu^{\prime}} \right\rangle_{\mathrm{CS}} . 
			\end{equation*}
			We change the summation domain in \eqref{eq:ZNf1knot} from $S_N$, the symmetric representations, to $\Lambda_{N}$, the antisymmetric representations, and use that $\nu \in S_{N}$ implies $\nu^{\prime} \in \Lambda_{N}$ to write 
			\begin{equation*}
				\frac{\mathcal{Z}^{\text{ferm.}} (q,t) }{ \mathcal{Z}_{\mathrm{CS}} (q) } = \sum_{\nu \in \Lambda_{N}} t^{\vert \nu \vert } \left\langle W_{\nu} \right\rangle_{\mathrm{CS}} .
			\end{equation*}
			The breakdown of the mirror symmetry for the unreduced unknot invariants can be seen explicitly at this level, since we obtain $\mathcal{Z}^{\text{ferm.}} (q,t) \ne \mathcal{Z} (q,t)$. The difference is that $\mathcal{Z}^{\text{ferm.}} (q,t)$ generates averages of elementary symmetric polynomials $\mathfrak{e}_{\nu}$ in the CS matrix model, cfr. \eqref{eq:genele}, whilst $\mathcal{Z} (q,t)$ generates averages of homogeneous symmetric polynomials $\mathfrak{h}_{\nu}$, cfr. \eqref{eq:genhomo}.\par
			\medskip
			Keeping the discussion on general grounds, consider fugacities $\left\{t_j \right\}$ for the torus $U(1)^{N_{\text{global}}}$ of a global symmetry and dynamical variables $\left\{ X_a \right\}$ as fugacities of a $U(1)^{N_{\text{gauge}}}$ gauge symmetry, parametrizing the holonomy of the gauge connection along a knot $\mathscr{K}$ embedded in $\mathbb{S}^3$. The sum 
			\begin{equation}
			\label{eq:CauchyGauge}
				\sum_{\nu} \mathfrak{s}_{\nu} (t_1, \dots, t_{N_{\text{global}}}) \mathfrak{s}_{\nu} (X_1, \dots, X_{N_{\text{gauge}}}) 
			\end{equation}
			inserted into the path integral of Chern--Simons theory gives a generating-like function of Wilson loops for the knot $\mathscr{K}$, coloured by representations $\nu$. This holds in general, although arbitrary knots do not admit a simple matrix model description. Therefore, a transposition map sending \eqref{eq:CauchyGauge} to 
			\begin{equation}
			\label{eq:CauchyGDual}
				\sum_{\nu} \mathfrak{s}_{\nu} (t_1, \dots, t_{N_{\text{global}}}) \mathfrak{s}_{\nu^{\prime}} (X_1, \dots, X_{N_{\text{gauge}}}) 
			\end{equation}
			switches from one Cauchy identity to the other, and is a remnant of the mirror map of \cite{GukovStosic1,GukovStosic2}. Since it maps Schur polynomials into Schur polynomials, this is a map of unreduced knot invariants, and is not a duality symmetry.

		 \section{Explicit Schur expansions for selected quivers}
		 \label{app:AbelianquiverSchurexp}
		 
		 In this Appendix we collect explicit Schur expansions for Abelian necklace quivers with $r+1$ nodes, described in Subsection \ref{sec:NecklaceSchurExp}. We adopt, as usual, the definition \eqref{defq} $q= e^{- i \frac{2 \pi }{k}}$.\par
		 When $r=3$, as in Figure \ref{fig:Mcrystal4}, and for $\vec{k}= (k,-k,k,-k)$ formula \eqref{AbelianneckScurformula} gives  
	{\small \begin{align*}
		- \frac{\mz_{\widehat{\mathsf{A}}_3} (\vec{k} , \vec{m})}{ \left( \mz_{\mathrm{CS} (1); k } \right)^2  \left( \mz_{\mathrm{CS} (1); -k } \right)^2 }  \stackrel{\text{pert.}}{=} & t_1^3 t_2^3 t_3^3 t_0^3+t_1^2 t_2^3 t_3^3 t_0^3+t_1 t_2^3 t_3^3 t_0^3+t_1^3 t_2^2 t_3^3 t_0^3 +\frac{t_1^2 t_2^2 t_3^3 t_0^3}{q}+\frac{t_1 t_2^2 t_3^3 t_0^3}{q^2} \\ 
   & +t_1^3 t_2 t_3^3 t_0^3  + \frac{t_1^2 t_2 t_3^3 t_0^3}{q^2}+ \frac{t_1 t_2 t_3^3 t_0^3}{q^4}+t_1^3 t_2^3 t_3^2 t_0^3+t_1^2 t_2^3 t_3^2 t_0^3+t_1 t_2^3 t_3^2 t_0^3  \\ 
   & +q  t_1^3 t_2^2 t_3^2 t_0^3+t_1^2 t_2^2 t_3^2 t_0^3 +\frac{t_1 t_2^2 t_3^2 t_0^3}{q} +q^2 t_1^3 t_2 t_3^2 t_0^3+t_1^2 t_2 t_3^2 t_0^3  \\ 
   & +\frac{t_1 t_2 t_3^2 t_0^3}{q^2} +t_1^3 t_2^3 t_3 t_0^3+t_1^2 t_2^3 t_3 t_0^3+t_1 t_2^3 t_3 t_0^3 +q^2 t_1^3 t_2^2 t_3 t_0^3  \\ 
   & +q t_1^2 t_2^2 t_3 t_0^3+t_1 t_2^2 t_3 t_0^3+q^4 t_1^3 t_2 t_3 t_0^3 +q^2 t_1^2 t_2 t_3 t_0^3+t_1 t_2 t_3 t_0^3  \\
   & +t_1^3 t_2^3 t_3^3 t_0^2+ q t_1^2 t_2^3 t_3^3 t_0^2 +q^2 t_1 t_2^3 t_3^3 t_0^2+t_1^3 t_2^2 t_3^3 t_0^2+t_1^2 t_2^2 t_3^3 t_0^2+t_1 t_2^2 t_3^3 t_0^2 \\
   & +t_1^3 t_2 t_3^3 t_0^2+\frac{t_1^2 t_2 t_3^3 t_0^2}{q}  +\frac{t_1 t_2 t_3^3 t_0^2}{q^2} +\frac{t_1^3 t_2^3 t_3^2 t_0^2}{q}+t_1^2 t_2^3 t_3^2 t_0^2 \\
   & + q t_1 t_2^3 t_3^2 t_0^2+t_1^3 t_2^2 t_3^2 t_0^2+t_1^2 t_2^2 t_3^2 t_0^2+t_1 t_2^2 t_3^2 t_0^2+q t_1^3 t_2 t_3^2 t_0^2+t_1^2 t_2 t_3^2 t_0^2 \\
   & +\frac{t_1 t_2 t_3^2 t_0^2}{q}+\frac{t_1^3 t_2^3 t_3 t_0^2}{q^2}+ \frac{t_1^2 t_2^3 t_3 t_0^2}{q} +t_1 t_2^3 t_3 t_0^2+t_1^3 t_2^2 t_3 t_0^2+t_1^2 t_2^2 t_3 t_0^2 \\ 
   & +t_1 t_2^2 t_3 t_0^2 +q^2 t_1^3 t_2 t_3 t_0^2+ q t_1^2 t_2 t_3 t_0^2+t_1 t_2 t_3 t_0^2+t_1^3 t_2^3 t_3^3 t_0 \\ 
   & + q^2 t_1^2 t_2^3 t_3^3 t_0+e q^4 t_1 t_2^3 t_3^3 t_0+t_1^3 t_2^2 t_3^3 t_0+ q t_1^2 t_2^2 t_3^3 t_0  + q^2 t_1 t_2^2 t_3^3 t_0 \\
   & +t_1^3 t_2 t_3^3 t_0+t_1^2 t_2 t_3^3 t_0+t_1 t_2 t_3^3 t_0+\frac{t_1^3 t_2^3 t_3^2 t_0}{q^2} +t_1^2 t_2^3 t_3^2 t_0 + q^2 t_1 t_2^3 t_3^2 t_0  \\
   & +\frac{t_1^3 t_2^2 t_3^2 t_0}{q} +t_1^2 t_2^2 t_3^2 t_0+ q t_1 t_2^2 t_3^2 t_0+t_1^3 t_2 t_3^2 t_0+t_1^2 t_2 t_3^2 t_0+t_1 t_2 t_3^2 t_0 \\
   & +\frac{t_1^3 t_2^3 t_3 t_0}{q^4} + \frac{t_1^2 t_2^3 t_3 t_0}{q^2} +t_1 t_2^3 t_3 t_0+\frac{t_1^3 t_2^2 t_3 t_0}{q^2}  \\ 
   & +\frac{t_1^2 t_2^2 t_3 t_0}{q} +t_1 t_2^2 t_3 t_0+t_1^3 t_2 t_3 t_0+t_1^2 t_2 t_3 t_0+t_1 t_2 t_3 t_0 \ + O \left( t_p ^4 \right) .
	\end{align*}}\par
	The same example, but with all equal CS levels $\vec{k}=(k,k,k,k)$ reads 
	{\small \begin{align*}
	  - \frac{\mz_{\widehat{\mathsf{A}}_3} (\vec{k} , \vec{m})}{ \left( \mz_{\mathrm{CS} (1); k } \right)^4 }  \stackrel{\text{pert.}}{=} &  t_0 t_1^3 t_2 t_3^3 q^4+t_0^3 t_1 t_2^3 t_3 q^4+t_0 t_1^2 t_2 t_3^3 q^3+t_0 t_1^3 t_2 t_3^2 q^3 +t_0^2 t_1 t_2^3 t_3 q^3 +t_0^3 t_1 t_2^2 t_3 q^3  \\ 
	  & +t_0 t_1 t_2 t_3^3 q^2  +t_0 t_1^2 t_2 t_3^2 q^2+t_0 t_1 t_2^3 t_3 q^2+t_0^2 t_1 t_2^2 t_3
   q^2 +t_0 t_1^3 t_2 t_3 q^2 \\
   & +t_0^3 t_1 t_2 t_3 q^2+t_0 t_1^3 t_2^2 t_3^3 q  +t_0 t_1^2 t_2^2 t_3^3 q+t_0 t_1 t_2^2 t_3^3
   q+t_0^2 t_1^3 t_2 t_3^3 q +t_0^2 t_1^2 t_2 t_3^3 q  \\
   & +t_0^2 t_1 t_2 t_3^3 q+t_0^3 t_1 t_2^3 t_3^2 q +t_0^2 t_1 t_2^3 t_3^2
   q+t_0 t_1 t_2^3 t_3^2 q+t_0 t_1^3 t_2^2 t_3^2 q +t_0 t_1^2 t_2^2 t_3^2 q \\
   & +t_0^3 t_1 t_2^2 t_3^2 q+t_0^2 t_1 t_2^2 t_3^2
   q+t_0 t_1 t_2^2 t_3^2 q +t_0^2 t_1^3 t_2 t_3^2 q+t_0^2 t_1^2 t_2 t_3^2 q+t_0^3 t_1 t_2 t_3^2 q \\
   & +t_0^2 t_1 t_2 t_3^2 q+t_0
   t_1 t_2 t_3^2 q+t_0^3 t_1^2 t_2^3 t_3 q  +t_0^2 t_1^2 t_2^3 t_3 q+t_0 t_1^2 t_2^3 t_3 q+t_0 t_1^3 t_2^2 t_3 q \\
   & +t_0^3 t_1^2
   t_2^2 t_3 q +t_0^2 t_1^2 t_2^2 t_3 q+t_0 t_1^2 t_2^2 t_3 q+t_0 t_1 t_2^2 t_3 q+t_0^2 t_1^3 t_2 t_3 q +t_0^3 t_1^2 t_2 t_3
   q \\
   & +t_0^2 t_1^2 t_2 t_3 q+t_0 t_1^2 t_2 t_3 q +t_0^2 t_1 t_2 t_3 q+t_0 t_1 t_2^3 t_3^3+t_0^2 t_1 t_2^2 t_3^3 +t_0^3 t_1 t_2 t_3^3 \\
   & +t_0 t_1^2 t_2^3 t_3^2+t_0^2 t_1^2 t_2^2 t_3^2+t_0^3 t_1^2 t_2 t_3^2  +t_0 t_1^3 t_2^3 t_3+t_0^2 t_1^3 t_2^2
   t_3+t_0^3 t_1^3 t_2 t_3+t_0 t_1 t_2 t_3 \\
   & +\frac{t_0 t_1^2 t_2^3 t_3^3}{q}+\frac{t_0^2 t_1 t_2^3 t_3^3}{q}+\frac{t_0^2
   t_1^2 t_2^2 t_3^3}{q} +\frac{t_0^3 t_1 t_2^2 t_3^3}{q}+\frac{t_0^3 t_1^2 t_2 t_3^3}{q}+\frac{t_0 t_1^3 t_2^3
   t_3^2}{q}  \\
   & +\frac{t_0^2 t_1^2 t_2^3 t_3^2}{q}  +\frac{t_0^2 t_1^3 t_2^2 t_3^2}{q}+\frac{t_0^3 t_1^2 t_2^2
   t_3^2}{q}+\frac{t_0^3 t_1^3 t_2 t_3^2}{q}+\frac{t_0^2 t_1^3 t_2^3 t_3}{q}+\frac{t_0^3 t_1^3 t_2^2 t_3}{q} \\ & +\frac{t_0
   t_1^3 t_2^3 t_3^3}{q^2}  +\frac{t_0^3 t_1 t_2^3 t_3^3}{q^2}+\frac{t_0^2 t_1^3 t_2^2 t_3^3}{q^2}+\frac{t_0^3 t_1^3 t_2
   t_3^3}{q^2}+\frac{t_0^3 t_1^2 t_2^3 t_3^2}{q^2} +\frac{t_0^3 t_1^3 t_2^3 t_3}{q^2}  \\ 
   & +\frac{t_0^2 t_1^2 t_2^3
   t_3^3}{q^3}+\frac{t_0^3 t_1^2 t_2^2 t_3^3}{q^3}+\frac{t_0^2 t_1^3 t_2^3 t_3^2}{q^3}  +\frac{t_0^3 t_1^3 t_2^2
   t_3^2}{q^3}+\frac{t_0^2 t_1^3 t_2^3 t_3^3}{q^5}+\frac{t_0^3 t_1^2 t_2^3 t_3^3}{q^5} \\
   & +\frac{t_0^3 t_1^3 t_2^2
   t_3^3}{q^5}+\frac{t_0^3 t_1^3 t_2^3 t_3^2}{q^5}+\frac{t_0^3 t_1^3 t_2^3 t_3^3}{q^8} \ + O \left( t_p ^4 \right) .
	\end{align*}}\par
	Going to one rank higher, the Abelian $\widehat{\mathsf{A}}_4$ theory with all equal CS levels has the Schur expansion 
	
		{\small \begin{align*}
		- \frac{\mz_{\widehat{\mathsf{A}}_3} (\vec{k} , \vec{m})}{ \left( \mz_{\mathrm{CS} (1); k } \right)^5  }  \stackrel{\text{pert.}}{=} & t_0 t_1 t_2^3 t_3 t_4^3 q^4+t_0 t_1^3 t_2 t_3 t_4^3 q^4+t_0 t_1^3 t_2 t_3^3 t_4 q^4+t_0^3 t_1 t_2 t_3^3 t_4 q^4+t_0^3 t_1
   t_2^3 t_3 t_4 q^4+t_0 t_1^3 t_2 t_3^2 t_4^3 q^3 \\ & +t_0 t_1^2 t_2^3 t_3 t_4^3 q^3+t_0^2 t_1 t_2^3 t_3 t_4^3 q^3+t_0 t_1^3
   t_2^2 t_3 t_4^3 q^3+t_0 t_1^2 t_2^2 t_3 t_4^3 q^3+t_0 t_1 t_2^2 t_3 t_4^3 q^3 + t_0 t_1^2 t_2 t_3 t_4^3 q^3\\ & +t_0 t_1^3 t_2
   t_3^3 t_4^2 q^3+t_0 t_1^3 t_2 t_3^2 t_4^2 q^3+t_0^3 t_1 t_2^3 t_3 t_4^2 q^3+t_0^2 t_1 t_2^3 t_3 t_4^2 q^3+t_0 t_1 t_2^3
   t_3 t_4^2 q^3+t_0 t_1^3 t_2 t_3 t_4^2 q^3 \\ & +t_0^3 t_1 t_2^2 t_3^3 t_4 q^3+t_0^2 t_1^3 t_2 t_3^3 t_4 q^3+t_0^3 t_1^2 t_2
   t_3^3 t_4 q^3+t_0^2 t_1^2 t_2 t_3^3 t_4 q^3+t_0 t_1^2 t_2 t_3^3 t_4 q^3+t_0^2 t_1 t_2 t_3^3 t_4 q^3 \\ & +t_0^3 t_1 t_2^3
   t_3^2 t_4 q^3+t_0^3 t_1 t_2^2 t_3^2 t_4 q^3+t_0 t_1^3 t_2 t_3^2 t_4 q^3+t_0^3 t_1 t_2 t_3^2 t_4 q^3+t_0^2 t_1 t_2^3 t_3
   t_4 q^3+t_0^3 t_1 t_2^2 t_3 t_4 q^3 \\ & +t_0 t_1^3 t_2 t_3^3 t_4^3 q^2+t_0 t_1^2 t_2 t_3^2 t_4^3 q^2+t_0 t_1^3 t_2^3 t_3
   t_4^3 q^2+t_0^3 t_1 t_2^3 t_3 t_4^3 q^2+t_0^2 t_1 t_2^2 t_3 t_4^3 q^2+t_0 t_1 t_2 t_3 t_4^3 q^2 \\ & +t_0 t_1^2 t_2 t_3^3
   t_4^2 q^2+t_0 t_1^2 t_2 t_3^2 t_4^2 q^2+t_0 t_1^2 t_2^3 t_3 t_4^2 q^2+t_0 t_1^3 t_2^2 t_3 t_4^2 q^2+t_0 t_1^2 t_2^2 t_3
   t_4^2 q^2+t_0^3 t_1 t_2^2 t_3 t_4^2 q^2 \\ & +t_0^2 t_1 t_2^2 t_3 t_4^2 q^2+t_0 t_1 t_2^2 t_3 t_4^2 q^2+t_0 t_1^2 t_2 t_3
   t_4^2 q^2+t_0^3 t_1 t_2^3 t_3^3 t_4 q^2+t_0^2 t_1 t_2^2 t_3^3 t_4 q^2+t_0^3 t_1^3 t_2 t_3^3 t_4 q^2 \\ & +t_0 t_1 t_2 t_3^3
   t_4 q^2+t_0^2 t_1 t_2^3 t_3^2 t_4 q^2+t_0^2 t_1 t_2^2 t_3^2 t_4 q^2+t_0^2 t_1^3 t_2 t_3^2 t_4 q^2+t_0^3 t_1^2 t_2 t_3^2
   t_4 q^2+t_0^2 t_1^2 t_2 t_3^2 t_4 q^2 \\ & +t_0 t_1^2 t_2 t_3^2 t_4 q^2+t_0^2 t_1 t_2 t_3^2 t_4 q^2+t_0 t_1 t_2^3 t_3 t_4
   q^2+t_0^2 t_1 t_2^2 t_3 t_4 q^2+t_0 t_1^3 t_2 t_3 t_4 q^2+t_0^3 t_1 t_2 t_3 t_4 q^2 \\ & +t_0 t_1^2 t_2 t_3^3 t_4^3 q+t_0 t_1
   t_2^3 t_3^2 t_4^3 q+t_0 t_1^3 t_2^2 t_3^2 t_4^3 q+t_0 t_1^2 t_2^2 t_3^2 t_4^3 q+t_0 t_1 t_2^2 t_3^2 t_4^3 q+t_0 t_1 t_2
   t_3^2 t_4^3 q \\ & +t_0^2 t_1^2 t_2^3 t_3 t_4^3 q+t_0^2 t_1^2 t_2^2 t_3 t_4^3 q+t_0^3 t_1 t_2^2 t_3 t_4^3 q+t_0^2 t_1^3 t_2
   t_3 t_4^3 q+t_0^2 t_1^2 t_2 t_3 t_4^3 q+t_0^2 t_1 t_2 t_3 t_4^3 q \\ & +t_0^2 t_1^3 t_2 t_3^3 t_4^2 q+t_0^2 t_1^2 t_2 t_3^3
   t_4^2 q+t_0^3 t_1 t_2 t_3^3 t_4^2 q+t_0^2 t_1 t_2 t_3^3 t_4^2 q+t_0 t_1 t_2 t_3^3 t_4^2 q+t_0^3 t_1 t_2^3 t_3^2 t_4^2
   q \\ & +t_0^2 t_1 t_2^3 t_3^2 t_4^2 q+t_0 t_1 t_2^3 t_3^2 t_4^2 q+t_0 t_1^3 t_2^2 t_3^2 t_4^2 q+t_0 t_1^2 t_2^2 t_3^2 t_4^2
   q+t_0^3 t_1 t_2^2 t_3^2 t_4^2 q+t_0^2 t_1 t_2^2 t_3^2 t_4^2 q \\ & +t_0 t_1 t_2^2 t_3^2 t_4^2 q+t_0^2 t_1^3 t_2 t_3^2 t_4^2
   q+t_0^2 t_1^2 t_2 t_3^2 t_4^2 q+t_0^3 t_1 t_2 t_3^2 t_4^2 q+t_0^2 t_1 t_2 t_3^2 t_4^2 q+t_0 t_1 t_2 t_3^2 t_4^2 q \\ & +t_0
   t_1^3 t_2^3 t_3 t_4^2 q+t_0^2 t_1^2 t_2^3 t_3 t_4^2 q+t_0^2 t_1^2 t_2^2 t_3 t_4^2 q+t_0^2 t_1^3 t_2 t_3 t_4^2 q+t_0^2
   t_1^2 t_2 t_3 t_4^2 q+t_0^3 t_1 t_2 t_3 t_4^2 q \\ & +t_0^2 t_1 t_2 t_3 t_4^2 q+t_0 t_1 t_2 t_3 t_4^2 q+t_0^2 t_1 t_2^3 t_3^3
   t_4 q+t_0 t_1^3 t_2^2 t_3^3 t_4 q+t_0^3 t_1^2 t_2^2 t_3^3 t_4 q+t_0^2 t_1^2 t_2^2 t_3^3 t_4 q \\ & +t_0 t_1^2 t_2^2 t_3^3 t_4
   q+t_0 t_1 t_2^2 t_3^3 t_4 q+t_0 t_1 t_2^3 t_3^2 t_4 q+t_0 t_1^3 t_2^2 t_3^2 t_4 q+t_0^3 t_1^2 t_2^2 t_3^2 t_4 q+t_0^2
   t_1^2 t_2^2 t_3^2 t_4 q  \\ & +t_0 t_1^2 t_2^2 t_3^2 t_4 q+t_0 t_1 t_2^2 t_3^2 t_4 q+t_0^3 t_1^3 t_2 t_3^2 t_4 q+t_0 t_1 t_2
   t_3^2 t_4 q+t_0^3 t_1^2 t_2^3 t_3 t_4 q+t_0^2 t_1^2 t_2^3 t_3 t_4 q \\ & +t_0 t_1^2 t_2^3 t_3 t_4 q+t_0 t_1^3 t_2^2 t_3 t_4
   q+t_0^3 t_1^2 t_2^2 t_3 t_4 q+t_0^2 t_1^2 t_2^2 t_3 t_4 q+t_0 t_1^2 t_2^2 t_3 t_4 q+t_0 t_1 t_2^2 t_3 t_4 q \\ & +t_0^2 t_1^3
   t_2 t_3 t_4 q +t_0^3 t_1^2 t_2 t_3 t_4 q+t_0^2 t_1^2 t_2 t_3 t_4 q+t_0 t_1^2 t_2 t_3 t_4 q+t_0^2 t_1 t_2 t_3 t_4 q+t_0
   t_1 t_2 t_3^3 t_4^3 \\ & +t_0 t_1^2 t_2^3 t_3^2 t_4^3 +t_0^2 t_1 t_2^3 t_3^2 t_4^3+t_0^2 t_1 t_2^2 t_3^2 t_4^3+t_0^2 t_1^3 t_2
   t_3^2 t_4^3+t_0^2 t_1^2 t_2 t_3^2 t_4^3+t_0^2 t_1 t_2 t_3^2 t_4^3 \\ & +t_0^2 t_1^3 t_2^2 t_3 t_4^3+t_0^3 t_1 t_2 t_3
   t_4^3+t_0 t_1^3 t_2^2 t_3^3 t_4^2+t_0 t_1^2 t_2^2 t_3^3 t_4^2+t_0^3 t_1 t_2^2 t_3^3 t_4^2+t_0^2 t_1 t_2^2 t_3^3
   t_4^2 \\ & +t_0 t_1 t_2^2 t_3^3 t_4^2+t_0^3 t_1^2 t_2 t_3^3 t_4^2+t_0 t_1^2 t_2^3 t_3^2 t_4^2+t_0^2 t_1^2 t_2^2 t_3^2
   t_4^2+t_0^3 t_1^2 t_2 t_3^2 t_4^2+t_0^3 t_1^2 t_2^3 t_3 t_4^2 \\ &  +t_0^2 t_1^3 t_2^2 t_3 t_4^2+t_0^3 t_1^2 t_2^2 t_3
   t_4^2+t_0^3 t_1^2 t_2 t_3 t_4^2+t_0 t_1 t_2^3 t_3^3 t_4+t_0^2 t_1^3 t_2^2 t_3^3 t_4+t_0^3 t_1^2 t_2^3 t_3^2 t_4 \\ & +t_0^2
   t_1^2 t_2^3 t_3^2 t_4+t_0 t_1^2 t_2^3 t_3^2 t_4+t_0^2 t_1^3 t_2^2 t_3^2 t_4+t_0 t_1^3 t_2^3 t_3 t_4+t_0^2 t_1^3 t_2^2
   t_3 t_4+t_0^3 t_1^3 t_2 t_3 t_4 +t_0 t_1 t_2 t_3 t_4 \\ & +\frac{t_0 t_1^3 t_2^2 t_3^3 t_4^3}{q}+\frac{t_0 t_1^2 t_2^2 t_3^3
   t_4^3}{q}+\frac{t_0 t_1 t_2^2 t_3^3 t_4^3}{q}+\frac{t_0^2 t_1^3 t_2 t_3^3 t_4^3}{q}+\frac{t_0^2 t_1^2 t_2 t_3^3
   t_4^3}{q}+\frac{t_0^2 t_1 t_2 t_3^3 t_4^3}{q} \\ & +\frac{t_0 t_1^3 t_2^3 t_3^2 t_4^3}{q}+\frac{t_0^3 t_1 t_2^3 t_3^2
   t_4^3}{q}+\frac{t_0^2 t_1^2 t_2^2 t_3^2 t_4^3}{q}+\frac{t_0^3 t_1 t_2^2 t_3^2 t_4^3}{q}+\frac{t_0^3 t_1 t_2 t_3^2
   t_4^3}{q}+\frac{t_0^2 t_1^3 t_2^3 t_3 t_4^3}{q}+\frac{t_0^3 t_1^2 t_2^3 t_3 t_4^3}{q} \\ & +\frac{t_0^3 t_1^2 t_2^2 t_3
   t_4^3}{q}+\frac{t_0^3 t_1^2 t_2 t_3 t_4^3}{q}+\frac{t_0^3 t_1 t_2^3 t_3^3 t_4^2}{q}+\frac{t_0^2 t_1 t_2^3 t_3^3
   t_4^2}{q}+\frac{t_0 t_1 t_2^3 t_3^3 t_4^2}{q}+\frac{t_0^2 t_1^2 t_2^2 t_3^3 t_4^2}{q}+\frac{t_0^3 t_1^3 t_2 t_3^3
   t_4^2}{q} \\ & +\frac{t_0 t_1^3 t_2^3 t_3^2 t_4^2}{q}+\frac{t_0^2 t_1^2 t_2^3 t_3^2 t_4^2}{q}+\frac{t_0^2 t_1^3 t_2^2 t_3^2
   t_4^2}{q}+\frac{t_0^3 t_1^2 t_2^2 t_3^2 t_4^2}{q}+\frac{t_0^3 t_1^3 t_2 t_3^2 t_4^2}{q}+\frac{t_0^2 t_1^3 t_2^3 t_3
   t_4^2}{q}+\frac{t_0^3 t_1^3 t_2 t_3 t_4^2}{q} \\ & +\frac{t_0^3 t_1^2 t_2^3 t_3^3 t_4}{q}+\frac{t_0^2 t_1^2 t_2^3 t_3^3
   t_4}{q}+\frac{t_0 t_1^2 t_2^3 t_3^3 t_4}{q}+\frac{t_0^3 t_1^3 t_2^2 t_3^3 t_4}{q}+\frac{t_0 t_1^3 t_2^3 t_3^2
   t_4}{q}+\frac{t_0^3 t_1^3 t_2^2 t_3^2 t_4}{q}+\frac{t_0^2 t_1^3 t_2^3 t_3 t_4}{q} \\ & +\frac{t_0^3 t_1^3 t_2^2 t_3
   t_4}{q}+\frac{t_0 t_1 t_2^3 t_3^3 t_4^3}{q^2}+\frac{t_0^2 t_1 t_2^2 t_3^3 t_4^3}{q^2}+\frac{t_0^3 t_1 t_2 t_3^3
   t_4^3}{q^2}+\frac{t_0^2 t_1^2 t_2^3 t_3^2 t_4^3}{q^2}+\frac{t_0^2 t_1^3 t_2^2 t_3^2 t_4^3}{q^2}+\frac{t_0^3 t_1^2 t_2
   t_3^2 t_4^3}{q^2} \\ & +\frac{t_0^3 t_1^3 t_2 t_3 t_4^3}{q^2}+\frac{t_0 t_1^2 t_2^3 t_3^3 t_4^2}{q^2}+\frac{t_0^2 t_1^3 t_2^2
   t_3^3 t_4^2}{q^2}+\frac{t_0^3 t_1^2 t_2^2 t_3^3 t_4^2}{q^2}+\frac{t_0^3 t_1^2 t_2^3 t_3^2 t_4^2}{q^2}+\frac{t_0^3 t_1^3
   t_2^2 t_3 t_4^2}{q^2}+\frac{t_0 t_1^3 t_2^3 t_3^3 t_4}{q^2} \end{align*}
   \begin{align*} & +\frac{t_0^2 t_1^3 t_2^3 t_3^2 t_4}{q^2}+\frac{t_0^3 t_1^3
   t_2^3 t_3 t_4}{q^2}+\frac{t_0 t_1^2 t_2^3 t_3^3 t_4^3}{q^3}+\frac{t_0^2 t_1 t_2^3 t_3^3 t_4^3}{q^3}+\frac{t_0^2 t_1^2
   t_2^2 t_3^3 t_4^3}{q^3}+\frac{t_0^3 t_1 t_2^2 t_3^3 t_4^3}{q^3}+\frac{t_0^3 t_1^2 t_2 t_3^3 t_4^3}{q^3} \\ & +\frac{t_0^3
   t_1^2 t_2^2 t_3^2 t_4^3}{q^3}+\frac{t_0^3 t_1^3 t_2 t_3^2 t_4^3}{q^3}+\frac{t_0^3 t_1^3 t_2^2 t_3 t_4^3}{q^3}+\frac{t_0
   t_1^3 t_2^3 t_3^3 t_4^2}{q^3}+\frac{t_0^2 t_1^2 t_2^3 t_3^3 t_4^2}{q^3}+\frac{t_0^2 t_1^3 t_2^3 t_3^2
   t_4^2}{q^3}+\frac{t_0^3 t_1^3 t_2^2 t_3^2 t_4^2}{q^3} \\ &  +\frac{t_0^3 t_1^3 t_2^3 t_3 t_4^2}{q^3}+\frac{t_0^2 t_1^3 t_2^3
   t_3^3 t_4}{q^3}+\frac{t_0^3 t_1^3 t_2^3 t_3^2 t_4}{q^3}+\frac{t_0 t_1^3 t_2^3 t_3^3 t_4^3}{q^4}+\frac{t_0^3 t_1 t_2^3
   t_3^3 t_4^3}{q^4}+\frac{t_0^2 t_1^3 t_2^2 t_3^3 t_4^3}{q^4}+\frac{t_0^3 t_1^3 t_2 t_3^3 t_4^3}{q^4} \\ &  +\frac{t_0^2 t_1^3
   t_2^3 t_3^2 t_4^3}{q^4}+\frac{t_0^3 t_1^2 t_2^3 t_3^2 t_4^3}{q^4}+\frac{t_0^3 t_1^3 t_2^3 t_3 t_4^3}{q^4}+\frac{t_0^3
   t_1^2 t_2^3 t_3^3 t_4^2}{q^4}+\frac{t_0^3 t_1^3 t_2^2 t_3^3 t_4^2}{q^4}+\frac{t_0^3 t_1^3 t_2^3 t_3^3
   t_4}{q^4}+\frac{t_0^2 t_1^2 t_2^3 t_3^3 t_4^3}{q^5} \\ &  +\frac{t_0^3 t_1^2 t_2^2 t_3^3 t_4^3}{q^5}+\frac{t_0^3 t_1^3 t_2^2
   t_3^2 t_4^3}{q^5}+\frac{t_0^2 t_1^3 t_2^3 t_3^3 t_4^2}{q^5}+\frac{t_0^3 t_1^3 t_2^3 t_3^2 t_4^2}{q^5}+\frac{t_0^2 t_1^3
   t_2^3 t_3^3 t_4^3}{q^7}+\frac{t_0^3 t_1^2 t_2^3 t_3^3 t_4^3}{q^7}+\frac{t_0^3 t_1^3 t_2^2 t_3^3 t_4^3}{q^7} \\ &  +\frac{t_0^3
   t_1^3 t_2^3 t_3^2 t_4^3}{q^7}+\frac{t_0^3 t_1^3 t_2^3 t_3^3 t_4^2}{q^7}+\frac{t_0^3 t_1^3 t_2^3 t_3^3 t_4^3}{q^{10}} \ + O \left( t_p ^4 \right) .
		\end{align*}}\par
		For arbitrary $r$ the Schur expansion of Abelian necklace quivers is easily obtained to very high order with a computer algebra.

	\clearpage
	\bibliographystyle{ourstyle}
	\bibliography{ExactCSbiblio}

\end{document}